\documentclass[3p,times,number,sort&compress]{elsarticle}

\usepackage{amsmath,amssymb,amsfonts}
\usepackage{xcolor}
\usepackage{array}
\usepackage{textcomp}
\usepackage{url}
\usepackage{graphicx}
\usepackage{booktabs}
\usepackage{tabularx}
\usepackage{ragged2e}
\usepackage{tikz}
\usetikzlibrary{positioning,matrix,arrows.meta,calc,fit,backgrounds}
\usepackage[hidelinks,pdfpagelabels=false]{hyperref}

\journal{Sustainable Computing: Informatics and Systems}

\newcolumntype{Y}{>{\raggedright\arraybackslash}X}
\newcolumntype{C}{>{\centering\arraybackslash}X}

\newcommand{\tblsetup}{%
  \footnotesize
  \setlength{\tabcolsep}{3.2pt}%
  \renewcommand{\arraystretch}{1.07}%
  \emergencystretch=1.5em
  \sloppy
}

\newcommand{\rev}[1]{#1}
\begin{document}

\begin{frontmatter}

\title{Energy- and Thermal-Aware Resource Management on Mobile IoT Endpoints: A Survey}

\author[yc]{Yuchao Chang\corref{cor1}}
\ead{yuchaoc@mail.ustc.edu.cn}
\author[hw]{Haoran Wei}
\ead{haoranwei@foxmail.com}
\cortext[cor1]{Corresponding author. Accepted Manuscript, to appear in Sustainable Computing: Informatics and Systems. DOI: https://doi.org/10.1016/j.suscom.2026.101508 License: CC BY‑NC‑ND, not the publisher‑formatted Journal Pre‑proof.}

\affiliation[yc]{organization={Performance and Power Group, OPPO},
            addressline={Shanghai},
            city={Shanghai},
            postcode={200030},
            country={China}}

\affiliation[hw]{organization={Rivian and Volkswagen Group Technologies},
            city={Palo Alto, CA},
            postcode={94304},
            country={USA}}

\begin{abstract}
Energy-aware and thermal-aware management of computing resources on portable and mobile Internet-of-Things (IoT) endpoints is still practiced pillar by pillar: radios, processors, and flash are optimized in isolation, yet sealed system-on-chip (SoC) products share sleep states, interconnects, and thermal limits. Pillar-local joule or temperature gains therefore often disappear after integration, and published savings are frequently incomparable. This survey asks how radio, compute, and storage should be coordinated so that deep sleep remains reachable, and which reporting rules make endpoint energy and thermal claims comparable. We synthesize literature from roughly 2000 to early 2026 with three instruments: a five-phase wireless power anatomy for power-aware networking; a four-branch taxonomy for collective power--performance optimization under mixed regimes; and eight evaluation gates that separate energy outcome, control cost, thermal feasibility, and evidence quality. We qualify the radio-limited premise with compute-, storage-, and thermal-dominated counterexamples, compare modeling and monitoring families, \rev{treat thermal behaviour as a shared constraint by separating sensing interfaces, compact models, throttling actuators, and the short-peak versus sustained-throttling regimes that reverse room-temperature rankings}, \rev{give an illustrative phase-budget example with a sensitivity check and a model-selection workflow}, and close with open problems on composability and reproducible measurement. The intended readers work on sustainable computing, mobile systems, and embedded energy management.
\end{abstract}

\begin{keyword}
sustainable computing \sep energy-aware computing \sep thermal-aware management \sep mobile IoT \sep power-aware networking \sep energy measurement
\end{keyword}

\end{frontmatter}

\section{Introduction}
\label{sec:intro}

Mobile Internet-of-Things (IoT) endpoints---smartphones, wearables, and gateway hubs---combine cellular, Wi-Fi, Bluetooth Low Energy (BLE), and low-power wide-area (LPWA) radios with heterogeneous processors and non-volatile storage under tight battery and thermal budgets~\cite{alfuqaha2015iot,andreev2015connectivity,shi2016edge}. Classic measurements already showed that radios, idle listening, and protocol tails often dominate daily energy~\cite{carroll2010smartphone,pathak2012whereenergy}. The contemporary difficulty is sharper: discontinuous reception (DRX), modem--application-processor (AP) handshakes, on-device inference, and over-the-air (OTA) maintenance interact, so a ``faster'' kernel or a quieter file-system API can \emph{increase} total joules by delaying deep sleep or aligning NAND program/erase with radio activity~\cite{mittal2014embedded,patterson2024energy,chang2025flashhost}. Without shared reporting norms, two papers can both claim ``energy efficiency'' while answering incomparable questions.

\textbf{Central thesis.} On sealed mobile SoCs, endpoint energy is a \emph{composition} problem: radio, compute, and storage share time bases, interconnects, and heat, so a mechanism is energy-efficient only if it preserves (or enlarges) eligibility for deep sleep under the dominant coordination regime~\cite{patterson2024energy,xie2021apollo,andreev2015connectivity}. We therefore treat ``low power'' as a claim about joint schedules and evidence quality, not about isolated PHY rates, MAC counts, or file-system bytes.

\textbf{Place in sustainable computing.} The object of study is energy-aware and thermal-aware management of computing resources on portable and mobile platforms. Four themes structure the review: (i)~power implications for portable and mobile computing; (ii)~power-aware networking as the temporal envelope of sleep; (iii)~power management in memory, disk, and other storage; and (iv)~models and monitoring tools for collective power--performance optimization under shared thermal limits. Radio is treated as a duty-cycle and tail constraint, not as a protocol tutorial.

\textbf{Problem.} Most surveys still silo the pillars. Wireless and LPWA reviews emphasize PHY--MAC timers~\cite{andreev2015connectivity,centenaro2016longrange,diane2025lpwan}; compute-power surveys target CPUs and accelerators~\cite{reda2012survey,zhai2024mlsurvey}; flash surveys address write amplification and host placement~\cite{chang2025flashhost,mittal2014embedded}; edge and MEC surveys stress offloading~\cite{mao2017mobileedge,shi2016edge,dong2024mecsurvey}. Closely related network-side surveys on ICN-IoT caching~\cite{zhang2023icniot} and edge orchestration~\cite{chiang2023edgeorch,kong2023edgeiot} improve platform efficiency but under-specify modem--application-processor (AP) tails, multi-radio contention on one die, and NAND maintenance that redefine ``idle'' on the handset itself. The resulting \textbf{coordination gap} is methodological: without a shared vocabulary for phases, regimes, and measurement gates, numerical comparisons across papers are often incomparable~\cite{xie2021apollo,patterson2024energy,reda2012survey}.

\rev{\textbf{The gap this survey fills.} Prior surveys each cover a pillar well and stop at that pillar's boundary. None of them supplies, in one place, (i)~a rule that assigns a mixed phone or wearable trace to a dominant coordination regime, (ii)~a gate list that separates energy outcome from control cost and from evidence quality, and (iii)~a reporting checklist for sealed devices on which per-rail counters are missing. That joint instrument, not another catalogue of DRX timers or DVFS knobs, is the gap. Table~\ref{tab:survey_compare} records where earlier surveys stop.}

\textbf{What this survey is (and is not).} It is a sustainable-computing survey of energy-aware and thermal-aware resource management on sealed mobile IoT endpoints: we organize known results, compare modeling families critically, and extract reusable evaluation instruments. It is \emph{not} a protocol tutorial, a chip datasheet digest, or an expanded conference paper. Where radio appears first in the narrative, it is because radio policy often sets the temporal envelope of sleep---the scarce resource that compute and storage must not violate; Section~\ref{sec:scope_limits} lists compute-, display-, sensing-, storage-, and thermal-dominated counterexamples.

\textbf{Approach.} Evidence is organized with (1)~a five-phase wireless power anatomy, (2)~four coordination branches with mixed-regime decision rules, and (3)~eight evaluation gates separating outcome, cost, and evidence quality. Running examples, an illustrative quantitative phase-budget example, and a model-family workflow show how to apply the instruments under sealed-device observability limits. \rev{Numerical illustrations are labeled measured, modeled, estimated, or illustrative so that they are not read as product experiments.}

\textbf{Contributions.}
\begin{itemize}
\item \textbf{Conceptual:} a qualified energy- and thermal-aware thesis with counterexamples, four-branch taxonomy, and eight gates that make joule and temperature claims falsifiable across papers (Sections~\ref{sec:logic_framework}--\ref{sec:evaldim}).
\item \textbf{Methodological:} a documented search and coding protocol and classification matrix (Section~\ref{sec:lit_method}, Tables~\ref{tab:lit_corpus}--\ref{tab:corpus_quant}).
\item \textbf{Critical synthesis:} matched-depth pillar surveys that emphasize joint failure modes---modem--AP tails, DMA/LLC contention, FTL/OTA sleep blocking---rather than catalogues (Sections~\ref{sec:lpcomm}--\ref{sec:lpstore},~\ref{sec:lessons}).
\item \textbf{Operational:} running examples, an illustrative phase-budget example with sensitivity ranges, a model-selection workflow, and an observation contract plus open-problem roadmap (Sections~\ref{sec:case_study}--\ref{sec:open_problems}).
\end{itemize}

\textbf{Historical remarks.} Endpoint energy accounting matured from early smartphone instrumentation~\cite{carroll2010smartphone,pathak2012whereenergy}, through LPWA/cellular IoT connectivity surveys~\cite{andreev2015connectivity,centenaro2016longrange}, to edge offloading, on-device learning, and host-side flash management~\cite{shi2016edge,zhai2024mlsurvey,chang2025flashhost}. Measurement practice improved; \emph{cross-pillar composition theory and reporting norms} did not keep pace---the gap this survey addresses.

\section{Outline of the Survey}
\label{sec:outline}

The remainder of the survey is organized as follows.
\begin{enumerate}
\item \textbf{Introduction} (Section~\ref{sec:intro}): motivation, problem statement, contributions, and historical remarks.
\item \textbf{Problem space and related surveys} (Section~\ref{sec:survey_positioning}): positioning against prior IoT, wireless, edge, compute-power, and flash surveys, and three linked research gaps.
\item \textbf{Literature methodology} (Section~\ref{sec:lit_method}): documented search, screening, and classification protocol with a representative matrix.
\item \textbf{Basic concepts, definitions, and instruments} (Sections~\ref{sec:logic_framework}--\ref{sec:evaldim}): coordination framework, scope limits and counterexamples, running examples, five-phase wireless power anatomy, and eight evaluation gates.
\item \textbf{Review of existing approaches} (Sections~\ref{sec:lpcomm}--\ref{sec:lpstore}): matched-depth surveys of power-aware networking, energy- and thermal-aware compute, and energy-aware storage, emphasizing joint failure modes rather than isolated catalogues.
\item \textbf{Tutorial application} (Sections~\ref{sec:case_study}--\ref{sec:model_decision}): illustrative quantitative phase-budget example and model-family selection workflow.
\item \textbf{Critical review and comparison} (Section~\ref{sec:lessons}): relevance of results, comparison of approaches, and joint-optimization lessons.
\item \textbf{Open problems} (Section~\ref{sec:open_problems}): observation contract and research roadmap.
\item \textbf{Conclusion} (Section~\ref{sec:conclusion}): closing synthesis for energy-aware and thermal-aware mobile computing.
\end{enumerate}

\section{Positioning Against Prior Surveys}
\label{sec:survey_positioning}

Prior surveys are strong within their scopes but leave a \textbf{coordination gap} for sealed mobile IoT, where modem policy, heterogeneous compute, and flash maintenance share time, energy, and thermal budgets. IoT enabling-technology surveys~\cite{alfuqaha2015iot,xia2012internet} catalog protocols and applications yet treat energy peripherally. \rev{The same is true of the foundational IoT literature that defined the field's scope: vision and architecture papers fixed the vocabulary of devices, gateways, and services~\cite{atzori2010socialiot,miorandi2012internet,gubbi2013iotvision}, deployment studies worked out what that architecture means for a city~\cite{zanella2014smartcities}, and research agendas set the open problems for the decade that followed~\cite{stankovic2014iotresearch}. Their cross-cutting concerns---context acquisition~\cite{perera2014contextsurvey} and security and trust~\cite{sicari2015securitysurvey}---are treated as functional requirements rather than as consumers of the endpoint's energy and sleep budget, which is the reading this survey applies to them.} \rev{The offload literature follows a parallel arc: cloudlets, fog, and standardized mobile-edge platforms progressively moved computation closer to the device~\cite{satyanarayanan2009casecloudlets,dastjerdi2016fogcomputing,hu2015mecwhite}, and partitioning systems made the split point adaptive at inference time~\cite{kang2017neurosurgeon,li2018edgent}. Those systems optimize latency and device compute under an assumed network cost. On a sealed endpoint the assumption is the difficulty: the uplink that carries the intermediate tensor opens a modem tail and blocks the sleep state that the offload was supposed to reach, so the same split can be a saving or a loss depending on the coordination regime and the declared boundary (Gates~2 and~8).} Edge-centric surveys on ICN-IoT caching~\cite{zhang2023icniot}, management and orchestration of edge computing for IoT~\cite{chiang2023edgeorch}, and edge-computing-driven IoT~\cite{kong2023edgeiot} deepen network and platform design but remain complementary to our \emph{endpoint SoC} focus: they do not systematically attribute modem--AP tails, multi-RAT contention, and NAND maintenance on the handset or wearable itself. Wireless and LPWA surveys~\cite{andreev2015connectivity,centenaro2016longrange,diane2025lpwan,islam2024lpwan,gu2020lpwatechsurvey} emphasize PHY--MAC and standards timers but rarely link DRX tails to AP retention or NAND program bursts. Mobile-edge surveys~\cite{mao2017mobileedge,porambage2018meceiot,shi2016edge,friha2024llmedge,dong2024mecsurvey} stress radio--compute tradeoffs at the network edge yet under-specify modem--AP coupling and on-device persistence. Green IoT computing surveys~\cite{alsharif2024massiveiot} broaden scheduling, offloading, and compression mechanisms but give endpoint radio states and NAND maintenance less weight than handset-class coordination requires. Compute power-model surveys~\cite{reda2012survey,kumar2023mlmicro,zhai2024mlsurvey} target processors without duty-cycled radios or FTL-induced sleep blocking. Flash surveys~\cite{chang2025flashhost,patterson2024energy,mittal2014embedded} address write amplification and placement but seldom analyze OTA or checkpoint storms against paging windows. \textbf{This survey} organizes energy-aware and thermal-aware endpoint management around \textbf{sleep-envelope coordination}: radio phases bound when compute and storage may act without breaking deep sleep or the thermal cap; eight gates test whether joule and temperature claims remain comparable; four branches name the temporal regime for each mechanism. Table~\ref{tab:survey_compare} summarizes the positioning.

\textbf{Explicit differentiation from recent complementary surveys.} Zhang et al.~\cite{zhang2023icniot} survey in-network caching for ICN-IoT: their unit of analysis is the network cache and content path, not sealed modem--AP sleep coupling or NAND program bursts on the UE. Chiang and Zhang~\cite{chiang2023edgeorch} survey management and orchestration of edge computing for IoT: they deepen control-plane and platform lifecycle issues at the edge, whereas we ask whether an endpoint-local mechanism preserves deep-sleep eligibility under multi-RAT and FTL activity. Kong et al.~\cite{kong2023edgeiot} survey edge-computing-driven IoT architectures and applications: their primary axis is edge--cloud functional placement; ours is on-device composition of radio, compute, and storage energy under shared hardware constraints. These lines are complementary: network/edge surveys answer \emph{where} to place functions; this survey answers \emph{how to manage and compare} endpoint energy and thermal claims when placement decisions collide with modem tails, flash maintenance, and shared heat.

{
\begin{table*}[!t]

\renewcommand{\arraystretch}{1.08}
\caption{Positioning against prior surveys (qualitative; $\checkmark$ = primary, $\circ$ = partial, -- = minimal). R/C/S = radio/compute/storage. X = cross-pillar coupling. Th = thermal modeling. OTA = maintenance/OTA. Sl = sleep-state coordination. Ms = measurement methodology. Rp = reproducibility framework. Tx = taxonomy stated and given an assignment procedure.}
\label{tab:survey_compare}
\centering
\tblsetup\scriptsize
\begin{tabular*}{\textwidth}{@{\extracolsep{\fill}}>{\raggedright\arraybackslash}p{0.20\textwidth}*{10}{>{\centering\arraybackslash}p{0.038\textwidth}}@{}}
\toprule
\textbf{Survey focus} & \textbf{R} & \textbf{C} & \textbf{S} & \textbf{X} & \textbf{Th} & \textbf{OTA} & \textbf{Sl} & \textbf{Ms} & \textbf{Rp} & \textbf{Tx} \\
\midrule
IoT enabling tech.~\cite{alfuqaha2015iot} & $\checkmark$ & $\circ$ & -- & -- & -- & -- & $\circ$ & -- & -- & -- \\
Wireless / LPWA~\cite{andreev2015connectivity,diane2025lpwan} & $\checkmark$ & -- & -- & $\circ$ & -- & -- & $\checkmark$ & $\circ$ & $\circ$ & $\circ$ \\
Mobile / MEC edge~\cite{mao2017mobileedge,shi2016edge,friha2024llmedge} & $\checkmark$ & $\checkmark$ & -- & $\circ$ & $\circ$ & -- & $\circ$ & $\circ$ & -- & -- \\
Compute power models~\cite{reda2012survey,zhai2024mlsurvey} & -- & $\checkmark$ & $\circ$ & -- & $\circ$ & -- & -- & $\checkmark$ & $\circ$ & -- \\
Host-side flash~\cite{chang2025flashhost} & -- & $\circ$ & $\checkmark$ & $\circ$ & -- & $\checkmark$ & $\circ$ & $\circ$ & -- & -- \\
Embedded efficiency~\cite{mittal2014embedded} & $\circ$ & $\checkmark$ & $\circ$ & $\circ$ & $\circ$ & $\circ$ & $\checkmark$ & $\circ$ & -- & -- \\
M2M MAC taxonomy~\cite{rajandekar2015m2mmac} & $\checkmark$ & -- & -- & -- & -- & -- & $\checkmark$ & -- & -- & $\checkmark$ \\
Green IoT / edge~\cite{alsharif2024massiveiot,dong2024mecsurvey} & $\circ$ & $\checkmark$ & -- & $\circ$ & $\circ$ & -- & $\circ$ & -- & -- & -- \\
ICN-IoT caching~\cite{zhang2023icniot} & $\circ$ & -- & $\checkmark$ & $\circ$ & -- & -- & -- & -- & -- & $\checkmark$ \\
Edge orchestration~\cite{chiang2023edgeorch} & $\circ$ & $\checkmark$ & -- & $\circ$ & -- & $\circ$ & -- & -- & -- & $\checkmark$ \\
Edge-driven IoT~\cite{kong2023edgeiot} & $\circ$ & $\checkmark$ & -- & $\circ$ & -- & -- & -- & -- & -- & -- \\
\textbf{This survey} & $\checkmark$ & $\checkmark$ & $\checkmark$ & $\checkmark$ & $\checkmark$ & $\checkmark$ & $\checkmark$ & $\checkmark$ & $\checkmark$ & $\checkmark$ \\
\bottomrule
\end{tabular*}
\end{table*}
}

\textbf{Research-gap chain.} Three linked gaps motivate the paper structure. \emph{Gap~1 (mechanism silos):} compute, radio, and storage optimizations are published separately, yet shipping SoCs share buses, sleep states, and thermal envelopes, so pillar-level gains often vanish after integration~\cite{patterson2024energy,xie2021apollo}. \emph{Gap~2 (measurement mismatch):} inconsistent sensors, boundaries, and phase segmentation undermine cross-study comparison~\cite{patterson2024energy,reda2012survey,hackenberg2013rapl,bocus2025superiotriot,bocus2025netiotenergy}. \emph{Gap~3 (regime blindness):} one device may traverse all four coordination branches within a day; averages without branch labels mis-rank policies~\cite{wunderlich2003simpoints,rajandekar2015m2mmac}. Sections~\ref{sec:comm_power}--\ref{sec:lpstore} address these gaps through shared wireless vocabulary, evaluation gates, and pillar synthesis under the four-branch taxonomy.

\rev{\textbf{Broader coordination literature, outside the endpoint.} A separate line of work coordinates mobile or transportable storage in hydrogen hubs, eco-industrial parks, city energy communities, and offshore clusters, often with Stackelberg games, digital twins, fractional-order control, or learning-based schedules~\cite{shayeghi2026dth2,zare2027eip,seyedshenava2026scooter,suscom2025freq,mohammadian2026fleet,zare2026interactive}. Those papers are about grid-side or hub-side flexibility and stakeholder welfare. They are not measurements of modem tails, NAND program energy, or sealed-SoC thermal coupling, and they are not used as such. The distinction is the unit of control: they coordinate fleets, parks, and energy communities; this survey coordinates radio, compute, and flash inside one sealed endpoint. Game-structured or digitally twinned hub coordination does not replace the branch, gate, and boundary checklist above.}

\section{Literature Search and Selection Methodology}
\label{sec:lit_method}

Because this survey spans wireless, compute, and storage from circa 2000 through early 2026, we document a \textbf{selection protocol} intended to support reconstruction. The goal is not bibliometric exhaustiveness---mobile IoT energy literature is fragmented across COMSOC, architecture, and systems venues---but a \textbf{traceable corpus} that supports the classification matrix in Table~\ref{tab:lit_matrix}.

\rev{\textbf{Article type.} This is a critical narrative survey with a documented search protocol. It is not a registered systematic review and not a meta-analysis: no protocol was pre-registered, no risk-of-bias instrument was applied to the included works, and no effect sizes are pooled---the primary outcomes in this literature (joules, milliwatts, temperature) are reported against device-specific and boundary-specific baselines that are not commensurable enough to average. Table~\ref{tab:prisma} and Figure~\ref{fig:prisma_flow} borrow the PRISMA \emph{flow presentation} because it is the clearest way to show which screening stages have a defendable count and which do not; that borrowing is a transparency device and should not be read as a claim of PRISMA compliance.}

\textbf{Databases and query window.} Primary searches used IEEE Xplore, ACM Digital Library, and Web of Science (January 2000--March 2026). Supplementary forward/backward citation chaining and Google Scholar alerts captured edge-computing, flash-management, and on-device LLM work published outside communications venues. arXiv and industry white papers were retained only when they introduced mechanisms not yet peer-reviewed; such entries are tagged \textbf{R} (preprint) or \textbf{I} (industry/standards) in Table~\ref{tab:lit_matrix}.

\textbf{Keyword clusters (Boolean OR within cluster; AND across clusters).}
\begin{itemize}
\item \emph{Device context:} mobile IoT, smartphone, wearable, gateway, M2M, NB-IoT, LTE-M, RedCap, LPWA, BLE.
\item \emph{Energy objective:} low power, energy efficiency, energy-aware, thermal-aware, battery life, duty cycle, DRX, eDRX, PSM, power model, PMIC.
\item \emph{Subsystem:} application processor, modem, NAND, flash, FTL, OTA, edge computing, on-device inference, SLM, TinyML.
\item \emph{Coordination:} cross-layer, joint scheduling, offload, contention, composability, power-aware networking, storage power management.
\end{itemize}

\rev{\textbf{Exact query and search record.} The same three-cluster Boolean string was issued on IEEE Xplore, the ACM Digital Library, and Web of Science. Coverage is 1~January 2000 through 31~March 2026; the last database update and the last citation-chaining pass were completed in March 2026. Google Scholar alerts over the same window were used only to surface items missed by the three databases, not as a fourth hit-count source.}

{\par\nopagebreak
\noindent\raggedright\footnotesize\ttfamily\emergencystretch=2em
("mobile IoT" OR smartphone OR wearable OR gateway OR M2M OR "NB-IoT" OR "LTE-M" OR RedCap OR LPWA OR BLE OR "Internet of Things")
AND ("low power" OR "energy efficiency" OR "energy-aware" OR "thermal-aware" OR "battery life" OR "duty cycle" OR DRX OR eDRX OR PSM OR "power model" OR PMIC)
AND (modem OR "application processor" OR NAND OR flash OR FTL OR OTA OR "edge computing" OR "on-device" OR composability OR "power-aware networking")
\par}

\rev{Database interfaces rewrite this string (field tags, stemming, phrase handling), so a later search will not reproduce a vendor's hit integer even with the same words. The same words were issued as an IEEE Xplore All-Metadata query, an ACM Digital Library advanced search, and a Web of Science Topic search. Final search date for all three was 31~March 2026. Deduplication used the DOI when present and otherwise a normalized title plus year; a journal version replaced its preprint, and each retained work received one category in Table~\ref{tab:lit_corpus}. Raw hit logs, including the count removed as duplicates, were not archived, so those integers are not imputed. The phrase ``several thousand candidate records'' in the previous version referred to that unlogged pool and is withdrawn as a funnel statistic. Table~\ref{tab:prisma} reports only stages whose counts we can defend.}

\textbf{Inclusion criteria.} (i)~Endpoint or gateway-class devices with sealed or product-like firmware; (ii)~energy, power, or battery-life as an explicit outcome; (iii)~at least one of compute, radio, or storage treated with enough detail to map to a pillar; (iv)~English-language journal, conference, standard, or widely cited technical report.

\textbf{Exclusion criteria.} Pure data-center energy without endpoint implications; PHY-only link budgets with no system sleep or maintenance context; single-knob accelerator papers with no connectivity or persistence coupling; duplicate preprint/journal versions (journal version kept).

\rev{\textbf{Screening workflow.} Inclusion and exclusion were applied in two passes (title/abstract, then full text). Three populations must not be conflated. The \emph{screened pool} is everything returned by the queries, alerts, and citation chaining; its vendor hit total was not archived (Table~\ref{tab:prisma}). The \textbf{core synthesis set} is the 186 category assignments in Table~\ref{tab:lit_corpus} ($58{+}47{+}31{+}28{+}22$) used to build the branch, gate, and model-family synthesis. The \emph{cited set} is the reference list (136 entries): core works that are discussed in the text, plus supporting citations for definitions, standards, and boundary context. Those 136 entries are not a head-count of the core set, which is why the bibliography is shorter than 186. Table~\ref{tab:lit_matrix} is an 18-paper slice of the core set, not a fourth population. Table~\ref{tab:corpus_quant} summarizes the 186 by category and the 18-row matrix by branch, gate, device class, and boundary. Figure~\ref{fig:prisma_flow} shows the same flow in PRISMA form, with the two unarchived stages marked as missing rather than filled in. Full branch and gate tallies for all 186 were not frozen as a separate spreadsheet; only the category assignment of the 186 and the double-coded 18-row matrix are reported.}

\rev{\textbf{Coding procedure and agreement.} Branch and gate labels were produced by combined coding. The asynchronous, synchronous, and frame-slotted seeds were taken deductively from prior MAC and rendezvous taxonomies~\cite{rajandekar2015m2mmac}; the multichannel-heterogeneous branch was added inductively because phone and gateway traces in the core set run parallel radios that those taxonomies treat as separate networks. Assignment then follows Table~\ref{tab:branch_decision}: one dominant branch per phase, with secondary regimes recorded as multi-label metadata. Both authors independently applied that procedure to the 18-paper representative matrix (Table~\ref{tab:lit_matrix}); disagreements were resolved by consensus before the table was frozen. The remaining core-set entries were coded by one author and checked against the same rules. The full set of 186 was not double-coded, and item-level disagreement logs were not retained, so we do not report Cohen's kappa or a percentage agreement. That is a limitation of the classification evidence. Another researcher can still reproduce a label by applying Table~\ref{tab:branch_decision} to a phase description. A gate label is assigned when the paper's primary claim, reported metric, or stated failure mode turns on that row of Table~\ref{tab:eval_gates}; a paper may carry more than one gate. On the 18-row matrix the two authors compared labels, listed disagreements, and kept a label only after consensus. No disagreement tally was stored, so the adjudication record is the frozen table, not a kappa.}

\begin{table}[!t]
\renewcommand{\arraystretch}{1.08}
\caption{Composition of the core synthesis set used in this survey (coded through March 2026; counts are category assignments of included works, not database hit totals).}
\label{tab:lit_corpus}
\centering
\tblsetup\begin{tabularx}{\columnwidth}{@{}>{\raggedright\arraybackslash}p{0.38\columnwidth}cc@{}}
\toprule
\textbf{Category} & \textbf{Included} & \textbf{In Table~\ref{tab:lit_matrix}} \\
\midrule
Low-power communication / LPWA / cellular IoT & 58 & 7 \\
Compute / accelerator / on-device ML & 47 & 4 \\
Storage / flash / OTA / host-side management & 31 & 1 \\
Cross-pillar / measurement / composability & 28 & 4 \\
Edge / MEC / split inference & 22 & 2 \\
\midrule
\textbf{Total (core synthesis set)} & \textbf{186} & \textbf{18} \\
\bottomrule
\end{tabularx}
\end{table}

\rev{The third column of Table~\ref{tab:lit_corpus} counts rows printed in Table~\ref{tab:lit_matrix} (earlier drafts used a planning target of 44; the printed matrix has 18 representative works). Category totals for the 186 remain unchanged.}

{
\begin{table}[!t]

\renewcommand{\arraystretch}{1.08}
\caption{Compact quantitative overview. Part~A: category distribution of the 186-work core set. Part~B: coded attributes of the 18-row representative matrix (a work may carry more than one gate; multi-label branches are counted under each named regime). Part~C: the branch and gate profile that each category carries by construction of the inclusion rules; Part~C is a category-level reading, not an item-level tally of the 186.}
\label{tab:corpus_quant}
\centering
\tblsetup\scriptsize
\begin{tabularx}{\columnwidth}{@{}>{\raggedright\arraybackslash}p{0.42\columnwidth}Y@{}}
\toprule
\textbf{Attribute} & \textbf{Count} \\
\midrule
\multicolumn{2}{@{}l}{\emph{A. Core set ($n{=}186$) by category}} \\
Low-power communication / LPWA / cellular IoT & 58 \\
Compute / accelerator / on-device ML & 47 \\
Storage / flash / OTA / host-side & 31 \\
Cross-pillar / measurement / composability & 28 \\
Edge / MEC / split inference & 22 \\
\midrule
\multicolumn{2}{@{}l}{\emph{B. Representative matrix ($n{=}18$)}} \\
Source type P / R / I / S & 13 / 3 / 1 / 1 \\
Dominant device: phone or phone/edge / wear.\ or module / other & 8 / 5 / 5 \\
Branch mentions: frame / multi-RAT / async / sync / all & 8 / 6 / 5 / 4 / 2 \\
Gate mentions: 1 / 2 / 3 / 4 / 5 / 6 / 7 / 8 & 12 / 7 / 3 / 7 / 8 / 4 / 2 / 6 \\
Boundary: device-local / split or edge-inclusive / other & 11 / 4 / 3 \\
\midrule
\multicolumn{2}{@{}l}{\emph{C. Category profile ($n{=}186$): branches and gates each category carries}} \\
Communication / LPWA / cellular IoT (58) & Frame, async; Gates 1, 2, 5 \\
Compute / accelerator / on-device ML (47) & Sync, frame; Gates 1, 6, 7 \\
Storage / flash / OTA / host-side (31) & Async, frame; Gates 1, 3, 4 \\
Cross-pillar / measurement / composability (28) & All four; Gates 4, 5, 8 \\
Edge / MEC / split inference (22) & Multi-RAT, sync; Gates 2, 8 \\
\bottomrule
\end{tabularx}
\end{table}
}

\rev{Part~C of Table~\ref{tab:corpus_quant} answers the request for a corpus-level view without overstating what the coding log supports. It reports, per category, the coordination regimes and gates that membership in that category implies under the inclusion rules of this section: a paper enters the LPWA and cellular-IoT category because its energy account turns on standards-imposed paging or duty cycles, which is what makes frame-slotted and Gates~1, 2, and~5 the defining stresses of that group. It is therefore a statement about the categories, not a per-item tally of 186 rows. The only item-level tallies we can defend are Part~A (one category per included work) and Part~B (the double-coded 18-row slice). We do not report a branch~$\times$~gate crosstab over all 186, because that spreadsheet was never frozen and reconstructing it after the fact would not be the coding that produced the synthesis.}

{
\begin{table}[!t]

\renewcommand{\arraystretch}{1.08}
\caption{Screening stages that have a defendable count. Vendor hit totals were not archived and are not imputed.}
\label{tab:prisma}
\centering
\tblsetup\begin{tabularx}{\columnwidth}{@{}>{\raggedright\arraybackslash}p{0.42\columnwidth}Y@{}}
\toprule
\textbf{Stage} & \textbf{What is reported} \\
\midrule
Databases and window & IEEE Xplore, ACM DL, Web of Science; 1 Jan 2000--31 Mar 2026; chaining and Scholar alerts supplementary \\
Query & Boolean string in Section~\ref{sec:lit_method}; identical words on all three databases \\
Records identified & Not archived; no imputed funnel total \\
Title/abstract exclusions & Count not archived (limitation) \\
Full-text exclusions & Count not archived (limitation) \\
Core synthesis set & 186 category assignments (Table~\ref{tab:lit_corpus}) \\
Representative matrix & 18 works (Table~\ref{tab:lit_matrix}), consensus-resolved \\
Cited in this manuscript & 136 reference-list entries \\
\bottomrule
\end{tabularx}
\end{table}
}

{
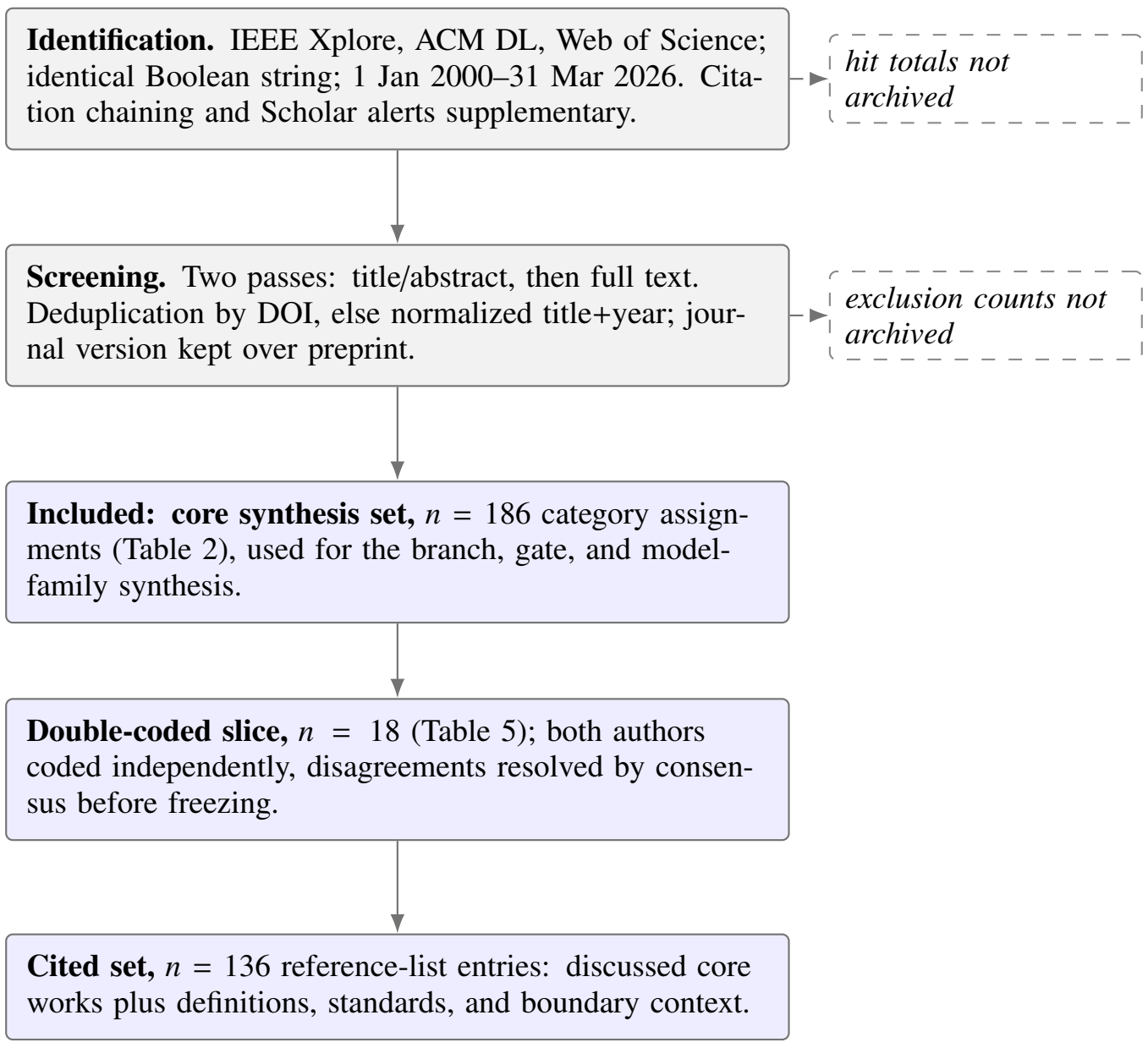
\begin{figure}[!t]
\centering
\begin{tikzpicture}[
  font=\scriptsize,
  box/.style={draw=black!55, rounded corners=1.5pt, align=left, text width=58mm, inner sep=1.6mm},
  stage/.style={box, fill=black!5},
  keep/.style={box, fill=blue!7},
  gap/.style={draw=black!45, dashed, rounded corners=1.5pt, align=left, text width=22mm, inner sep=1.2mm, font=\scriptsize\itshape},
  ar/.style={-{Latex[length=1.5mm]}, black!55}
]
\node[stage] (id) at (0,0)
  {\textbf{Identification.} IEEE Xplore, ACM DL, Web of Science; identical Boolean string;
   1 Jan 2000--31 Mar 2026. Citation chaining and Scholar alerts supplementary.};
\node[stage] (sc) at (0,-1.85)
  {\textbf{Screening.} Two passes: title/abstract, then full text. Deduplication by DOI, else
   normalized title${+}$year; journal version kept over preprint.};
\node[keep]  (co) at (0,-3.7)
  {\textbf{Included: core synthesis set, $n=186$} category assignments (Table~\ref{tab:lit_corpus}),
   used for the branch, gate, and model-family synthesis.};
\node[keep]  (mx) at (0,-5.4)
  {\textbf{Double-coded slice, $n=18$} (Table~\ref{tab:lit_matrix}); both authors coded
   independently, disagreements resolved by consensus before freezing.};
\node[keep]  (ci) at (0,-7.1)
  {\textbf{Cited set, $n=136$} reference-list entries: discussed core works plus definitions,
   standards, and boundary context.};
\node[gap] (x1) at (4.6,0)     {hit totals not archived};
\node[gap] (x2) at (4.6,-1.85) {exclusion counts not archived};
\draw[ar] (id) -- (sc);
\draw[ar] (sc) -- (co);
\draw[ar] (co) -- (mx);
\draw[ar] (mx) -- (ci);
\draw[ar, dashed] (id.east) -- (x1.west);
\draw[ar, dashed] (sc.east) -- (x2.west);
\end{tikzpicture}
\caption{\rev{Screening and coding flow. Solid boxes are stages with a count we can defend; the dashed
annotations mark the two stages whose raw logs were not archived and are therefore reported as
missing rather than imputed.}}
\label{fig:prisma_flow}
\end{figure}
}

\rev{\textbf{Evidence classes used throughout.} Every numerical statement in this survey carries one of four classes, and the class is named wherever the number appears. \emph{Measured:} an instrumented reading reported by the cited work on real hardware (for example, the tail current of Pathak et al.~\cite{pathak2012whereenergy} or the component split of Carroll and Heiser~\cite{carroll2010smartphone}). \emph{Modeled:} the output of an analytical or simulated power model (Wattch-, McPAT-, gem5-, or DRAMPower-class~\cite{brooks2000wattch,li2009mcpat,binkert2011gem5,chandrasekar2012drampower}). \emph{Estimated:} a quantity derived by the cited authors from partial observation, such as a PMIC total apportioned to subsystems. \emph{Illustrative:} a magnitude chosen by us to demonstrate the instruments, with a literature anchor but no device measurement; Table~\ref{tab:case_budget} and the analytical family tables (Tables~\ref{tab:lpcomm_anal}, \ref{tab:lpcompute_anal}, \ref{tab:lpstore_anal}) are entirely of this class. Captions repeat the class so that a table can be read out of context.}

\rev{\textbf{Worked coding example.} Because no agreement statistic is available, the assignment rules themselves must carry the reproducibility burden, so we show them applied. Take Pathak et al.~\cite{pathak2012whereenergy}: the reported phases are user-initiated transfers with 3G tails, no fixed paging epoch governs them, and the tail is the largest single share, so Table~\ref{tab:branch_decision} yields \emph{asynchronous} as dominant with \emph{multi-RAT} as a secondary label (Wi-Fi and cellular both appear); the primary claim is a joule attribution obtained from fine-grained instrumentation, which tags Gates~1, 4, and~6. Take Andreev et al.~\cite{andreev2015connectivity}: the energy account is organized around DRX and paging cycles imposed by the network, so the dominant branch is \emph{frame-slotted}, and because the paper trades sleep depth against reachability under standards-visible timers, the gate labels are 1, 2, and~5. A third reader applying the same two tables to the same two papers should recover these labels; where they would not, the rule rather than the label is what needs revision.}

\noindent\textbf{Source-type legend (Table~\ref{tab:lit_matrix}).} \textbf{P} = peer-reviewed journal or conference; \textbf{S} = standard or specification (e.g., 3GPP); \textbf{I} = industry report or vendor measurement brief; \textbf{R} = preprint not yet peer-reviewed at survey cutoff.

\begin{table*}[!t]
\renewcommand{\arraystretch}{1.06}
\caption{Representative literature mapped to device class, workload, wireless technology, subsystem focus, measurement boundary, coordination branch, and evaluation gates (type codes: P/S/I/R per Section~\ref{sec:lit_method}).}
\label{tab:lit_matrix}
\centering
\tblsetup\scriptsize
\begin{tabularx}{\textwidth}{@{}
>{\raggedright\arraybackslash}p{0.13\textwidth}
c
>{\raggedright\arraybackslash}p{0.09\textwidth}
>{\raggedright\arraybackslash}p{0.11\textwidth}
>{\raggedright\arraybackslash}p{0.10\textwidth}
>{\raggedright\arraybackslash}p{0.13\textwidth}
>{\raggedright\arraybackslash}p{0.10\textwidth}
>{\raggedright\arraybackslash}p{0.09\textwidth}
>{\raggedright\arraybackslash}p{0.11\textwidth}
@{}}
\toprule
\textbf{Work} & \textbf{T} & \textbf{Device} & \textbf{Workload} & \textbf{Wireless} & \textbf{Focus} & \textbf{Boundary} & \textbf{Branch} & \textbf{Gates} \\
\midrule
Carroll \& Heiser~\cite{carroll2010smartphone} & P & Phone & Daily use & Cell.+Wi-Fi & Radio & Device PMIC & Multi-RAT & 1,4,8 \\
Pathak et al.~\cite{pathak2012whereenergy} & P & Phone & App mix & Multi-RAT & Radio+CPU & Fine-grained & Async./multi & 1,4,6 \\
Andreev et al.~\cite{andreev2015connectivity} & P & M2M/phone & Telemetry & LTE-M/NB-IoT & Radio & Device & Frame & 1,2,5 \\
Rajandekar \& Prasad~\cite{rajandekar2015m2mmac} & P & WSN/gateway & Periodic & Duty-cycled MAC & Radio & Node & Async./sync. & 1,3,6 \\
Mittal~\cite{mittal2014embedded} & P & Embedded SoC & Mixed & Multi-RAT & Compute+stor. & Device & All & 1,5,7 \\
Chen et al. (Eyeriss)~\cite{chen2016eyeriss} & P & Accelerator & Vision DNN & N/A (coupled) & Compute & Chip/sim. & Sync./framed & 1,6,8 \\
Shi et al.~\cite{shi2016edge} & P & Phone/gateway & Edge infer. & Wi-Fi/cell. & Compute+radio & Split device/edge & Multi-RAT & 2,8 \\
Patterson et al.~\cite{patterson2024energy} & {R} & Phone/edge & Mixed & Multi-RAT & Cross-pillar & Declared boundary & All & 1,4,5 \\
Chang et al.~\cite{chang2025flashhost} & P & Phone/edge & OTA/logs & Wi-Fi/cell. & Storage & Host+flash & Async./framed & 1,4,5 \\
Bocus et al.~\cite{bocus2025superiotriot} & {R} & Wear./gateway & Sensing+BLE & BLE+LPWA & Cross-pillar & Network+node & Async. & 1,4,8 \\
Lu et al.~\cite{lu2024slm} & R & Phone/wear. & On-device LLM & Wi-Fi/cell. & Compute+DRAM & Device & Sync./multi & 1,7,8 \\
Veedu \& Pekka~\cite{veedu2022redcap} & I & Module/wear. & Video IoT & 5G RedCap & Radio & Lab module & Frame & 1,2,5 \\
Joerke et al.~\cite{joerke2025redcap} & P & Wear./indust. & Telemetry/video & RedCap/eRedCap & Radio & Device/module & Frame & 1,2,4 \\
Centenaro et al.~\cite{centenaro2016longrange} & P & M2M & Sparse uplink & LPWA/cell. & Radio & Device & Frame/async. & 1,2,3 \\
Xie et al. (APOLLO)~\cite{xie2021apollo} & P & Phone SoC & Mixed & Multi-RAT & Cross-pillar & Bus/contention & Multi-RAT & 5,6 \\
Reda et al.~\cite{reda2012survey} & P & CPU/server & Benchmark & N/A & Compute model & Sim./silicon & Sync. & 4,5,6 \\
Dong et al.~\cite{dong2024mecsurvey} & P & Edge+UE & Offload & Multi-RAT & Compute+radio & Edge-inclusive & Multi-RAT & 2,8 \\
3GPP~\cite{3gpp24501,3gpp36304,3gpp36331} & S & UE & Idle telemetry & LTE/NB-IoT/NR & Radio policy & Standard timers & Frame & 2,3,5 \\
\bottomrule
\end{tabularx}
\end{table*}

\section{Basic Concepts: Energy-Aware Coordination Framework}
\label{sec:logic_framework}

\begin{figure*}[!t]
\centering
\resizebox{\textwidth}{!}{% Fig. logic_framework — inline TikZ (included from manuscript.tex)
\begin{tikzpicture}[
  font=\footnotesize,
  >=Stealth,
  arr/.style={->, draw=black!45, line width=0.45pt},
  badge/.style={circle, minimum size=6.5mm, inner sep=0pt,
                font=\scriptsize\bfseries, draw=black!45, line width=0.4pt},
  badgeL1/.style={badge, fill=blue!14},
  badgeL2/.style={badge, fill=orange!18},
  badgeL3/.style={badge, fill=blue!10},
  badgeL4/.style={badge, fill=gray!16},
  laybox/.style={rectangle, rounded corners=2.5pt, draw=black!50,
                 line width=0.4pt, align=center, inner sep=5pt},
  hubbox/.style={laybox, fill=orange!11, text width=4.2cm, minimum height=2.0cm},
  sidebox/.style={laybox, fill=green!7, text width=3.45cm, minimum height=1.8cm},
  spinebox/.style={laybox, fill=blue!5, text width=3.45cm, minimum height=1.7cm},
  banner/.style={laybox, fill=blue!6, text width=15.6cm, minimum height=1.0cm},
  footbox/.style={laybox, fill=gray!9, text width=15.6cm, minimum height=0.9cm},
  laytitle/.style={font=\small\bfseries, text=black!80, align=center}
]

% ----- Layer 1 -----
\node[badgeL1] (b1) at (0,0) {1};
\node[banner, right=2.5mm of b1, anchor=west] (l1) {%
  {\small\bfseries Layer 1: Mobile IoT context}\\[1pt]
  {\scriptsize Phones, wearables, gateway hubs \textbullet\ multi-RAT SoC \textbullet\ LPDDR \textbullet\ NAND}\\[0.5pt]
  {\scriptsize Sparse / always-on workloads \textbullet\ recharge, standby drift, thermal limits}};

% ----- Layer 2 -----
\node[laytitle, anchor=north] (t2) at ($(l1.south)+(0,-5mm)$)
  {{Layer 2: Default envelope, not a universal one}};

\node[hubbox, anchor=north] (comm) at ($(t2.south)+(0,-2mm)$) {%
  \textbf{Low-power communication}\\[1pt]
  {\scriptsize {Often the default envelope}}\\[0.5pt]
  {\scriptsize Sleep, tails, modem--AP coupling}};

\node[sidebox, left=2.5mm of comm] (comp) {%
  \textbf{Low-power compute}\\[1pt]
  {\scriptsize Coordination partner}\\[0.5pt]
  {\scriptsize Batching, DVFS, accelerators}};

\node[sidebox, right=2.5mm of comm] (stor) {%
  \textbf{Low-power storage}\\[1pt]
  {\scriptsize Coordination partner}\\[0.5pt]
  {\scriptsize NAND, FTL, OTA windows}};

\node[badgeL2] (b2) at (b1 |- comm) {2};

% ----- Layer 3 -----
\node[laytitle, anchor=north] (t3) at ($(comm.south)+(0,-5mm)$)
  {Layer 3: Shared methodological spine};

\node[spinebox, anchor=north] (tax) at ($(t3.south)+(0,-2mm)$) {%
  \textbf{Four-branch taxonomy}\\[1pt]
  {\scriptsize Async, sync}\\[0.5pt]
  {\scriptsize Frame-slotted, multi-RAT}};

\node[spinebox, left=2.5mm of tax] (ana) {%
  \textbf{Power anatomy}\\[1pt]
  {\scriptsize {Sec.~\ref{sec:comm_power}}}\\[0.5pt]
  {\scriptsize Sleep, idle, RX/TX, tails}};

\node[spinebox, right=2.5mm of tax] (gate) {%
  \textbf{Eight gates}\\[1pt]
  {\scriptsize {Sec.~\ref{sec:evaldim}}}\\[0.5pt]
  {\scriptsize Outcome, cost, truth}};

\node[badgeL3] (b3) at (b1 |- tax) {3};

% ----- Layer 4 -----
\node[footbox, anchor=north] (l4) at ($(tax.south)+(0,-5.5mm)$) {%
  {\small\bfseries Layer 4: Mechanism and evidence synthesis}\\[1pt]
  {\scriptsize {Pillar surveys (Secs.~\ref{sec:lpcomm}--\ref{sec:lpstore}): communication \textbullet\ compute \textbullet\ storage
  \ $\rightarrow$\ roadmap and conclusion (Secs.~\ref{sec:open_problems}--\ref{sec:conclusion})}}};

\node[badgeL4] (b4) at (b1 |- l4) {4};

% ----- Vertical flow -----
\coordinate (c1) at ($(l1.south)!0.5!(t2.north)$);
\draw[arr] (l1.south) -- (c1) -- (t2.north);
\draw[arr] (t2.south) -- (comm.north);

\coordinate (c2) at ($(comm.south)!0.5!(t3.north)$);
\draw[arr] (comm.south) -- (c2) -- (t3.north);

\coordinate (c3) at ($(t3.south)!0.5!(tax.north)$);
\draw[arr] (t3.south) -- (c3) -- (tax.north);

\draw[arr] (tax.south) -- (l4.north);

% ----- Cross-pillar couplings (Layer 2) -----
\draw[arr, bend left=12]
  (comp.east) to node[above, font=\scriptsize, inner sep=1pt, fill=white]
  {align} (comm.west);
\draw[arr, bend left=12]
  (stor.west) to node[above, font=\scriptsize, inner sep=1pt, fill=white]
  {persist} (comm.east);

% ----- Layer backgrounds -----
\begin{scope}[on background layer]
  \node[fit=(b1)(l1), fill=blue!4, rounded corners=3.5pt, inner sep=5pt, draw=none] {};
  \node[fit=(b2)(comp)(stor)(t2), fill=orange!4, rounded corners=3.5pt, inner sep=5pt, draw=none] {};
  \node[fit=(b3)(ana)(gate)(t3), fill=blue!3, rounded corners=3.5pt, inner sep=5pt, draw=none] {};
  \node[fit=(b4)(l4), fill=gray!5, rounded corners=3.5pt, inner sep=5pt, draw=none] {};
\end{scope}

\end{tikzpicture}}
\caption{Unified logic framework: Layer~1 domain; Layer~2 energy-first coordination under thermal limits; Layer~3 shared instruments (anatomy, branches, gates); Layer~4 pillar synthesis and lessons.}
\label{fig:logic_framework}
\end{figure*}
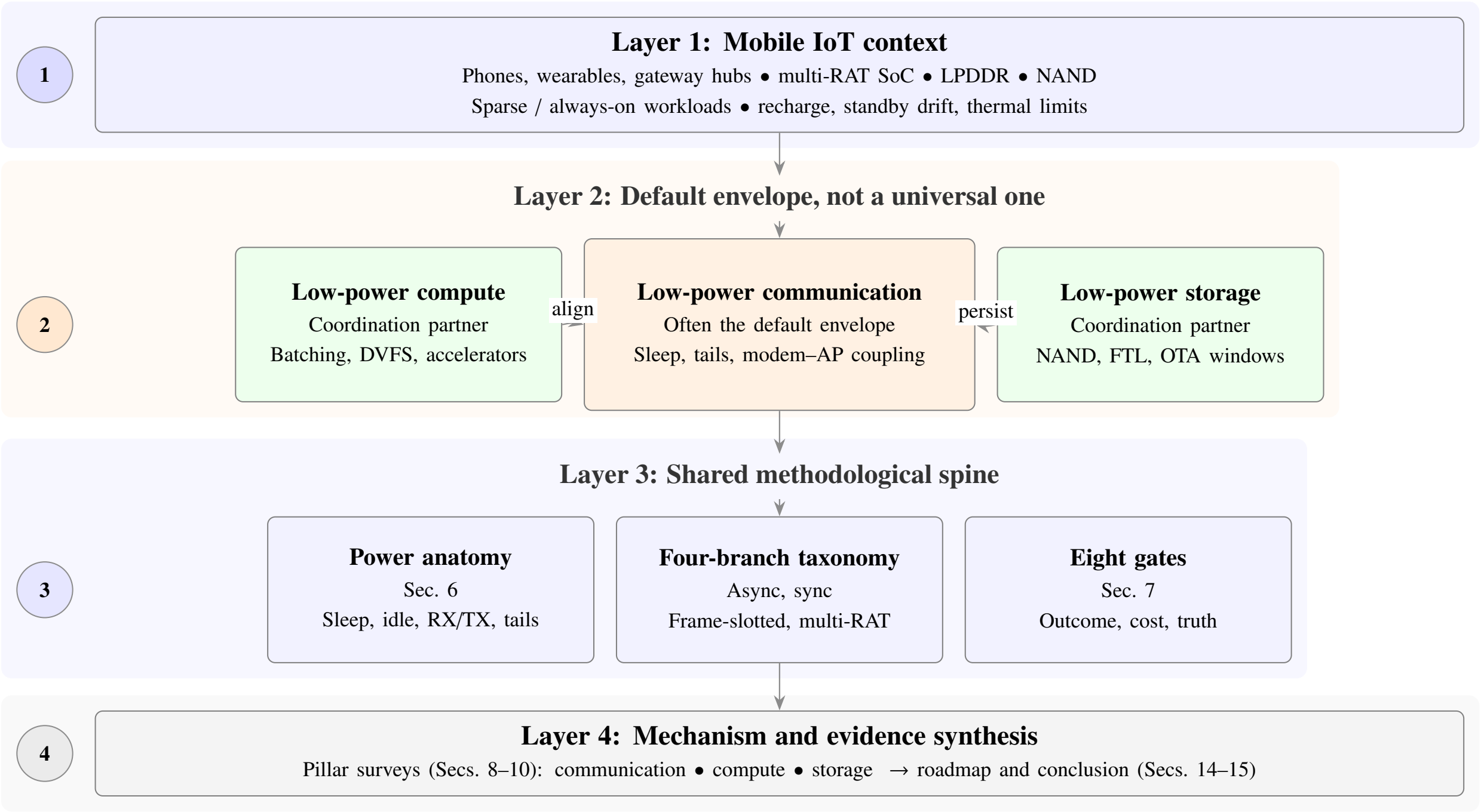

We organize the survey as a \textbf{four-layer logic chain} (Fig.~\ref{fig:logic_framework}, Table~\ref{tab:logic_framework}). Lower layers inherit upper-layer constraints. \rev{Table~\ref{tab:logic_framework} states each layer once, together with the failure mode that follows when it is skipped; only Layer~2 is developed in prose here, because its couplings are what the pillar sections and the evaluation gates both depend on. The domain of Layer~1 is battery-powered phones, wearables, and gateways whose success metric is recharge interval and thermal comfort rather than peak benchmark score~\cite{alfuqaha2015iot,patterson2024energy,skadron2004hotspot}.}

\textbf{Layer~2 --- Constraint graph.} Communication, compute, and storage form a graph, not a stack (Fig.~\ref{fig:logic_framework}). \rev{Communication is often the default temporal envelope} (sleep depth, listen/scan, RX/TX, tails, modem--AP handshakes), \rev{not a universal one: Table~\ref{tab:scope_limits} names phases owned by compute, storage, display, sensing, or thermal limits.} Compute fills or violates that envelope via batching, DVFS, and accelerators. Storage governs persistence and program/erase intervals that can block deep sleep~\cite{andreev2015connectivity,mittal2014embedded,chang2025flashhost}. Couplings include compute$\rightarrow$radio (compression vs.\ airtime), radio$\rightarrow$compute (timers, DMA), storage$\rightarrow$sleep (FTL/GC), and storage$\rightarrow$radio (OTA bursts)~\cite{xie2021apollo,jin2020review}.

\begin{table*}[!t]
\renewcommand{\arraystretch}{1.12}
\caption{Four-layer logic framework for energy-aware mobile IoT energy.}
\label{tab:logic_framework}
\centering
\tblsetup\begin{tabularx}{\textwidth}{@{}>{\centering\arraybackslash}p{0.03\textwidth}YYYY@{}}
\toprule
\textbf{Lay.} & \textbf{Question answered} & \textbf{Core objects} & \textbf{Paper anchor} & \textbf{If omitted, reviews tend to...} \\
\midrule
1 & What is the problem domain? & Phones, wearables, gateways; multi-RAT SoC; workloads & Sec.~\ref{sec:intro} & Treat motes, phones, and edge servers as interchangeable \\
2 & Who orchestrates whom? & Communication envelope; compute batching; NAND/OTA; couplings & Fig.~\ref{fig:logic_framework} & Report pillar-only savings that vanish after tails or FTL bursts \\
3 & How are claims compared? & Anatomy; four branches; eight gates & Secs.~\ref{sec:comm_power}--\ref{sec:evaldim} & Mix incompatible sensors and boundaries \\
4 & What mechanisms exist? & \rev{Pillar surveys; illustrative example; lessons; roadmap} & Secs.~\ref{sec:lpcomm}--\ref{sec:conclusion} & List techniques without applicability or measurement rules \\
\bottomrule
\end{tabularx}
\end{table*}

\textbf{Four-branch taxonomy (summary).} Branches sort designs by how tightly time, spectrum, and on-chip activity align~\cite{rajandekar2015m2mmac,patterson2024energy}: \emph{asynchronous}, no shared cluster clock and event-driven wakeups; \emph{synchronous}, repeating epochs that align sensing, inference, and uplink; \emph{frame-slotted}, named slots inside those epochs; \emph{multichannel heterogeneous}, parallel radio access technologies and policies on one SoC. \rev{Each branch is defined here, its scenarios and modeling risks are tabulated once in Table~\ref{tab:fourway}, and its assignment rule is given once in Table~\ref{tab:branch_decision}; none of the three is restated elsewhere.} Table~\ref{tab:branch_decision} covers mixed-regime traces (e.g., sync application frames over framed DRX, or multi-RAT days). Products routinely migrate among branches within a day; evaluations must name the \textbf{dominant branch per phase}~\cite{wunderlich2003simpoints}.

\rev{\textbf{How the four branches were derived.} The taxonomy is a combined coding scheme, not a purely deductive partition and not a cluster discovered by an unsupervised pass. Three branch names (asynchronous, synchronous, frame-slotted) were seeded from published MAC and rendezvous classifications~\cite{rajandekar2015m2mmac}. The fourth (multichannel heterogeneous) was added because core-set phone and gateway studies describe concurrent RATs on one SoC, a situation those classifications place outside a single network. Section~\ref{sec:lit_method} states the coding and agreement limitation. Table~\ref{tab:branch_decision} is the reproducible assignment procedure: apply the rows in order to a phase, keep one dominant label, and record secondary regimes as multi-label metadata.}

\subsection{Applicability Scope and Limiting Cases}
\label{sec:scope_limits}

The \textbf{energy-aware} lens is an \emph{organizing principle}, not a claim that every joule on every SKU is radio-limited. Table~\ref{tab:scope_limits} states \textbf{boundary cases} where compute, storage, display, sensing, or thermal envelopes dominate---cases the survey explicitly covers but does not subsume under a single bottleneck narrative.

\begin{table}[!t]
\renewcommand{\arraystretch}{1.10}
\caption{When mobile IoT energy is \emph{not} primarily communication-limited (counterexamples to the default thesis).}
\label{tab:scope_limits}
\centering
\tblsetup\begin{tabularx}{\columnwidth}{@{}>{\raggedright\arraybackslash}p{0.22\columnwidth}X@{}}
\toprule
\textbf{Dominant limiter} & \textbf{Representative scenario and implication} \\
\midrule
Compute & Sustained on-device vision, SLM/LLM decode-heavy phases, continuous audio DSP: DRAM/NPU traffic and shallow CPU retention during inference can exceed radio tails when connectivity is quiescent~\cite{lu2024slm,xu2024devicellm,chen2016eyeriss,husom2025sustainablellm}. \\
Storage & OTA install days, encrypted log storms, checkpoint rotation: FTL program/erase and verify paths block deep sleep regardless of DRX settings~\cite{patterson2024energy,chang2025flashhost,sultan2014processor}. \\
Display / UI & Active phone UI, always-on wearable displays: panel and touch backlights dominate over paging when the screen is on~\cite{carroll2010smartphone,bocus2025superiotriot}. \\
Sensing & High-rate IMU/mic pipelines without aggressive duty cycling: sensor hub and DMA keep the AP awake independent of uplink schedule~\cite{mittal2014embedded,heydari2025tinyml}. \\
Thermal & Skin-contact wearables, outdoor hubs in summer: throttling caps DVFS and RF duty cycle, re-ranking policies that looked optimal at room temperature~\cite{skadron2004hotspot,roy2003leakage}. \\
\bottomrule
\end{tabularx}
\end{table}

\textbf{Decision logic.} For each workload \emph{phase}, ask: (1)~If the radio were ideal (zero tails, instant sleep), would battery life change materially? If \textbf{no}, the phase is not communication-limited. (2)~Which pillar's activity blocks the deepest reachable sleep state? That pillar sets the \textbf{coordination envelope} for the phase even when another pillar consumes more instantaneous power. (3)~Report the dominant limiter per phase in traces; daily averages that blend display-on, OTA, and paging-idle phases without segmentation mis-rank policies~\cite{wunderlich2003simpoints,patterson2024energy}. The four-branch taxonomy and eight gates apply regardless of which pillar limits a phase---they specify \emph{how} to compare evidence, not \emph{which} pillar always wins.

\rev{\textbf{Operational test on a trace (proposed reporting cut, not a fitted constant).} Segment the trace by modem state or at boundaries of at least one second. Call the phase \emph{communication-limited} only if a radio-off segment, or a counterfactual that removes idle/listen and protocol tails, changes that phase's energy by at least 20\%. Otherwise name the non-radio pillar with the largest remaining share as the limiter. Separately name the \emph{sleep-envelope owner}: the pillar whose activity blocks the deepest reachable sleep state, even when its joules are smaller. Example~A is communication-limited when BLE listen plus tails are most of the display-off window, and the sensor hub is the sleep-envelope owner only if sampling keeps the AP awake after the radio has returned to sleep. Example~B on an OTA night is storage-limited if program/erase energy exceeds the 20\% radio-removal test, while paging can still be the sleep-envelope owner between erase bursts. The 20\% cut is a checklist convention so two papers use the same sentence; it is not a measured universal threshold.}

\subsection{Running Examples}
\label{sec:running_examples}

Two recurring product archetypes anchor the abstract framework in concrete phase sequences. They reappear in Sections~\ref{sec:comm_power}--\ref{sec:lpstore} when illustrating anatomy, branches, and gates.

\textbf{Example~A --- Wearable health sensor (periodic upload).} A wrist device samples vitals every minute, buffers on UFS, and uploads via BLE to a phone gateway every 15~min; cellular is absent. \emph{Anatomy:} deep sleep between samples; shallow retention during BLE connection events; short TX bursts; tail energy from link-layer supervision. \emph{Branch:} predominantly \textbf{frame-slotted} (connection interval) with \textbf{async} sensor interrupts. \emph{Gates:} Gate~1 (tail fraction during ``idle'' display-off), Gate~4 (PMIC cannot split BLE vs.\ sensor hub), Gate~7 (skin-contact thermal cap). \emph{Coordination lesson:} doubling inference accuracy on-device is worthless if it delays the BLE window and widens shallow retention~\cite{bocus2025superiotriot,mittal2014embedded}.

\textbf{Example~B --- Phone/gateway (local inference + logging + OTA).} A hub runs a quantized SLM for voice commands, encrypts logs to eMMC/UFS, and receives a nightly OTA over Wi-Fi while maintaining cellular paging. \emph{Anatomy:} concurrent idle/listen (paging), RX/TX (OTA), decode-heavy compute bursts, NAND program/erase during verify/swap. \emph{Branch:} \textbf{multichannel heterogeneous} within the hour; \textbf{frame-slotted} during OTA window. \emph{Gates:} Gate~5 (modem DMA vs.\ NPU LLC contention), Gate~8 (device-only SLM joules vs.\ cloud-inclusive), Gate~1 (maintenance day dominates screen-off mean). \emph{Coordination lesson:} scheduling OTA during a planned quiet window fails if FTL GC or federated checkpoint upload collides with paging---storage sets the feasible radio schedule, not the reverse~\cite{chang2025flashhost,lu2024slm,xie2021apollo,shi2016edge}.

Readers can apply the same template to other classes by filling Table~\ref{tab:lit_matrix} columns for their workload and declaring the dominant branch per phase before comparing mechanisms.

\rev{\textbf{Operational definition: sealed mobile IoT endpoint.} A sealed mobile IoT endpoint is a battery-powered product whose modem sleep ladder and flash-translation firmware are not fully open to the experimenter, and whose radio, application processor, DRAM, and NAND share one package, interconnect, and thermal envelope. \emph{Included:} smartphones; wearables (wrist and hearable); and gateway hubs that terminate at least one of cellular, Wi-Fi, BLE, or LPWA and persist data on flash. \emph{Included with an explicit qualifier:} industrial IoT nodes and modem modules when the measurement is of that sealed modem--AP--NAND assembly rather than of a rack. \emph{Excluded as primary objects:} warehouse servers and edge-rack hosts (cited only when a paper's energy boundary crosses into them~\cite{patterson2024energy,barroso2023dasc}); open single-radio sensor motes, which inform duty-cycle theory but are not the reference class~\cite{rajandekar2015m2mmac,chang2017sensors,dunkels2004contiki,chang2019wsnaccess,akyildiz2002wsn}. Smartphones are inside the scope. An ``edge device'' is in scope only for the joules attributed on the endpoint itself.}

\rev{Standards are discussed where they define experimentally settable timers and power-saving modes, not as exhaustive protocol tutorials~\cite{andreev2015connectivity,larsen2023greeneran}.}

\begin{table*}[!t]

\renewcommand{\arraystretch}{1.10}

\caption{Main acronyms used throughout this paper.}

\label{tab:acronyms}

\centering
\tblsetup\begin{tabularx}{\textwidth}{@{}p{0.07\textwidth}Xp{0.07\textwidth}X@{}}
\toprule
\textbf{Acronym} & \textbf{Definition} & \textbf{Acronym} & \textbf{Definition} \\
\midrule
AP    & application processor                                       & MCU    & microcontroller unit \\
ASIC  & application-specific integrated circuit                     & ML/DL  & machine learning / deep learning \\
BLE   & Bluetooth Low Energy                                        & NAND   & NAND flash memory \\
DMA   & direct memory access                                        & NB-IoT & narrowband Internet of Things \\
DRAM  & dynamic random-access memory                                & NPU    & neural processing unit \\
DRX   & discontinuous reception                                     & OS     & operating system \\
DSP   & digital signal processing                                   & OTA    & over-the-air (update) \\
DTIM  & delivery traffic indication message                         & P/E    & program/erase (flash cycle) \\
DVFS  & dynamic voltage and frequency scaling                       & PHY    & physical (layer) \\
eDRX  & extended discontinuous reception                            & PMC    & performance monitoring counter \\
FTL   & flash translation layer                                     & PMIC   & power-management integrated circuit \\
GC    & garbage collection                                          & PSM    & power-saving mode \\
GPU   & graphics processing unit                                    & RAPL   & running average power limit \\
I/O   & input/output                                                & RAT    & radio access technology \\
IoT   & Internet of Things                                          & RF     & radio frequency \\
ISP   & image signal processing                                     & RRC    & radio resource control \\
LLC   & last-level cache                                            & RTL    & register-transfer level \\
LPDDR & low-power double data rate (SDRAM)                          & SKU    & stock-keeping unit \\
LPWA  & low-power wide-area                                         & SoC    & system on chip \\
LTE-M & Long-Term Evolution for machine-type communication          & WSN    & wireless sensor network \\
MAC   & medium access control (link); MAC (accelerators) & M2M & machine-to-machine \\
\bottomrule
\end{tabularx}
\end{table*}

\begin{table*}[!t]
\renewcommand{\arraystretch}{1.08}
\caption{Four-branch energy-aware coordination taxonomy for mobile IoT: pillar readings per branch. \rev{Branch definitions are given once in Section~\ref{sec:logic_framework} and the assignment rules once in Table~\ref{tab:branch_decision}; neither is repeated here.}}
\label{tab:fourway}
\centering
\tblsetup\begin{tabularx}{\textwidth}{@{}>{\raggedright\arraybackslash}p{0.11\textwidth}YYYY@{}}
\toprule
\textbf{Branch} & \textbf{Representative scenarios} & \textbf{Dominant wireless phase} & \textbf{Compute / storage pattern} & \textbf{Primary modeling risk} \\
\midrule
Asynchronous & BLE ads; paging; LPWA listen windows; sporadic sensing & Idle/listen; tails; discovery & Interrupt-driven inference; flash suspend without global schedule & Mean power misleads; tail and retry bursts hidden \\
Synchronous & Audio/display frames; clustered WSN rounds; hub batch upload & Idle + short RX/TX bursts within epoch & Periodic DVFS; ring-buffer commits; aligned hub peaks & Aligned thermal/current peaks; stragglers stretch epoch \\
Frame-slotted & Cellular DRX/eDRX; BLE connection interval; planned OTA & Framed idle; bounded RX/TX; protocol tails & Compute batching in quiet windows; scheduled maintenance & Stack mismatch splits energy across tails \\
Multichannel heterogeneous & Handset multi-RAT; gateway Wi-Fi+cellular+BLE; scan matrices & Concurrent idle/scan; cross-RAT tails & Modem DMA vs.\ NPU/LLC; FTL vs.\ RF heat & Additive PHY or CPU models fail \\
\bottomrule
\end{tabularx}
\end{table*}

\begin{table*}[!t]
\renewcommand{\arraystretch}{1.10}
\caption{Decision rules for assigning the dominant four-branch coordination regime (including mixed-regime systems).}
\label{tab:branch_decision}
\centering
\tblsetup
\begin{tabularx}{\textwidth}{@{}
>{\raggedright\arraybackslash\noindent}p{0.45\textwidth}
X
>{\raggedright\arraybackslash\noindent}p{0.4\textwidth}
@{}}
\toprule
\textbf{If the phase exhibits...} & \textbf{Assign branch (dominant)} & \textbf{Mixed-regime note} \\
\midrule
Event-driven wakeups without shared epoch; paging/ads/LPWA listen on demand; sporadic sensing & \textbf{Asynchronous} & Tag coexisting \textbf{frame} sub-phases (e.g., BLE connection events inside async day) \\
Repeating epoch aligning sensing, inference, and uplink (audio/display/industrial frame) & \textbf{Synchronous} & If epoch contains DRX slots, record \textbf{sync} outer + \textbf{frame} inner \\
Standards-visible slots: DRX/eDRX/PSM, BLE conn.\ interval, planned OTA window & \textbf{Frame-slotted} & Modem frame may override compute batching---split tails across pillars \\
$\geq$2 active RATs or parallel scan/offload policies on one SoC & \textbf{Multichannel heterogeneous} & Attribute per-RAT energy separately; do not sum isolated PHY sheets \\
Conflicting signals (e.g., async sensing + framed DRX + Wi-Fi scan) & \textbf{Multi-label} & Report \emph{dominant} branch by energy share; list secondary in metadata \\
\bottomrule
\end{tabularx}
\end{table*}

\rev{\textbf{Quantitative mixed-regime rule (synthesis recommendation).} Inside one phase, compute each regime's share of phase energy. The dominant branch is the regime with the largest share. If the second regime is within 10 percentage points of the first, keep a multi-label and do not drop the secondary. Sub-phases keep an inner label: an outer synchronous epoch that contains DRX slots is sync-outer plus frame-inner. Worked assignment, illustrative: in Example~A, if BLE connection events plus tails exceed the energy of async sensor interrupts, the dominant branch is frame-slotted and async is secondary; if the two shares differ by less than 10 points, report both. In Example~B, concurrent Wi-Fi OTA and cellular paging are multichannel whenever neither interface leads by more than 10 points; the OTA window itself remains an inner frame-slotted label. These cuts are for comparable reporting. They are not a claim that 10 points is an empirically optimal boundary.}

\begin{table*}[!t]
\renewcommand{\arraystretch}{1.10}
\caption{Taxonomy of power and energy model \textbf{classes} mapped to mobile IoT subsystems (see pillar tables \ref{tab:lpcomm_models}--\ref{tab:lpstore_models} for family-level detail).}
\label{tab:taxonomy}
\centering
\tblsetup
\begin{tabularx}{\textwidth}{@{}
>{\raggedright\arraybackslash}p{0.19\textwidth}
>{\raggedright\arraybackslash}p{0.17\textwidth}
>{\raggedright\arraybackslash}p{0.15\textwidth}
>{\raggedright\arraybackslash}p{0.28\textwidth}
>{\raggedright\arraybackslash}p{0.17\textwidth}
@{}}
\toprule
\textbf{Class} & \textbf{Primary inputs} & \textbf{Pillar} & \textbf{Typical mobile IoT use} & \textbf{Gate most stressed} \\
\midrule
Analytical $\mu$arch & Config, event counts & Compute & Early SoC / open-ISA exploration & Interpretability; portability \\
RTL / gate-level & Toggle, netlist & Compute, accel. & Sensor-hub ASIC sign-off & Fidelity; observability \\
Hybrid calibration & Sim + silicon traces & Compute + DRAM & Product McPAT / PowerTrain alignment & Portability; composability \\
PMC / OS runtime & HW counters, time & Compute (if exposed) & Governor tuning on dev boards & Observability; boundary \\
ML / DL surrogate & Events, traces & Compute, I/O proxy & Per-SKU prediction on sealed phones & Fidelity vs.\ interpretability \\
Thermal coupled & Power + package & Cross-pillar & Wearable skin-temperature caps & Thermal; energy \\
Radio link / packet & PHY params, traces & Communication & Coverage, duty-cycle planning & Energy; delivery \\
Standards-timer / DRX & RRC/PSM/eDRX tables & Communication & NB-IoT, LTE-M, RedCap studies & Delivery; overhead \\
Trace / drive-test fit & Field logs, PMIC & Communication & Product modem tuning & Observability; boundary \\
Duty-cycle / Markov MAC & Sleep--wake stats & Communication & LPWA / WSN metaphors & Energy (tails); branch context \\
File-system / byte proxy & FS API writes & Storage & Quick what-if (weak for NAND) & Boundary; fidelity \\
FTL / P/E analytical & Program/erase, WA & Storage & Maintenance planning & Composability; observability \\
Coupled OTA--radio burst & OTA schedule + RF & Storage + comm. & Firmware campaign modeling & Energy; delivery \\
Contention-aware joint & Shared-bus traces & Cross-pillar & Phone-class SoC integration & Composability; observability \\
\bottomrule
\end{tabularx}
\end{table*}

\begin{table*}[!t]
\renewcommand{\arraystretch}{1.08}
\caption{Cross-pillar coupling effects in mobile IoT integration (qualitative); branch column names regimes where the effect is most visible.}
\label{tab:cross}
\centering
\tblsetup
\begin{tabularx}{\textwidth}{@{}
>{\raggedright\arraybackslash}p{0.16\textwidth}
>{\raggedright\arraybackslash}p{0.27\textwidth}
>{\raggedright\arraybackslash}p{0.36\textwidth}
>{\raggedright\arraybackslash}p{0.18\textwidth}
@{}}
\toprule
\textbf{Coupling} & \textbf{Typical trigger} & \textbf{Energy / system consequence} & \textbf{Sensitive branch} \\
\midrule
Radio $\rightarrow$ CPU & Scanning, missed DTIM, modem IRQ & Shallow AP retention; leakage dominates mean & Async.; multi-RAT \\
CPU $\rightarrow$ radio & Small frequent uplinks, early flush & Protocol tails exceed payload~\cite{andreev2015connectivity} & Async.; framed \\
Compute $\rightarrow$ DRAM & Large tiles, on-device LLM decode & Memory-bound phases; RF heat + bandwidth & Sync.; framed \\
Storage $\rightarrow$ sleep & Erase, FTL GC, secure wipe & Deep sleep blocked; ``idle'' misleading & Async.; maintenance day \\
Storage $\rightarrow$ radio & OTA, checkpoint upload & Airtime + flash peaks align & Framed; multi-RAT \\
Radio $\rightarrow$ storage & Deferred sync after connectivity & Write storms when link returns & Async. $\rightarrow$ framed \\
Edge $\rightarrow$ radio+storage & Split inference, cloud cache~\cite{shi2016edge} & Uplink and wear shift; boundary changes totals~\cite{jin2020review} & Multi-RAT \\
Thermal $\rightarrow$ all pillars & Sustained RF or NPU use & Throttling changes schedule and tails~\cite{skadron2004hotspot} & Sync.; multi-RAT \\
Security $\rightarrow$ all pillars & Attestation, encrypted logs & Overhead + blocked sleep~\cite{sultan2014processor} & Framed; async. \\
\bottomrule
\end{tabularx}
\end{table*}

\rev{\textbf{Published measurements of non-additivity.} The couplings above are not only qualitative. \emph{Measured.} Pathak et al.~\cite{pathak2012whereenergy} account a 3G tail of 187\,$\mu$Ah, 59.6\% of a 314\,$\mu$Ah microbenchmark, with the radio left in the tail state for about 6\,s after the transfer; separating one send into two operations raised the same workload to 520\,$\mu$Ah because a second tail was incurred. In their application study, most apps spent roughly 50--90\% of energy in I/O rather than in the CPU. \emph{Measured.} Carroll and Heiser~\cite{carroll2010smartphone} instrumented a smartphone at component level and showed cellular and display power, not application-processor cycles, dominating several everyday scenarios, so a CPU-only budget mis-ranks the day. \emph{Measured / system.} Xie et al.~\cite{xie2021apollo} show modem DMA and accelerator traffic sharing LLC and DRAM, which is why isolated CPU and radio sheets cannot be summed (Gate~5). \emph{Illustrative implication, not a new measurement.} A NAND program/erase or OTA burst can block the deep sleep that the modem would otherwise have entered~\cite{patterson2024energy,mittal2014embedded,chang2025flashhost}; CPU counters on that day still look idle. These anchors are why subsystem joules are treated as non-additive unless the trace shows otherwise.}

\rev{\textbf{When a local gain raises another pillar.} Shared DRAM, LLC, DMA, a single PMIC rail, and one thermal cap are the paths. A larger NPU tile increases DRAM and LLC traffic; modem DMA then misses its quiet window and the AP stays in shallow retention~\cite{xie2021apollo}. Shortening DRX to save latency wakes the AP more often, so CPU idle and DRAM self-refresh rise even if the PHY payload shrinks~\cite{pathak2012whereenergy,andreev2015connectivity}. A log format with fewer file-system bytes can start FTL garbage collection, occupy the bus, and block the deep sleep that PSM would otherwise enter~\cite{patterson2024energy,chang2025flashhost}. A radio or NPU burst that is acceptable at room temperature becomes sustained throttling against the skin, which lengthens both the job and the tail~\cite{skadron2004hotspot}. Because those events share a PMIC rail, a pillar-local saving can disappear in the package total.}

\rev{\textbf{Magnitude of non-additivity (measured).} In the Pathak et al.\ microbenchmark, one 3G transfer costs 314\,$\mu$Ah and the same logical send split into two operations costs 520\,$\mu$Ah~\cite{pathak2012whereenergy}. The extra 206\,$\mu$Ah is a second tail, not the sum of two independent transfers. CPU work placed inside an already open tail stretches that high-power radio state, so a CPU-only model plus a radio-idle model does not equal the measured total. We do not restate a separate LLC double-counting percentage from Xie et al.; their result is the system evidence that modem DMA and accelerators share LLC and DRAM~\cite{xie2021apollo}.}

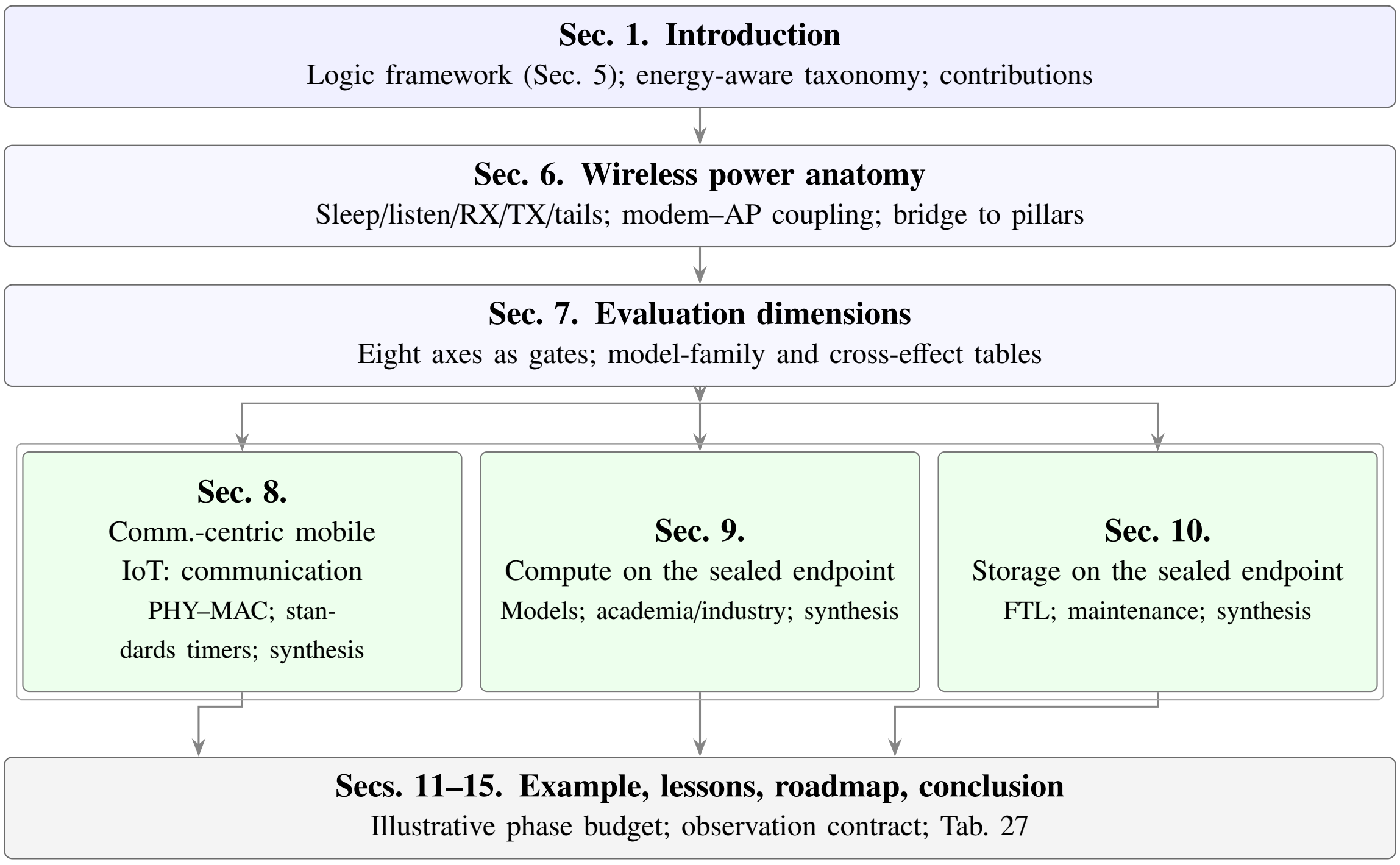
\begin{figure*}[!t]
\centering
\begin{tikzpicture}[
  font=\normalsize,
  >={Stealth[length=2.1mm]},
  arr/.style={-Stealth, line width=0.58pt, draw=black!48},
  spine/.style={rectangle, rounded corners=2.5pt, draw=black!58, inner sep=4.2pt, align=center,
                text width=0.90\textwidth, minimum height=11mm},
  tierA/.style={spine, fill=blue!6},
  tierB/.style={spine, fill=blue!3},
  tierC/.style={rectangle, rounded corners=2pt, draw=black!52, inner sep=3.4pt, align=center,
                text width=0.275\textwidth, minimum height=26mm, fill=green!7},
  tierD/.style={spine, fill=gray!9, minimum height=11mm}
]
% Section numbers via \ref match the document counter (Sec.~1, Sec.~2, ... under IEEEtran journal).
\node[tierA] (L1) {\textbf{Sec.~\ref{sec:intro}. Introduction}\\[0pt]{\small Logic framework (Sec.~\ref{sec:logic_framework}); energy-aware taxonomy; contributions}};
\node[tierB, below=4mm of L1] (L1b) {\textbf{Sec.~\ref{sec:comm_power}. Wireless power anatomy}\\[0pt]{\small Sleep/listen/RX/TX/tails; modem--AP coupling; bridge to pillars}};
\node[tierB, below=4mm of L1b] (L2) {\textbf{Sec.~\ref{sec:evaldim}. Evaluation dimensions}\\[0pt]{\small Eight axes as gates; model-family and cross-effect tables}};
\coordinate (fork) at ($(L2.south)+(0,-1.8mm)$);
\node[tierC, anchor=north] at ($(fork)+(-0.302\textwidth,-5.2mm)$) (L3a) {\textbf{Sec.~\ref{sec:lpcomm}.}\\[0pt]{\small Comm.-centric mobile IoT: communication}\\[0pt]{\footnotesize PHY--MAC; standards timers; synthesis}};
\node[tierC, anchor=north] at ($(fork)+(0,-5.2mm)$) (L3b) {\textbf{Sec.~\ref{sec:lpcompute}.}\\[0pt]{\small Compute on the sealed endpoint}\\[0pt]{\footnotesize Models; academia/industry; synthesis}};
\node[tierC, anchor=north] at ($(fork)+(0.302\textwidth,-5.2mm)$) (L3c) {\textbf{Sec.~\ref{sec:lpstore}.}\\[0pt]{\small Storage on the sealed endpoint}\\[0pt]{\footnotesize FTL; maintenance; synthesis}};
\node[tierD, below=7mm of L3b.south, anchor=north] (L4) {\textbf{Secs.~\ref{sec:case_study}--\ref{sec:conclusion}. Example, lessons, roadmap, conclusion}\\[0pt]{\small Illustrative phase budget; observation contract; Tab.~\ref{tab:roadmap}}};
% Arrows: linear spine then fork / merge
\draw[arr] (L1) -- (L1b);
\draw[arr] (L1b) -- (L2);
\draw[arr] (L2.south) -- (fork);
\draw[arr] (fork) -| (L3a.north);
\draw[arr] (fork) -- (L3b.north);
\draw[arr] (fork) -| (L3c.north);
\draw[arr] (L3a.south) -- ++(0,-1.6mm) -| ($(L4.north west)!0.28!(L4.north)$);
\draw[arr] (L3b.south) -- (L4.north);
\draw[arr] (L3c.south) -- ++(0,-1.6mm) -| ($(L4.north east)!0.72!(L4.north)$);
% Frame: parallel core = Secs.~\ref{sec:lpcomm}--\ref{sec:lpstore} (four-branch detail inside)
\draw[draw=black!35, line width=0.35pt, rounded corners=1.5pt]
  ($(L3a.north west)+(-0.6mm,0.8mm)$) rectangle ($(L3c.south east)+(0.6mm,-0.8mm)$);
\end{tikzpicture}
\caption{Organization of the present paper.}
\label{fig:paper_structure}
\end{figure*}

As shown in Fig.~\ref{fig:paper_structure}, the paper proceeds from motivation and positioning (Section~\ref{sec:intro}), through the coordination framework (Section~\ref{sec:logic_framework}), wireless power anatomy (Section~\ref{sec:comm_power}), and evaluation gates (Section~\ref{sec:evaldim}), to matched pillar surveys (Sections~\ref{sec:lpcomm}--\ref{sec:lpstore}). \rev{Sections~\ref{sec:case_study}--\ref{sec:lessons} apply the framework with an illustrative phase-budget example, a model-selection workflow, and joint-optimization lessons.} Section~\ref{sec:conclusion} consolidates the observation contract and roadmap. Table~\ref{tab:acronyms} lists main acronyms.

\section{Basic Concepts: Wireless Power Anatomy}

\label{sec:comm_power}

Mobile IoT energy is hard to read because \textbf{radios draw power in named states} set by protocols, firmware sleep ladders, and PMIC rails---not only by how many bytes an app sends~\cite{andreev2015connectivity,patterson2024energy,mittal2014embedded}. Teams need a short split---deep vs.\ shallow sleep, idle/listen vs.\ RX/TX, tails---before judging compute or flash tweaks. This section gives that shared vocabulary; Sections~\ref{sec:evaldim}--\ref{sec:lpstore} reuse it.

\textbf{Why this split matters.} Early smartphone studies showed cellular, Wi-Fi, and idle dominate many days---more than naive CPU-centric guesses~\cite{carroll2010smartphone}---and that fair attribution needs finer phases than one package counter or app box~\cite{pathak2012whereenergy}. Mobile IoT is the same pattern with tighter budgets: the labels below match how engineers cut traces (modem state, connection events, rails), not PHY spreadsheets alone~\cite{andreev2015connectivity,patterson2024energy}.

\textbf{Device vs.\ network.} Cellular power-saving ties to radio-resource control and network load~\cite{hasan2011greencellular}; DRX/eDRX/PSM settings also depend on the network and operator. Release~17/18 RedCap and eRedCap extend the same story to mid-rate wearables and industrial links, where beam management, wake-up signals (WUS), and RRC state choice materially change shallow-retention time~\cite{veedu2022redcap,joerke2025redcap,raja2026redcapsurvey}. Read tail and idle shares here as \emph{measured under a stated network setup}, not as pure PHY facts~\cite{andreev2015connectivity,larsen2023greeneran}.

\textbf{Standards-based power-saving modes (3GPP-class).} Table~\ref{tab:drx_psm} summarizes citeable timers for NB-IoT, LTE-M, and RedCap studies. DRX in RRC\_CONNECTED schedules periodic monitoring occasions while allowing micro-sleep between them; \textbf{eDRX} extends the cycle for latency-tolerant M2M at the cost of downlink reachability delay; \textbf{PSM} (power-saving mode) in RRC\_IDLE detaches the device to a deep sleep with periodic tracking-area updates, minimizing idle draw for sparse uplink~\cite{centenaro2016longrange,rao2018nbiotoverview,andreev2015connectivity}. RedCap/eRedCap devices trade peak eMBB throughput for simpler RF chains but inherit the same RRC/DRX vocabulary---experiments must report which state (connected vs.\ idle), which timer set, and whether the core network honors the requested eDRX/PSM parameters~\cite{veedu2022redcap,joerke2025redcap,raja2026redcapsurvey}. Wake-up signal (Release~16+) and on-demand paging reduce false wakeups but add configuration overhead that appears in Gate~3 (mechanism overhead) rather than Gate~1 alone~\cite{larsen2023greeneran}.

\rev{\textbf{Standards behavior versus measured energy.} Numbered specifications are now in the reference list. TS~24.501 defines NAS behavior relevant to 5GS power saving, including how a UE requests and is granted low-power parameters~\cite{3gpp24501}. TS~36.331 defines connected-mode DRX as a pattern of on-duration and opportunity for sleep~\cite{3gpp36331}. TS~36.304 defines idle-mode procedures, including extended DRX and wake-up signaling, as reachability rules rather than as joules~\cite{3gpp36304}. Those documents specify timers and state machines. They do not measure package energy. A row in Table~\ref{tab:drx_psm} is therefore a standards-defined lever. An energy number attached to that lever is an empirical observation and must name the release, the granted timers, and the sensor path. The two must not be cited as if they were the same kind of evidence.}

\begin{table}[!t]
\renewcommand{\arraystretch}{1.08}
\caption{Standards-visible cellular power-saving levers for mobile IoT (illustrative; exact ranges are release- and band-specific---cite 3GPP TS for experiments).}
\label{tab:drx_psm}
\centering
\tblsetup\footnotesize
\begin{tabularx}{\columnwidth}{@{}
>{\raggedright\arraybackslash}p{0.14\columnwidth}
X
>{\raggedright\arraybackslash}p{0.30\columnwidth}
@{}}
\toprule
\textbf{Mode} & \textbf{System effect on anatomy phases} & \textbf{Typical IoT use} \\
\midrule
DRX (connected) & Short idle/listen between active RX/TX; tail after release & Voice-adjacent, moderate telemetry \\
eDRX (idle/ext.) & Longer idle windows; higher downlink latency & LPWA-class sparse downlink \\
PSM & Deep sleep + periodic TAU; minimal idle energy & Sensor uplink-only, months-scale battery \\
RedCap / eRedCap & Mid-tier NR with DRX/WUS; simpler RF than eMBB & Wearable video, industrial monitor \\
\bottomrule
\end{tabularx}
\end{table}

\subsection{Sleep, Retention, and Wake Transitions}

\textbf{Deep sleep} minimizes leakage by powering down clocks and large SRAM domains but pays \textbf{wake latency} and sometimes cold-cache penalties when sensing, inference, or uplink deadlines arrive~\cite{mittal2014embedded,sultan2014processor}. \textbf{Shallow retention} keeps orchestration state alive for modem rendezvous: the application processor may appear idle while volatile domains needed for radio handshakes remain energized~\cite{andreev2015connectivity,mittal2014embedded}. Coarse ``screen off'' averages therefore hide the decisive question---how often the stack \emph{must} climb out of deep sleep to honor paging, advertising, or connection supervision timers~\cite{andreev2015connectivity,centenaro2016longrange}.

\subsection{Idle, Listen, Scan, and Discovery}

Even without payload reception, \textbf{listening} for preambles, \textbf{scanning} for Wi-Fi/BLE neighbors, and \textbf{paging} windows in cellular modes burn time-aligned energy that scales with channel bandwidth, duty cycle, and coexistence policies~\cite{andreev2015connectivity,larsen2023greeneran}. Discovery and neighbor maintenance are classically lumped into ``idle power,'' yet on mobile IoT they are often the \textbf{dominant} shallow-drain mechanisms because they pin the digital baseband and sometimes the AP in retention while user-visible traffic is near zero~\cite{mittal2014embedded}.

\subsection{Receive and Transmit}

\textbf{RX} energy includes RF front-end and analog-to-digital conversion, baseband decoding, security processing, and memory copies into OS buffers; \textbf{TX} adds power-amplifier and backoff behavior under regulatory masks and link adaptation~\cite{andreev2015connectivity,centenaro2016longrange}. Effective energy per delivered bit therefore couples to channel quality, retransmissions, and batching policies---not only to modulation and coding choices visible in PHY spreadsheets~\cite{andreev2015connectivity}.

\subsection{Tail Energy and Modem--Application Processor Coupling}

\textbf{Tail states} arise when the radio remains powered while upper layers finish acknowledgments, security contexts, or transport timers; cellular discontinuous-reception (DRX) cycles, random-access bursts, and IEEE~802.11-style power-save listen intervals extend intervals where energy is spent without throughput~\cite{andreev2015connectivity,patterson2024energy}. Critically, tails often \textbf{couple the modem to the AP}: shallow CPU retention, DMA traffic, and interrupt storms continue after the channel looks ``quiet,'' so subsystem models that stop at the PHY boundary under-attribute energy~\cite{mittal2014embedded,xie2021apollo}.

\subsection{Bridge to Compute--Storage Coordination}

Compute-side batching changes \emph{when} uplinks run; flash maintenance changes \emph{whether} deep sleep is allowed before the next paging window. The four-branch taxonomy (Section~\ref{sec:logic_framework}) says \emph{which} effect dominates a phase; the eight gates (Section~\ref{sec:evaldim}) ask if reported energy includes the tails and edges above. With this vocabulary set, Sections~\ref{sec:lpcomm}--\ref{sec:lpstore} treat radio, compute, and persistence as partners, not as three unrelated literatures.

\section{Basic Concepts: Evaluation Gates for Energy and Thermal Claims}

\label{sec:evaldim}

Following low-power networking and endpoint measurement practice~\cite{patterson2024energy,rajandekar2015m2mmac}, evaluation is treated as \textbf{multi-dimensional}. Section~\ref{sec:comm_power} decomposes wireless power into sleep versus retention, idle/listen versus RX/TX, tails, and modem--AP coupling. Subsections~\ref{sec:evaldim:energy}--\ref{sec:evaldim:boundary} apply the same eight axes to \textbf{compute}, \textbf{communication}, and \textbf{storage}~\cite{alfuqaha2015iot,reda2012survey}. Table~\ref{tab:eval_gates} summarizes the gates; Tables~\ref{tab:taxonomy} and~\ref{tab:cross} map model classes and cross-pillar couplings to observability limits. Figure~\ref{fig:evaldim_logic} shows the layout: eight axes applied together, then linked to Sections~\ref{sec:lpcomm}--\ref{sec:lpstore}, where the four branches are unpacked per pillar.

{
\begin{figure*}[!t]
\centering
\begin{tikzpicture}[
  font=\scriptsize,
  gate/.style={draw=black!55, rounded corners=1.2pt, align=center, text width=17mm, minimum height=9mm, inner sep=1.2pt},
  outg/.style={gate, fill=blue!7},
  costg/.style={gate, fill=orange!12},
  evg/.style={gate, fill=green!9},
  pillar/.style={draw=black!55, rounded corners=2pt, align=center, text width=40mm, minimum height=7.5mm, fill=black!5},
  synth/.style={draw=black!55, rounded corners=2pt, align=center, text width=140mm, minimum height=8mm, fill=yellow!14},
  ar/.style={-{Latex[length=1.5mm]}, black!55}
]
\node[outg]  (g1) at (0,0)    {\textbf{G1}\\Energy and\\power};
\node[outg]  (g2) at (2.0,0)  {\textbf{G2}\\Data\\delivery};
\node[outg]  (g7) at (4.0,0)  {\textbf{G7}\\Thermal /\\form factor};
\node[costg] (g3) at (6.0,0)  {\textbf{G3}\\Mechanism\\overhead};
\node[costg] (g8) at (8.0,0)  {\textbf{G8}\\System\\boundary};
\node[evg]   (g4) at (10.0,0) {\textbf{G4}\\Observability\\/ ground truth};
\node[evg]   (g5) at (12.0,0) {\textbf{G5}\\Portability /\\composability};
\node[evg]   (g6) at (14.0,0) {\textbf{G6}\\Interpretability\\vs.\ fidelity};
\begin{scope}[on background layer]
  \node[draw=blue!55, dashed, rounded corners=2pt, inner sep=2.4mm, fit=(g1)(g7),
        label={[blue!55!black, font=\scriptsize\bfseries]above:Outcome}] {};
  \node[draw=orange!75!black, dashed, rounded corners=2pt, inner sep=2.4mm, fit=(g3)(g8),
        label={[orange!65!black, font=\scriptsize\bfseries]above:Cost}] {};
  \node[draw=green!45!black, dashed, rounded corners=2pt, inner sep=2.4mm, fit=(g4)(g6),
        label={[green!40!black, font=\scriptsize\bfseries]above:Evidence quality}] {};
\end{scope}
\node[pillar] (pc) at (2.3,-2.4)  {Communication\\(Section~\ref{sec:lpcomm})};
\node[pillar] (pp) at (7.0,-2.4)  {Compute\\(Section~\ref{sec:lpcompute})};
\node[pillar] (ps) at (11.7,-2.4) {Storage\\(Section~\ref{sec:lpstore})};
\node[synth]  (sy) at (7.0,-4.0)
  {Branch-wise synthesis: assign the dominant coordination regime (Tables~\ref{tab:fourway},~\ref{tab:branch_decision}),\\
   then report the phase with the template of Table~\ref{tab:report_template}};
\draw[ar] (g1.south) -- (pc.north);
\draw[ar] (g7.south) -- (pp.north);
\draw[ar] (g4.south) -- (ps.north);
\draw[ar] (pc.south) -- (sy.north);
\draw[ar] (pp.south) -- (sy.north);
\draw[ar] (ps.south) -- (sy.north);
\node[align=left, text width=150mm, font=\scriptsize] at (7.0,-5.1)
  {All eight gates are applied to each pillar. The arrows mark only the gate that each pillar
   fails most often: tails and idle for communication, throttling for compute, and the
   file-system-bytes proxy for storage (Section~\ref{sec:evaldim}).};
\end{tikzpicture}
\caption{\rev{Eight evaluation gates grouped into outcome, cost, and evidence quality, applied to communication, compute, and storage, and carried into branch-wise synthesis. Redrawn for this revision so that all section pointers are live cross-references.}}
\label{fig:evaldim_logic}
\end{figure*}
}

\textbf{Three groups organize the eight axes.} Energy/power, data delivery, and thermal/form factor are \textbf{outcome} dimensions; mechanism overhead and system-boundary accounting are \textbf{cost} dimensions; observability, portability/composability, and interpretability/fidelity are \textbf{measurement-validity (evidence)} dimensions. Grouping exposes tradeoffs that single-metric plots obscure: richer traces improve evidence quality at higher measurement cost; high-fidelity learned models may sacrifice standards-relevant knobs; wide DRX can lower average power while hiding uplink tails~\cite{patterson2024energy,reda2012survey,zhai2024mlsurvey}. Each subsection states a \textbf{gate} question and applies the axis to all three pillars. Table~\ref{tab:eval_gates} provides a cross-pillar summary.

\rev{\textbf{Why these eight gates?} A single energy number is not a comparable claim on a sealed endpoint. The eight axes are a \emph{proposed compact checklist} that separates three questions reviewers otherwise conflate: (i)~whether the claimed saving changes an outcome (energy, delivery, thermal feasibility), (ii)~what cost was paid (control overhead, a wider system boundary), and (iii)~whether the evidence can be compared (sensor path, portability, interpretability)~\cite{patterson2024energy,reda2012survey,bocus2025netiotenergy}. They are not validated as a unique minimal set over the full corpus. Alternatives were considered and folded in rather than added. Monetary cost and embodied carbon are downstream of Gates~1 and~8 and are not reported consistently in this corpus. User-perceived quality of experience is part of Gate~2 (latency, loss, freshness). Security assurance is not an energy outcome; attestation, encryption, and secure erase appear as Gate~3 overhead and as sleep blocking. A separate ``carbon'' or ``security'' gate would have split papers that do not measure those quantities. Mapping onto the literature is by the stress each gate was built to catch: Gate~1, component and tail measurements~\cite{carroll2010smartphone,pathak2012whereenergy,andreev2015connectivity}; Gate~2, delivery constraints that forbid longer sleep~\cite{rajandekar2015m2mmac,centenaro2016longrange}; Gate~3, scans, calibration, and journaling~\cite{reda2012survey,sultan2014processor}; Gate~4, PMIC and counter opacity~\cite{patterson2024energy,hackenberg2013rapl}; Gate~5, shared LLC/DRAM and firmware drift~\cite{xie2021apollo,chandrasekar2012drampower}; Gate~6, analytical versus learned models~\cite{zhai2024mlsurvey}; Gate~7, package and skin limits~\cite{skadron2004hotspot,roy2003leakage}; Gate~8, device versus cloud boundaries~\cite{shi2016edge,jin2020review,barroso2023dasc}. Sections~\ref{sec:lpcomm}--\ref{sec:lpstore} cite these gates by number instead of restating the definitions.}

\begin{table*}[!t]
\renewcommand{\arraystretch}{1.10}
\caption{Eight evaluation \textbf{gates} for energy-aware mobile IoT (Section~\ref{sec:evaldim}); Grp.\ = outcome/cost/evidence group; Out.\ = outcome; used when reading Tables~\ref{tab:lpcomm_models}--\ref{tab:lpstore_models}.}
\label{tab:eval_gates}
\centering
\tblsetup
\begin{tabularx}{\textwidth}{@{}
c
>{\raggedright\arraybackslash}p{0.16\textwidth}
>{\raggedright\arraybackslash}p{0.05\textwidth}
>{\raggedright\arraybackslash}p{0.21\textwidth}
>{\raggedright\arraybackslash}p{0.27\textwidth}
Y
@{}}
\toprule
\textbf{\#} & \textbf{Gate (dimension)} & \textbf{Grp.} & \textbf{Communication stress} & \textbf{Compute stress} & \textbf{Storage stress} \\
\midrule
1 & Energy and power & Out. & Tails, DRX, time-on-air & DRAM/NPU dominance; shallow idle & P/E bursts block deep sleep \\
2 & Data delivery & Out. & Latency, loss, freshness & Deadline misses $\rightarrow$ radio retries & Commit latency vs.\ log freshness \\
3 & Mechanism overhead & Cost & Scans, CQI, discovery traffic & Attestation, calibration sweeps & GC, journaling, secure erase \\
4 & Observability & Truth & PMIC-only modem totals & Counter drift; hidden accelerators & FS bytes $\neq$ NAND energy \\
5 & Portability / composability & Truth & Baseband sleep-ladder changes & $\mu$arch model per SKU & FTL policy per vendor \\
6 & Interpretability vs.\ fidelity & Truth & DRX sweeps vs.\ drive traces & Analytical vs.\ learned proxies & Analytical WA vs.\ bus traces \\
7 & Thermal / form factor & Out. & RF heat, skin limits & NPU/GPU throttle & Package heating during OTA \\
8 & System boundary & Cost & Device vs.\ cloud-inclusive & Saved MACs vs.\ uplink joules & Edge cache vs.\ local wear \\
\bottomrule
\end{tabularx}
\end{table*}

\subsection{Energy and Power}
\label{sec:evaldim:energy}
\textbf{Gate~1.} Which workload phase produced the reported energy number, and are tail/idle fractions reported separately?
Energy/power covers average draw, tails, and task-normalized joules~\cite{andreev2015connectivity,patterson2024energy,carroll2010smartphone,pathak2012whereenergy}. Compute is often memory-bound~\cite{chen2016eyeriss,parashar2019timeloop}; communication is often paging/discovery/DRX-dominated~\cite{andreev2015connectivity,centenaro2016longrange}; storage program/erase and FTL GC can block deep sleep~\cite{patterson2024energy,mittal2014embedded}. Always read energy in branch context (Section~\ref{sec:logic_framework}).

\subsection{Data-Delivery Performance}
\label{sec:evaldim:qos}
\textbf{Gate~2.} Alongside energy, are tail latency, retransmissions, or freshness violations reported under the same workload?
Delivery spans latency, throughput after retries, reliability, and freshness~\cite{mittal2014embedded,rajandekar2015m2mmac}. Compute deadlines can force radio retries~\cite{shi2016edge}; communication access/HARQ shapes tails~\cite{andreev2015connectivity}; storage commits can delay acknowledgments~\cite{patterson2024energy}.

\subsection{Mechanism Overhead and Control Cost}
\label{sec:evaldim:overhead}
\textbf{Gate~3.} Would the claimed saving survive without offline calibration, and are control-plane wakeups/bytes counted?
Overhead includes scans, CQI, attestation, calibration, journaling, and secure erase~\cite{reda2012survey,rajandekar2015m2mmac,patterson2024energy,sultan2014processor}. Omitting it overstates deployability on sealed products~\cite{alfuqaha2015iot}. \rev{Security work is the clearest case. Handshakes, key rotation, attestation, and encrypted logging are mandatory on shipped products and are almost never charged to the mechanism that requires them; surveys of energy-aware security for IoT and of cross-layer secure and efficient design treat that cost as a first-class budget item rather than as a rounding error~\cite{chien2024energysecurityiot,benosman2024crosslayeriot}. On an endpoint the cost is often not the cryptography itself but its placement: a rekey that wakes the modem outside its paging window, or a verify step that holds the flash bus open, is charged under Gate~1 as blocked sleep even when the cipher's own joules are negligible.}

\subsection{Observability and Ground Truth}
\label{sec:evaldim:observe}
\textbf{Gate~4.} Through which interface were joules observed, and what does that interface miss on the SKU?
Interfaces include PMIC totals, on-die sensors, and PMC proxies~\cite{patterson2024energy,hackenberg2013rapl,schone2015energy}. Modem and NAND rails are often opaque; FS bytes are not NAND energy~\cite{andreev2015connectivity,mittal2014embedded}.

\subsection{Portability and Composability of Models}
\label{sec:evaldim:compose}
\textbf{Gate~5.} Does the recipe transfer across SKUs/firmware, and can sub-models be summed without double counting?
Portability fails under counter/firmware drift; composability fails under shared LLC/DRAM/modem DMA~\cite{reda2012survey,xie2021apollo,chandrasekar2012drampower,mittal2014embedded}. Non-additivity should be reported as a result.

\subsection{Interpretability Versus Fidelity}
\label{sec:evaldim:interp}
\textbf{Gate~6.} Are the reported knobs standards-interpretable, or only high-fidelity black-box fits?
Interpretable DRX/analytical knobs aid comparison; drive traces and learned proxies fit products but transfer poorly~\cite{reda2012survey,zhai2024mlsurvey,larsen2023greeneran,patterson2024energy}.

\subsection{Thermal and Form-Factor Feasibility}
\label{sec:evaldim:thermal}
\textbf{Gate~7.} Do thermal/skin limits change the schedule that produced the energy number?
RF heat, NPU/GPU throttle, and OTA package heating re-rank policies that look optimal at room temperature~\cite{skadron2004hotspot,roy2003leakage,andreev2015connectivity,mittal2014embedded}.

\rev{\textbf{What ``thermal-aware'' can mean on a sealed endpoint.} Thermal coverage in this survey is a constraint shared by all three pillars rather than a fourth pillar with its own device stack, and Gate~7 is where that constraint is tested. Four elements have to be named before a temperature claim is comparable: the \emph{sensing interface}, the \emph{model} used to reason beyond the sensor, the \emph{actuator} that the controller is allowed to move, and the \emph{regime} that the workload sustains. Table~\ref{tab:thermal_evidence} collects the families that appear in the corpus under those four headings.}

\rev{\textbf{Sensing.} Handsets and wearables expose a small set of firmware-filtered thermal zones---one or two junction sensors near the SoC, a skin or back-cover estimate, sometimes a battery and a charger sensor---sampled at a fraction of a hertz and already smoothed by the vendor's governor. That interface is not comparable to the die-resolved maps of architectural thermal studies~\cite{skadron2004hotspot,huang2006hotspot}, and the mismatch matters: a paper reporting ``no throttling observed'' may simply have read a zone that never crosses its trip point while a neighbouring block is already being clipped. A thermal claim should therefore name the zone, the sampling period, and whether the reading is raw or governor-filtered, in the same way Gate~4 asks for the energy sensor path.}

\rev{\textbf{Models.} Compact resistance--capacitance networks remain the workhorse: a handful of thermal resistances and capacitances mapped onto floorplan blocks, driven by a power trace, and solved for a temperature trajectory~\cite{skadron2004hotspot,huang2006hotspot}. Their attraction on endpoints is that they can be fitted from a few externally observable transients; their limit is that mobile packages add a passive heat-spreader and a user's hand as boundary conditions that no published floorplan captures. Coupled power--thermal--reliability models extend the same machinery to aging and to the leakage feedback loop~\cite{brooks2007power,roy2003leakage}: leakage grows with temperature, so a block that is already hot draws more static power, which heats it further. That loop is the reason a thermal margin cannot be treated as a fixed derate of a room-temperature energy figure. Dark-silicon analyses make the budget explicit at the chip level---only part of the die can be driven at full speed within the power envelope, so allocation, not peak throughput, is the design variable~\cite{esmaeilzadeh2011dark}.}

\rev{\textbf{Actuators and their cost.} Throttling is the actuator that turns a temperature into a schedule change: DVFS steps on CPU, GPU, and NPU; RF duty-cycle or transmit-power limits on the modem; deferral or cancellation of maintenance work such as OTA staging and garbage collection; and, in the worst case, a forced quiet window. Each carries a Gate~3 cost that is routinely omitted. Clipping the NPU lengthens an inference that was holding a radio tail open; clipping the modem stretches a transfer and therefore the total time the package spends warm; deferring OTA moves the program burst into a later phase where the paging pattern may be less favourable~\cite{mittal2014embedded,chang2025flashhost}. Reporting ``energy after throttling'' without reporting which actuator moved leaves the reader unable to tell an efficiency gain from a deferred cost.}

\rev{\textbf{Thermal--energy trade-offs by branch.} The four coordination branches stress heat differently. \emph{Asynchronous} wake storms create short peaks that a cooldown can absorb, so the day's dominant branch need not change. \emph{Synchronous} epochs align RF and accelerator heat by construction, so an epoch that is energy-optimal at room temperature can hit sustained throttling and should be lengthened even though the lengthening looks like a loss on a joules-per-task plot. \emph{Frame-slotted} quiet windows are the natural place to park OTA staging and garbage collection, precisely so that a program burst does not ride on top of a paging peak. \emph{Multichannel} days share one package budget across radios and accelerators, so a Wi-Fi OTA download and an NPU decode can throttle each other even when each is comfortable alone---the cross-pillar case that a per-pillar thermal budget cannot express.}

\rev{\textbf{Short peaks versus sustained throttling.} The single most useful distinction is between the two regimes. A short peak rises and falls inside one phase; the schedule change is local, and deferring the next burst or inserting a cooldown is sufficient. Sustained throttling holds the skin or junction limit for minutes and caps DVFS and RF duty cycle simultaneously. A synchronous ``finish inside the epoch'' policy then becomes worse than a longer frame-slotted policy that stretches the epoch and lowers duty cycle, which is a reversal of the room-temperature ranking rather than a small correction to it~\cite{skadron2004hotspot,esmaeilzadeh2011dark}. Gate~7 therefore asks which of the two regimes produced the joule number, not merely whether a temperature sensor was present.}

{
\begin{table*}[!t]

\renewcommand{\arraystretch}{1.12}
\caption{Thermal evidence families on sealed mobile endpoints: what each provides, what it cannot settle, and the coordination regime in which it is most often decisive. Evidence class of the right-hand column: illustrative analytical judgment, not new measurement.}
\label{tab:thermal_evidence}
\centering
\tblsetup\begin{tabularx}{\textwidth}{@{}>{\raggedright\arraybackslash}p{0.15\textwidth}YYY@{}}
\toprule
\textbf{Family} & \textbf{What it provides} & \textbf{What it cannot settle} & \textbf{Where it is decisive} \\
\midrule
Firmware thermal zones (junction, skin, battery)~\cite{skadron2004hotspot} & The only interface available on a shipped product; trip points that the governor actually acts on & Block-level attribution; a quiet zone does not prove an unthrottled die & Any product-level claim; pairs with Gate~4 sensor path \\
Compact RC / HotSpot-class models~\cite{skadron2004hotspot,huang2006hotspot} & Temperature trajectories from a power trace; transient behaviour that a steady-state derate misses & Hand contact, spreader, and case boundary conditions of a sealed phone & Deciding whether a burst is a short peak or the onset of sustained throttling \\
Leakage--temperature feedback~\cite{roy2003leakage,brooks2007power} & Why static power grows once a block is hot, and why margins are not linear & Per-SKU leakage parameters, which vendors do not publish & Long synchronous epochs and sustained inference \\
Power-envelope / dark-silicon allocation~\cite{esmaeilzadeh2011dark} & The framing that only part of the silicon can run fast at once & Which allocation a specific governor will choose & Multichannel days where radios and accelerators compete \\
Throttling as observed actuator~\cite{mittal2014embedded,sultan2014processor} & The schedule change that actually produced the measured energy & Whether the deferred work was cancelled or merely moved & Frame-slotted maintenance windows; OTA and GC placement \\
\bottomrule
\end{tabularx}
\end{table*}
}

\subsection{System-Boundary Accounting}
\label{sec:evaldim:boundary}
\textbf{Gate~8.} Is the boundary device-only, gateway-inclusive, or cloud-inclusive?
Offload and edge caches can invert ``green'' rankings once uplink tails and infrastructure joules are counted~\cite{shi2016edge,jin2020review,barroso2023dasc}. Declare the boundary with every claim.

\rev{\textbf{Three boundaries, used with these names throughout.} \emph{Device-only:} joules inside the endpoint battery or PMIC, excluding the gateway and the cloud. \emph{Gateway-inclusive:} the endpoint plus the phone or hub that terminates the short-range link, still excluding the cloud. \emph{Cloud-inclusive:} device-only joules plus uplink tails plus the infrastructure energy of the offload target. Tables~\ref{tab:case_budget} and~\ref{tab:approach_fail} use these names. The illustrative phase budget is device-only. An offload paper that reports only the endpoint is not comparable to a cloud-inclusive total until both are restated on the same boundary.}

\section{Review of Approaches: Power-Aware Networking}

\label{sec:lpcomm}

\textbf{Low-power communication} on mobile Internet-of-Things (IoT) devices means moving data reliably while most joules go to \emph{time on air}, paging, discovery, security, and DRX tails---often more than to the payload itself~\cite{andreev2015connectivity,patterson2024energy,rajandekar2015m2mmac}. Phones, wearables, and gateways \textbf{run} several stacks (short-range, Wi-Fi, cellular) whose scans, coexistence rules, and sleep ladders keep the AP in shallow sleep even when user traffic is tiny~\cite{mittal2014embedded,larsen2023greeneran}. NB-IoT, LTE-M, and LPWA extend the same story to sparse telemetry: link margin and repeats drive energy, and mobility or interference change which settings stay valid~\cite{andreev2015connectivity,centenaro2016longrange}. Claims about the radio should pass Section~\ref{sec:evaldim} and be read with Section~\ref{sec:lpcompute} when air time and host work are co-tuned~\cite{shi2016edge,jin2020review,barroso2023dasc}. We treat the radio as a \emph{policy-heavy} block: timers from standards (connection interval, DRX, power-saving classes) plus opaque modem firmware set the energy~\cite{andreev2015connectivity,larsen2023greeneran}.

\textbf{Radio stacks instantiate Section~\ref{sec:comm_power}.} Licensed cellular behavior makes the anatomy operational: radio resource control (RRC) \emph{connected} intervals concentrate energy in synchronization, uplink grants, random access, and hybrid automatic repeat request activity---the receive/transmit-heavy phases separated in Section~\ref{sec:comm_power}---whereas RRC \emph{idle} camping trades sleep depth for paging and mobility readiness, mapping to idle/listen, discovery, and shallow retention~\cite{andreev2015connectivity,centenaro2016longrange,andrews20145g}. Discontinuous reception (DRX), extended DRX (eDRX), and power-saving mode (PSM) should be read as \emph{timed wake contracts}: they slot modem activity relative to core-network timers rather than merely lowering a PHY rate~\cite{centenaro2016longrange,rao2018nbiotoverview}; NB-IoT and LTE-M practice reinforces that repetition, coverage extension, and operator-tuned idle behaviors often dominate telemetry energy through procedure time, not payload bits~\cite{andreev2015connectivity,rao2018nbiotoverview}. Short-range stacks repeat the pattern at finer granularity---Bluetooth Low Energy connection events and Wi-Fi delivery traffic indication map (DTIM) periods structure listen windows; neighbor discovery and coexistence scans inflate Section~\ref{sec:comm_power}'s ``idle'' when user-visible throughput is near zero~\cite{andreev2015connectivity,mittal2014embedded}. In all cases, transport and security tails couple the modem to the application processor through direct memory access, interrupts, and retention domains, i.e., the modem--application processor coupling emphasized in Section~\ref{sec:comm_power}~\cite{mittal2014embedded,xie2021apollo}.

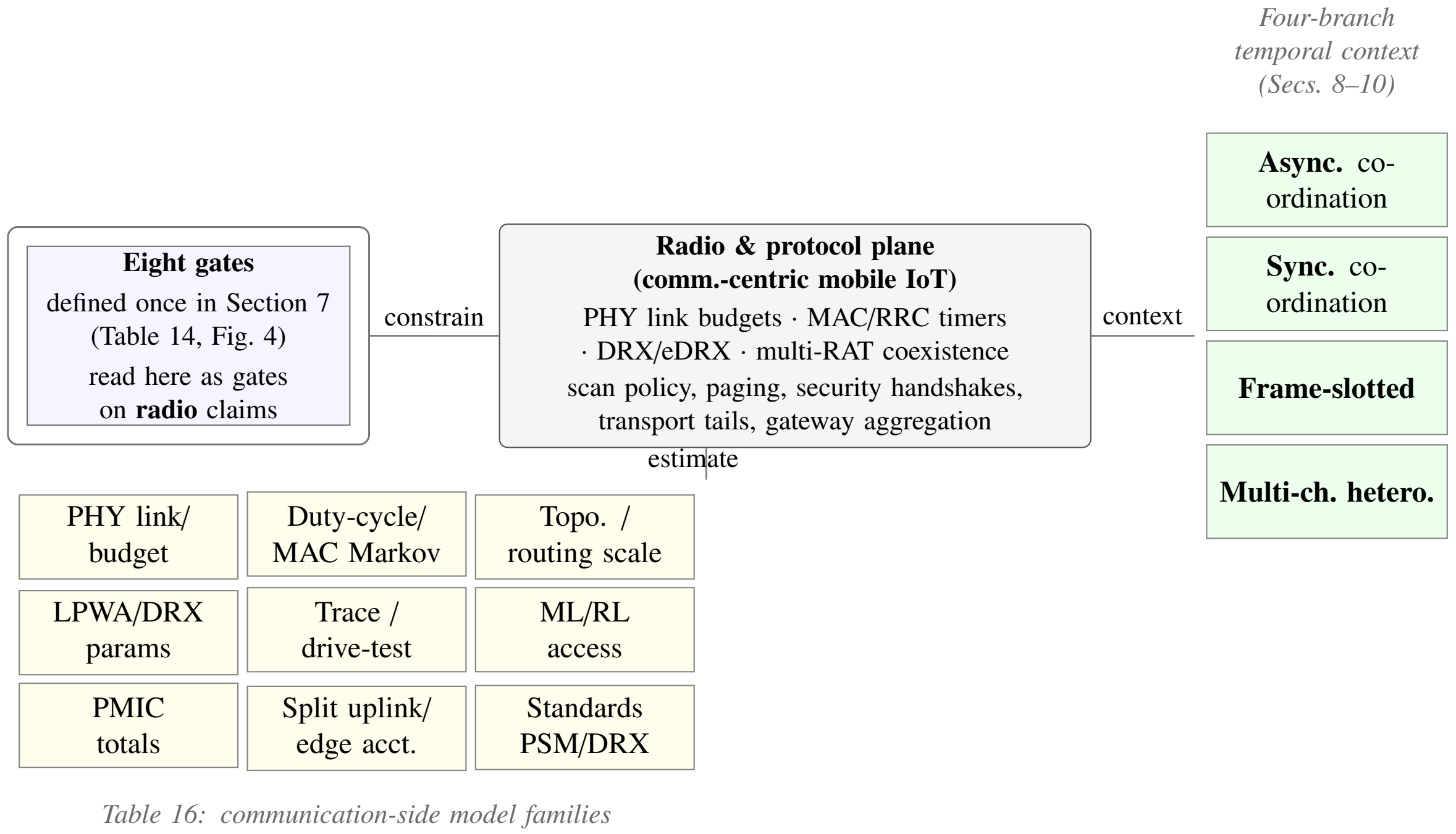
\begin{figure*}[!t]
\centering
\begin{tikzpicture}[
  scale=0.92, transform shape,
  font=\small,
  >={Stealth[length=2.2mm]},
  dim/.style={rectangle, draw=black!55, fill=blue!4, inner sep=2.2pt, minimum height=10mm, text width=24mm, align=center},
  main/.style={rectangle, draw=black!68, fill=gray!8, rounded corners=2.5pt, inner sep=4pt, text width=66mm, align=center},
  mod/.style={rectangle, draw=black!48, fill=yellow!9, inner sep=2pt, text width=22mm, align=center, minimum height=9mm},
  br/.style={rectangle, draw=black!48, fill=green!7, inner sep=2.5pt, text width=24mm, align=center, minimum height=10mm},
  arr/.style={draw=black!50, line width=0.55pt},
  dframe/.style={draw=black!55, line width=0.55pt, rounded corners=3pt, inner sep=2mm},
  fcap/.style={font=\small\itshape, text=black!58, align=center}
]
\node[dim, text width=36mm, minimum height=20mm, anchor=south west] (D2) at (0,0)
  {\textbf{Eight gates}\\[0.8mm]{\small defined once in Section~\ref{sec:evaldim}\\(Table~\ref{tab:eval_gates}, Fig.~\ref{fig:evaldim_logic})}\\[0.8mm]{\small read here as gates on \textbf{radio} claims}};
\node[dframe, fit=(D2)] (D2box) {};

% ========== 原来 +5mm → +12mm，拉大间距，箭头横线变长 ==========
\node[main, anchor=west] (R) at ($(D2box.east)+(15mm,0mm)$) {\textbf{Radio \& protocol plane (comm.-centric mobile IoT)}\\[0.6mm]
{\small PHY link budgets $\cdot$ MAC/RRC timers $\cdot$ DRX/eDRX $\cdot$ multi-RAT coexistence}\\[0.4mm]
{\small scan policy, paging, security handshakes, transport tails, gateway aggregation}};

\draw[arr] (D2box.east) -- (R.west);
\node[font=\small] at ($(D2box.east)!0.5!(R.west)+(0,2.2mm)$) {constrain};

\matrix (M2) [matrix of nodes, below=4mm of D2box.south west, anchor=north west, column sep=2.5pt, row sep=2.2pt, nodes={mod}]
{
  {PHY link/\\budget} & {Duty-cycle/\\MAC Markov} & {Topo. /\\routing scale} \\
  {LPWA/DRX\\params} & {Trace /\\drive-test} & {ML/RL\\access} \\
  {PMIC\\totals} & {Split uplink/\\edge acct.} & {Standards\\PSM/DRX} \\
};
\node[below=1.2mm of M2.south, font=\small\itshape, text=black!58, align=center, text width=72mm] {Table~\ref{tab:lpcomm_models}: communication-side model families};

\draw[arr] (M2.north east) -- ($(R.south west)!0.35!(R.south east)$);
\node[font=\small] at ($(M2.north east)!0.42!($(R.south west)!0.35!(R.south east)$)+(-1.5mm,1.0mm)$) {estimate};

% ========== 原来 +5mm → +12mm，右侧箭头同步拉长 ==========
\matrix (B2) [matrix of nodes, anchor=west, column sep=0pt, row sep=2.8pt, nodes={br}] at ($(R.east)+(12mm,0mm)$)
{
  \textbf{Async.} coordination \\
  \textbf{Sync.} coordination \\
  \textbf{Frame-slotted} \\
  \textbf{Multi-ch.\ hetero.} \\
};
\node[fcap, above=1.5mm of B2.north, text width=28mm] {Four-branch temporal context (Secs.~\ref{sec:lpcomm}--\ref{sec:lpstore})};

\draw[arr] (R.east) -- (B2.west);
\node[font=\small] at ($(R.east)!0.5!(B2.west)+(0,2.2mm)$) {context};
\end{tikzpicture}
\caption{Energy-aware mobile IoT: low-power communication map (evaluation dimensions, model families, four-branch context).}
\label{fig:lpcomm_logic}
\end{figure*}

Figure~\ref{fig:lpcomm_logic} maps the communication pillar: the eight dimensions gate radio claims, modeling families approximate the protocol plane, and the four coordination branches name the temporal regimes in which traces must be read. The exposition follows three blocks---academia and industry (Subsection~\ref{sec:lpcomm:status}); relations to evaluation dimensions and branches (Subsection~\ref{sec:lpcomm:relations}); open problems, classic models, and a comparative table (Subsection~\ref{sec:lpcomm:models})---using the same pillar template as Sections~\ref{sec:lpcompute} and~\ref{sec:lpstore}. Section~\ref{sec:lpcompute} develops the compute pillar next; Section~\ref{sec:lpstore} applies the same template to non-volatile storage and maintenance workloads.

\subsection{State of the Art in Academia and Industry}
\label{sec:lpcomm:status}

\textbf{Academia.} Duty-cycled MAC, preamble sampling, and rendezvous models still explain paging and advertising energy~\cite{rajandekar2015m2mmac,chang2017sensors,akyildiz2002wsn,ye2002sensormac,polastre2005telos}. Topology and aggregation studies explain sink/gateway hotspots~\cite{shi2016edge,chang2018wsntopology,chang2017wsnrouting}. Connectivity surveys document cellular IoT modes, DRX/power-saving classes, and LPWA tradeoffs~\cite{andreev2015connectivity,centenaro2016longrange,raza2017lpwanoverview,sun2022loracomprehensive,haxhibeqiri2018lorawanSurvey,diane2025lpwan,islam2024lpwan}; RedCap/eRedCap extend mid-tier IoT options~\cite{veedu2022redcap,joerke2025redcap,raja2026redcapsurvey}. \rev{The LPWA branch of that literature is worth separating, because it is where per-packet energy was studied most carefully and where the limits of a link-centric account are clearest. Range and coverage campaigns established what a single uplink costs in the field~\cite{petajarvi2015loracoverage,lauridsen2016coverageiot,augustin2016lorastudy}; scaling analyses then showed that the same uplink becomes expensive once collisions and repeats are counted, so a duty-cycle figure quoted for an isolated node does not survive network density~\cite{bor2016lorascaling,georgiou2017lorascale,adelantado2017lorawanlimits}. Comparative studies place those numbers next to NB-IoT and SigFox rather than treating one technology in isolation~\cite{mekki2019lpwancomparative,finnegan2018lpwasurvey,du2020loranetworkingsurvey}, and lightweight scheduling work shows that the energy ranking of two LPWA configurations can invert once the network coordinates access~\cite{reynders2018lorasched}. On the cellular side, narrowband IoT primers and random-access analyses make the same point about signalling: for sparse telemetry the attach, paging, and access procedures, not the payload, set the energy scale~\cite{wang2017nbiotprimer,laya2014lteaccessm2m}. Each of these results is a branch-dependent statement in the vocabulary of Section~\ref{sec:logic_framework}---they hold for frame-slotted access and say little about an asynchronous phone day---which is exactly the context that headline duty-cycle comparisons tend to drop.} Learning-based access control adapts under nonstationary interference at the cost of certifiability~\cite{chang2025dqngran,chang2023drlmultibs}. \rev{A related shift is \emph{what} is transmitted, not only \emph{when}. Zhang et al.~\cite{zhang2024semcomgraph} argue that graph-structured data---common in IoT topologies, contact graphs, and sensing relations---carry semantic information that is more abstract than images or text, and they propose topology-aware graph semantic coding with a graph information-bottleneck joint semantic--channel scheme (GIB-JSCC) that drops task-irrelevant structure, with particular gains at low SNR. For sealed endpoints that is not a free radio win: filtering bits can shorten RX/TX, but the encoder runs on the AP or NPU and still lands inside the same DRX, scan, and modem--AP tail envelope. Gate~1 must therefore count encode energy plus tails, Gate~8 must declare whether the claim is device-only, and Gate~2 must keep the graph-classification or delivery metric that justified the semantic cut. The paper is a SemCom design for graph payloads, not a PMIC measurement of phone sleep; we cite it as the communication-side counterpart of the survey's compute$\rightarrow$radio coupling (compress or distill before airtime).} The bottleneck has shifted from ``how to sleep'' to \textbf{how to attribute} tails and multi-RAT activity when protocol traces, PMIC totals, and AP wakeups disagree~\cite{patterson2024energy,reda2012survey}.

\textbf{Industry.} Products ship as coupled modem--AP stacks: scan budgets, carrier policy, and coexistence change energy without changing application bytes~\cite{andreev2015connectivity,mittal2014embedded,larsen2023greeneran,chang2018wsntopology}. Module measurements now ground RedCap spreadsheet models~\cite{joerke2025redcap,veedu2022redcap}. Field practice relies on drive tests and proprietary tools rather than fully open PHY--MAC stacks~\cite{andreev2015connectivity,patterson2024energy}, motivating declared observability and branch context before trusting headline air-interface numbers~\cite{alfuqaha2015iot,reda2012survey}.

\subsection{Reading Communication Work Through the Gates and Branches}
\label{sec:lpcomm:relations}

\rev{\textbf{Gate and branch crosswalk (not restated).} Gate questions and branch definitions stay in Sections~\ref{sec:evaldim} and~\ref{sec:logic_framework} (Tables~\ref{tab:eval_gates}, \ref{tab:fourway}, \ref{tab:branch_decision}). Communication-specific stresses are Gate~1, when tails and discovery dominate payload bits~\cite{pathak2012whereenergy,andreev2015connectivity}; Gate~4, when one PMIC rail hides the active RAT~\cite{patterson2024energy}; and Gate~5, when Wi-Fi and cellular share the front end~\cite{xie2021apollo}. Segment PMIC-only traces with the anatomy in Section~\ref{sec:comm_power}. Declare the dominant branch before ranking a radio knob.}

\subsection{Scenarios, Open Issues, and Modeling Families}
\label{sec:lpcomm:models}

\textbf{Open problems} recur in both surveys and product post-mortems~\cite{patterson2024energy,reda2012survey}. (1)~\textbf{Reproducible radio ground truth} on sealed products: identical byte patterns can yield different energy under different firmware builds and carrier configurations~\cite{andreev2015connectivity,mittal2014embedded}. (2)~\textbf{Multi-RAT attribution} when PMIC totals obscure which interface dominated a mixed trace~\cite{patterson2024energy,xie2021apollo}. (3)~\textbf{Mobility and interference nonstationarity}, which break steady Markov abstractions used in classical duty-cycle analyses~\cite{chang2018wsntopology,chang2025dqngran}. Classic \textbf{modeling families} include link-budget and rate--power abstractions, duty-cycle and sleep--wake Markov models from the WSN literature, topology- and routing-aware scaling laws, standards-parameterized DRX and LPWA models, trace- and drive-test regressions, learning-based schedulers, and split-uplink system models~\cite{andreev2015connectivity,centenaro2016longrange,patterson2024energy,shi2016edge,rajandekar2015m2mmac,larsen2023greeneran,chang2018wsntopology,jin2020review,chang2025dqngran,sun2022loracomprehensive,haxhibeqiri2018lorawanSurvey,islam2024lpwan,jafari2024iotrouting}.

Table~\ref{tab:lpcomm_models} summarizes these families with the same intent as Table~\ref{tab:lpcompute_models}: scenario fit, strengths, weaknesses, and where each family aligns or clashes with the four branches~\cite{andreev2015connectivity,rajandekar2015m2mmac}.

\begin{table*}[!t]
\renewcommand{\arraystretch}{1.15}
\caption{Classic communication energy and modeling families for mobile IoT: scenario fit, strengths, weaknesses, and energy-aware four-branch alignment.}
\label{tab:lpcomm_models}
\centering
\tblsetup\begin{tabularx}{\textwidth}{@{}>{\raggedright\arraybackslash}p{0.10\textwidth}YYYY@{}}
\toprule
\textbf{Family} & \textbf{Scenario fit} & \textbf{Strengths} & \textbf{Weaknesses} & \textbf{Four-branch note} \\
\midrule
PHY link-budget / rate--power~\cite{andreev2015connectivity} & Coverage planning, static channel assumptions & Interpretable SNR--rate tradeoffs & Misses MAC tails, mobility, multi-RAT & Useful in \textbf{frame}-structured planning; weak under \textbf{async} mobility bursts \\
WSN duty-cycle / sleep--wake Markov~\cite{rajandekar2015m2mmac,chang2017sensors} & Preamble-style MAC, low-rate telemetry & Captures idle listening vs.\ sleep tradeoffs & Poor fit to cellular state machines & Strong for \textbf{async} rendezvous metaphors \\
Topology- / routing-aware scaling~\cite{chang2018wsntopology,chang2017wsnrouting} & Aggregation trees, gateway hotspots & Explains sink-adjacent congestion & Less calibrated to cellular PHY details & Aligns with \textbf{sync} cluster and hub bottlenecks \\
LPWA / cellular DRX \& PSM parameterization~\cite{andreev2015connectivity,centenaro2016longrange,larsen2023greeneran,veedu2022redcap} & NB-IoT, LTE-M, RedCap, power-saving classes & Names standards-visible knobs & Vendor firmware still hidden & Natural fit to \textbf{frame-slotted} regimes \\
Trace \& drive-test regression~\cite{patterson2024energy,mittal2014embedded} & Product tuning, operator acceptance & Matches shipped behavior & Weak transfer across SKUs & Best for \textbf{multi}-radio days with rich logging \\
Learning-based access / resource control~\cite{chang2025dqngran,chang2023drlmultibs} & Dense, uncertain interference settings & Adapts to nonstationary traffic & Data-hungry; hard to certify & Promising under \textbf{async} / bursty traffic if retrained \\
PMIC / rail totals without PHY detail~\cite{patterson2024energy,schone2015energy} & End-to-end budgeting when modem is opaque & Honest about observability limits & No structural lever for MAC designers & Common baseline for all branches \\
Split uplink / edge accounting~\cite{shi2016edge,jin2020review,barroso2023dasc} & Offload and compression pipelines & Surfaces boundary shifts & Needs infrastructure-side data & Cross-cuts \textbf{multi}-RAT and OTA-heavy phases \\
\bottomrule
\end{tabularx}
\end{table*}

{
\begin{table*}[!t]

\renewcommand{\arraystretch}{1.08}
\caption{Analytical comparison of communication model families (illustrative judgments, not new measurements). Complements Table~\ref{tab:lpcomm_models}. Port.\ = portability across SKUs; Therm.\ = thermal awareness; Comp.\ = cross-pillar composability.}
\label{tab:lpcomm_anal}
\centering
\tblsetup\scriptsize
\begin{tabularx}{\textwidth}{@{}>{\raggedright\arraybackslash}p{0.13\textwidth}YYYYYccc@{}}
\toprule
\textbf{Family} & \textbf{Scenario} & \textbf{Assumptions} & \textbf{Observability} & \textbf{Strength} & \textbf{Weakness} & \textbf{Port.} & \textbf{Therm.} & \textbf{Comp.} \\
\midrule
PHY link budget & Static coverage & SNR maps to rate; no MAC state & PHY parameters & Interpretable & Misses tails & Med & Low & Low \\
Duty-cycle Markov & Preamble MAC & Stationary sleep/wake & Sleep statistics & Idle vs.\ sleep & Poor cellular fit & Med & Low & Low \\
Topology scaling & Gateway trees & Aggregation dominates & Routing graph & Sink hotspots & Weak cellular PHY & Low & Low & Med \\
DRX / PSM tables & NB-IoT, LTE-M & Network honors timers & Standards timers & Replayable knobs & Firmware hidden & Med & Low & Med \\
Drive-test regression & Product tuning & Trace matches deploy & PMIC + modem log & Matches shipping & Weak transfer & Low & Low & Med \\
Learned access & Nonstationary interference & Retrain on shift & Rich traces & Adapts & Hard to certify & Low & Low & Low \\
PMIC totals & Opaque modem & Rail is the ground truth & Package current & Honest limit & No MAC lever & High & Low & Low \\
Split uplink & Offload & Cloud joules countable & Device + edge & Boundary shift & Needs infra data & Med & Low & High \\
\bottomrule
\end{tabularx}
\end{table*}
}

Beyond Table~\ref{tab:lpcomm_models}, three \textbf{scenario patterns} recur in mobile IoT communication studies~\cite{shi2016edge,alfuqaha2015iot}. \textbf{Always-on adjacency} (peripheral BLE, wearable tethering) stresses advertising, connection supervision timeouts, and coexistence scans~\cite{andreev2015connectivity,mittal2014embedded}. \textbf{Sparse licensed uplink} (NB-IoT / LTE-M telemetry) stresses repetitions, DRX tradeoffs, and mobility-induced reselection~\cite{centenaro2016longrange,chang2018wsntopology}. \textbf{Hub-centric aggregation} concentrates downlink and scanning load on gateways that also run application logic, coupling RF heat to CPU scheduling~\cite{shi2016edge,skadron2004hotspot,chang2018wsntopology}. Choosing a modeling family is again choosing which failure modes are visible early: PHY spreadsheets expose coverage--rate levers but may miss transport tails; trace models fit products but obscure where energy concentrates inside the modem~\cite{patterson2024energy,reda2012survey}.

\par

\textbf{Communication-specific synthesis.} On mobile IoT, \emph{policy}---DRX, scans, power-saving classes, modem sleep ladders---often matters more than raw PHY curves~\cite{andreev2015connectivity,centenaro2016longrange,larsen2023greeneran}. Each named timer is an instance of Section~\ref{sec:comm_power}'s idle vs.\ tail tradeoff: longer DRX can cut average power but hurt freshness; faster paging can improve latency but keep shallow sleep~\cite{patterson2024energy,rajandekar2015m2mmac}. Two lessons vs.\ compute (Section~\ref{sec:lpcompute}): report \textbf{standards-visible parameters} (connection interval, eDRX, paging window) with mean power and tail share, or others cannot replay your setup~\cite{andreev2015connectivity,rajandekar2015m2mmac}; treat \textbf{multi-RAT} as real energy---Wi-Fi/BLE work during ``cellular idle'' can hold RF and LPDDR, so PMIC splits beat summing PHY sheets~\cite{patterson2024energy,xie2021apollo}. \rev{Semantic graph coding~\cite{zhang2024semcomgraph} sits on the same ladder: it can change payload size and task fidelity, but it does not retire the need to report tails, branch, and boundary.} Read Table~\ref{tab:lpcomm_models} as a ladder from intent (PHY/Markov) to reproducible knobs (standards) to what users pay for (PMIC / split-uplink). Section~\ref{sec:lpstore} does the same for flash, with FTL phases in the role of DRX~\cite{patterson2024energy,mittal2014embedded}.

\section{Review of Approaches: Energy- and Thermal-Aware Compute}

\label{sec:lpcompute}

\begin{figure*}[!t]
\centering
\begin{tikzpicture}[
  scale=0.92, transform shape,
  font=\small,
  >={Stealth[length=2.2mm]},
  dim/.style={rectangle, draw=black!55, fill=blue!4, inner sep=2.2pt, minimum height=10mm, text width=24mm, align=center},
  main/.style={rectangle, draw=black!68, fill=gray!8, rounded corners=2.5pt, inner sep=4pt, text width=66mm, align=center},
  mod/.style={rectangle, draw=black!48, fill=yellow!9, inner sep=2pt, text width=22mm, align=center, minimum height=9mm},
  br/.style={rectangle, draw=black!48, fill=green!7, inner sep=2.5pt, text width=24mm, align=center, minimum height=10mm},
  arr/.style={draw=black!50, line width=0.55pt},
  dframe/.style={draw=black!55, line width=0.55pt, rounded corners=3pt, inner sep=2mm},
  fcap/.style={font=\small\itshape, text=black!58, align=center}
]
\node[dim, text width=36mm, minimum height=20mm, anchor=south west] (D) at (0,0)
  {\textbf{Eight gates}\\[0.8mm]{\small defined once in Section~\ref{sec:evaldim}\\(Table~\ref{tab:eval_gates}, Fig.~\ref{fig:evaldim_logic})}\\[0.8mm]{\small read here as gates on \textbf{compute} claims}};
\node[dframe, fit=(D)] (Dbox) {};

\node[main, anchor=west] (C) at ($(Dbox.east)+(15mm,0mm)$) {\textbf{Low-power compute plane (sealed mobile IoT SoC)}\\[0.6mm]
{\small CPU clusters $\cdot$ always-on islands $\cdot$ DSP/ISP $\cdot$ NPU/accelerators $\cdot$ security blocks}\\[0.4mm]
{\small LLC, interconnect, DMA toward modem/storage, LPDDR bandwidth vs.\ arithmetic}};
\path[arr] (Dbox.east) -- node[above, font=\small, yshift=0.3mm] {constrain} (C.west);

\matrix (M) [matrix of nodes, below=4mm of Dbox.south west, anchor=north west, column sep=2.5pt, row sep=2.2pt, nodes={mod}]
{
  {Wattch/\\McPAT-class} & {Hybrid\\calibration} & {gem5-\\class} \\
  {Accel-Sim/\\Wattch} & {PMC/\\RAPL-class} & {ML\\surrogate} \\
  {RTL/\\gate} & {DRAM\\cmd trace} & {SimPoint/\\phases} \\
};
\node[below=1.2mm of M.south, font=\small\itshape, text=black!58, align=center, text width=72mm] {Table~\ref{tab:lpcompute_models}: model families (fidelity vs.\ cost vs.\ observability)};
\path[arr] (M.north east) -- ($(C.south west)!0.35!(C.south east)$) node[pos=0.42, left, font=\small, xshift=-0.5mm] {estimate};

\matrix (B) [matrix of nodes, anchor=west, column sep=0pt, row sep=2.8pt, nodes={br}] at ($(C.east)+(12mm,0mm)$)
{
  \textbf{Async.} coordination \\
  \textbf{Sync.} coordination \\
  \textbf{Frame-slotted} \\
  \textbf{Multi-ch.\ hetero.} \\
};
\node[fcap, above=1.5mm of B.north, text width=28mm] {Four-branch temporal context (Secs.~\ref{sec:lpcomm}--\ref{sec:lpstore})};
\path[arr] (C.east) -- node[above, font=\small, yshift=0.3mm] {context} (B.west);
\end{tikzpicture}
\caption{Energy-aware mobile IoT: low-power compute map (evaluation dimensions, model families, four-branch context).}
\label{fig:lpcompute_logic}
\end{figure*}
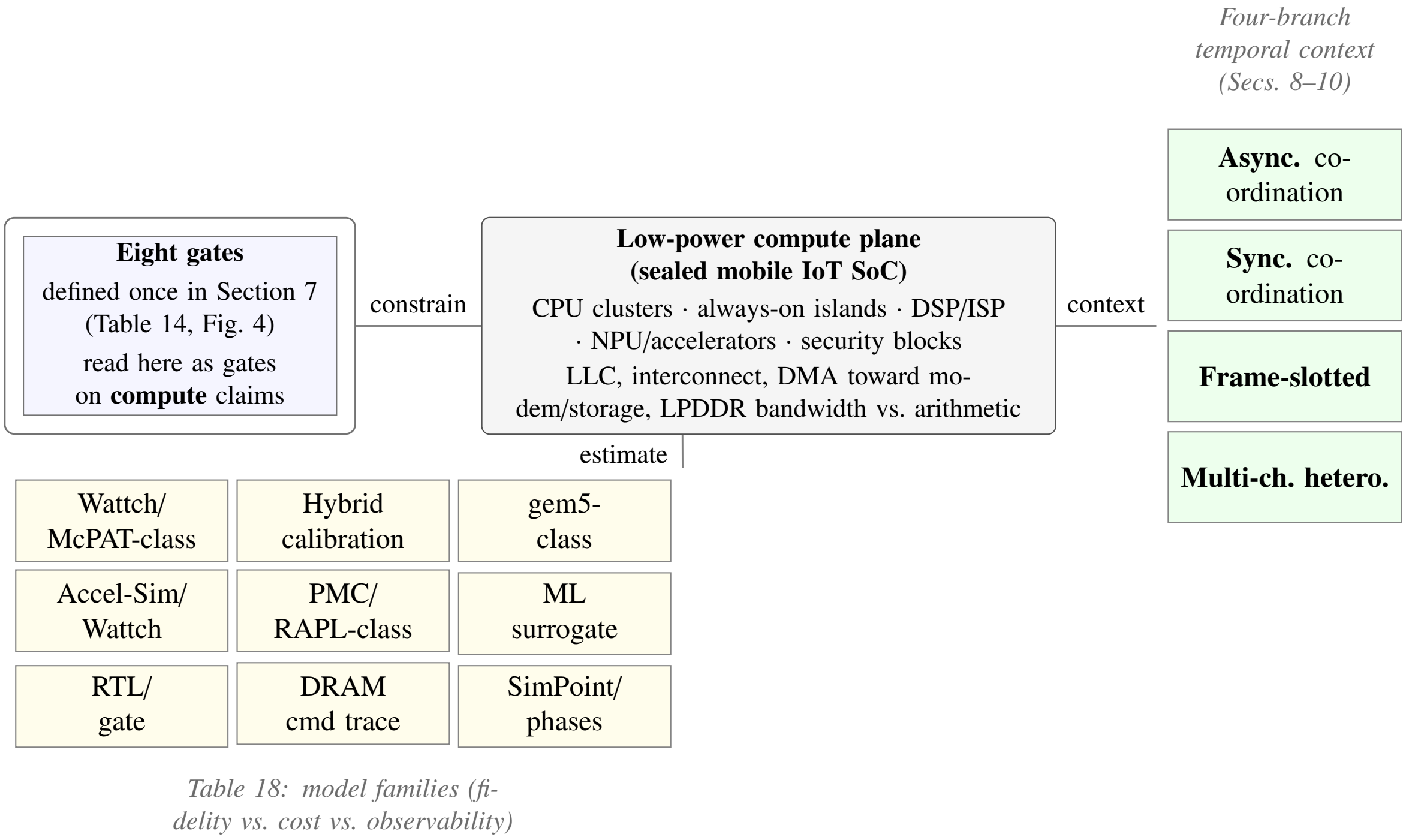

\textbf{Low-power compute} on mobile Internet-of-Things (IoT) endpoints is the art of sustaining sensing, control, and local intelligence under milliwatt-to-watt envelopes where \emph{useful work per joule} is dominated as often by data movement, sleep-state depth, and thermal headroom as by arithmetic throughput~\cite{mittal2014embedded,hennessy2019book,chen2016eyeriss,parashar2019timeloop}. Handheld hubs, wearables, and industrial gateways combine heterogeneous application processors, digital signal processing (DSP) and imaging pipelines, increasingly general-purpose neural accelerators, and security islands whose power states are orchestrated by firmware and operating-system (OS) governors that were not designed for academic microbenchmarks alone~\cite{mittal2014embedded,sultan2014processor}. Because the same silicon must coexist with duty-cycled radios and with non-volatile storage stacks that can block deep sleep, compute-centric optimizations---dynamic voltage and frequency scaling (DVFS), aggressive clock and power gating, precision reduction, operator fusion, and tiling for locality---must be judged against \textbf{end-to-end} budgets grounded in Section~\ref{sec:comm_power}'s wireless phases (idle/listen, RX/TX, tails, modem--application processor coupling) and gated by Section~\ref{sec:evaldim}~\cite{andreev2015connectivity,patterson2024energy,skadron2004hotspot}. This section therefore treats compute neither as an isolated central processing unit (CPU) curve nor as a single accelerator kernel, but as a \emph{time-varying allocation} of activity across cores, accelerators, caches, and dynamic random-access memory (DRAM) traffic that is coupled to connectivity and persistence~\cite{shi2016edge,chandrasekar2012drampower,xie2021apollo}. Figure~\ref{fig:lpcompute_logic} sketches the same relationships in compact graphical form: Section~\ref{sec:evaldim} dimensions as \emph{gates} on compute claims, classic modeling families as estimators of the compute plane, and the energy-aware four coordination branches as the temporal context in which traces and models must be interpreted. The exposition is organized in three blocks: a decade-spanning snapshot of academic and industrial practice (Subsection~\ref{sec:lpcompute:status}); a cross-dimensional analysis that ties compute to the eight evaluation axes and to the energy-aware four coordination branches (Subsection~\ref{sec:lpcompute:relations}); and a problem- and model-oriented synthesis with a comparative table of classic modeling families, their best-fit scenarios, and their limitations (Subsection~\ref{sec:lpcompute:models}).

\subsection{State of the Art in Academia and Industry}
\label{sec:lpcompute:status}

\textbf{Academia.} Measurement discipline and model composability now rival arithmetic innovations~\cite{patterson2024energy,reda2012survey}. Analytical (Wattch/McPAT), simulation (gem5), hybrid calibration, and accelerator/trace tools (Timeloop, Accel-Sim/AccelWattch) form the interpretable backbone~\cite{brooks2000wattch,li2009mcpat,binkert2011gem5,lowe2020gem5,walker2018gem5,zhai2022mcpatcalib,lee2015powertrain,chen2016eyeriss,parashar2019timeloop,mahajan2020accelsim,kandiah2021accelwattch}. \rev{Regression from microarchitectural events to power predates the learned surrogates now in fashion and already exposed their central weakness, namely that a fit is only valid inside the configuration space it was trained on~\cite{lee2006accurate}. The energy floor that all of these tools describe is set by data movement rather than arithmetic: an operand fetched from off-chip memory costs orders of magnitude more than the operation consuming it~\cite{horowitz2014computingenergy}, which is why on a sealed endpoint a model that counts only compute activity will mis-rank two kernels with the same operation count and different memory traffic.} On-device SLM/LLM and TinyML benchmarks make joules/token and memory traffic first-class~\cite{lu2024slm,xu2024devicellm,heydari2025tinyml,husom2025sustainablellm}. \rev{The efficiency techniques those benchmarks evaluate came from the mobile-vision line of work---compact architectures, pruning with trained quantization, and integer-only inference~\cite{howard2017mobilenets,iandola2016squeezenet,han2016deepcompression,jacob2018intquant}, surveyed together with the dataflow considerations that decide whether a reduction in parameters becomes a reduction in joules~\cite{sze2017efficientdnn}. Standardized embedded suites make those claims comparable across devices~\cite{banbury2021mlperftiny}, but they are still reported at the level of one inference: the endpoint question of whether a smaller model preserved deep-sleep eligibility, or merely moved work into an open radio tail, falls outside the benchmark boundary and is exactly what Gate~8 asks for.} Learned surrogates raise fidelity but hurt interpretability~\cite{kumar2023mlmicro,zhai2024mlsurvey,zhai2023aspdac}; APOLLO-class contention breaks naive additivity~\cite{chandrasekar2012drampower,xie2021apollo}.

\textbf{Industry.} Purpose-partitioned hubs, opaque sleep ladders, and OS-integrated assistants push batching, precision, and decode-core choice into default paths that couple DRAM, flash checkpoints, and radio offload~\cite{mittal2014embedded,lu2024slm,xu2024devicellm,friha2024llmedge,shi2016edge}. Thermal and skin limits often dominate peak TOPS marketing~\cite{esmaeilzadeh2011dark,skadron2004hotspot}. Vendor counters and PMICs remain the practical ground truth~\cite{hackenberg2013rapl,schone2015energy,singh2020granular}, which is why Section~\ref{sec:evaldim} treats observability as a gate.

\subsection{Reading Compute Work Through the Gates and Branches}
\label{sec:lpcompute:relations}

\rev{\textbf{Gate and branch crosswalk (not restated).} Definitions stay in Sections~\ref{sec:evaldim} and~\ref{sec:logic_framework}. Compute-specific stresses are Gate~1, when fewer active cycles increase wake--sleep or shallow retention for the modem~\cite{mittal2014embedded}; Gate~5, when a new NPU or LLC partition, or modem DMA, breaks an otherwise portable CPU model~\cite{xie2021apollo,chandrasekar2012drampower}; Gate~6, when a learned surrogate drops the standards-visible knob~\cite{zhai2024mlsurvey}; and Gate~7, when sustained inference throttles the schedule~\cite{skadron2004hotspot}. Name the branch per phase, separate core, accelerator, LLC, and DRAM in the boundary, and do not publish accelerator joules without that contention context~\cite{reda2012survey}.}

\subsection{Scenarios, Open Issues, and Modeling Families}
\label{sec:lpcompute:models}

Three intertwined \textbf{open problems} recur in the literature and in post-mortems of commercial mobile IoT products~\cite{patterson2024energy,reda2012survey}. (1)~\textbf{Reproducible ground truth} on sealed products: performance counter semantics drift across firmware, and PMIC-integrated current hides subsystem attribution~\cite{patterson2024energy,mittal2014embedded,hackenberg2013rapl}. (2)~\textbf{Composable end-to-end models} that avoid double counting DRAM or LLC when CPU, modem, and accelerator proxies overlap~\cite{chandrasekar2012drampower,xie2021apollo}. (3)~\textbf{Workload phase stationarity}: local learning, voice, and mobility break the long stationary traces that classical microarchitectural modeling assumed~\cite{wunderlich2003simpoints,kumar2023mlmicro,zhai2024mlsurvey}. Classic \textbf{algorithms} in the modeling sense include event-to-energy aggregation (Wattch-style), analytical configuration scaling (McPAT-style), statistical calibration to traces (PowerTrain / McPAT-Calib), cycle-accurate simulation (gem5), trace-driven accelerator simulation (Accel-Sim), and learned mapping from hardware counters to power (ML surrogates)~\cite{kumar2023mlmicro,brooks2000wattch,li2009mcpat,binkert2011gem5,lowe2020gem5,zhai2022mcpatcalib,lee2015powertrain,mahajan2020accelsim}. Each occupies a different point in the accuracy--cost--interpretability space; none is universally dominant for mobile IoT without additional assumptions~\cite{reda2012survey}.

Table~\ref{tab:lpcompute_models} consolidates these families for quick comparison. The ``scenario fit'' column is intentionally pragmatic: it names the evaluation contexts where a family is \emph{most} informative, not exclusive domains. The ``four-branch note'' column links each family to coordination regimes where its assumptions are strongest or weakest~\cite{andreev2015connectivity,rajandekar2015m2mmac}. Readers should treat the strengths and weaknesses as design-time heuristics: shipping systems often combine families (for example, gem5 exploration plus silicon calibration) rather than deploying one in isolation~\cite{patterson2024energy,lee2015powertrain,walker2018gem5}.

\begin{table*}[!t]
\renewcommand{\arraystretch}{1.15}
\caption{Classic compute power-modeling families for mobile IoT: scenario fit, strengths, weaknesses, and energy-aware four-branch alignment.}
\label{tab:lpcompute_models}
\centering
\tblsetup\begin{tabularx}{\textwidth}{@{}>{\raggedright\arraybackslash}p{0.10\textwidth}YYYY@{}}
\toprule
\textbf{Family} & \textbf{Scenario fit} & \textbf{Strengths} & \textbf{Weaknesses} & \textbf{Four-branch note} \\
\midrule
Event-based analytical (Wattch-style)~\cite{brooks2000wattch} & Early architecture exploration when event streams are known & Interpretable per-structure energy; fast iteration & Misses firmware sleep, accelerators, and I/O; weak for sealed SoCs & Strong under \textbf{synchronous} steady phases; brittle under \textbf{async} wake storms \\
Config-driven analytical (McPAT-class)~\cite{li2009mcpat,zhai2022mcpatcalib} & Tape-out planning and technology scaling studies & Technology knobs; block-level budgeting & Requires accurate config; may not track vendor gating & Useful when \textbf{frame} timing fixes activity; weak when \textbf{multi}-DMA dominates \\
Hybrid calibration (PowerTrain / McPAT-Calib)~\cite{zhai2022mcpatcalib,lee2015powertrain} & Aligning analytical stacks to silicon traces & Bridges model gap to measured totals & Calibration effort per SKU; sensitive to trace coverage & Best when \textbf{sync}/\textbf{frame} phases are repeatable \\
Full-system simulation (gem5)~\cite{binkert2011gem5,lowe2020gem5,walker2018gem5} & Contention and OS effects before silicon & Contention-sensitive what-if; reproducible & Slow; incomplete radio/storage fidelity & Strong for \textbf{multi}-resource contention stories if models exist \\
GPU / accel.\ trace-driven (Accel-Sim + AccelWattch)~\cite{mahajan2020accelsim,kandiah2021accelwattch} & Vision and inference kernels on spatial accelerators & Captures memory traffic effects on energy & Setup cost; GPU-centric assumptions & Fits framed ``quiet window'' batching; less for tiny MCU-class \\
PMC / RAPL-class runtime~\cite{hackenberg2013rapl,schone2015energy,singh2020granular} & On-endpoint profiling when counters exist & Fast; correlates with user traces & Coarse domains; semantics change by SKU & Good \textbf{async} telemetry if tails are separated \\
ML / DL power surrogates~\cite{kumar2023mlmicro,zhai2024mlsurvey,zhai2023aspdac} & Many workloads per SKU with rich counters & Accuracy potential across workloads & Data hunger; opaque failure modes & Risky under \textbf{async} regime shifts without retrain \\
On-device LLM / SLM benchmarks~\cite{lu2024slm,xu2024devicellm,husom2025sustainablellm,heydari2025tinyml} & Assistant, voice, edge-gateway inference & Reports tokens/s, memory, joules & Rapid model churn; OS/runtime opaque & Decode-heavy; couples to \textbf{frame} and \textbf{multi}-radio days \\
RTL / gate inference (PRIMAL / Simmani)~\cite{zhou2019primal,kim2019simmani} & Open hubs and sign-off flows & High fidelity when netlists/toggles exist & Costly; vendor IP limits & Strong for \textbf{sync} blocks; rare for full \textbf{multi}-radio days \\
DRAM command-trace models~\cite{chandrasekar2012drampower} & Memory-bandwidth-bound inference & Separates DRAM from core attribution & Trace availability; composability with core tools & Critical when \textbf{frame} compute batches stress LPDDR \\
Phase segmentation (SimPoint-style)~\cite{reda2012survey,wunderlich2003simpoints} & Measurement and modeling methodology & Prevents single-number averages from lying & Requires representative traces & Essential for \textbf{async} and OTA-heavy days \\
\bottomrule
\end{tabularx}
\end{table*}

{
\begin{table*}[!t]

\renewcommand{\arraystretch}{1.08}
\caption{Analytical comparison of compute model families (illustrative judgments). Complements Table~\ref{tab:lpcompute_models}.}
\label{tab:lpcompute_anal}
\centering
\tblsetup\scriptsize
\begin{tabularx}{\textwidth}{@{}>{\raggedright\arraybackslash}p{0.14\textwidth}YYYYYccc@{}}
\toprule
\textbf{Family} & \textbf{Scenario} & \textbf{Assumptions} & \textbf{Observability} & \textbf{Strength} & \textbf{Weakness} & \textbf{Port.} & \textbf{Therm.} & \textbf{Comp.} \\
\midrule
Wattch-style & Early exploration & Events known; no firmware sleep & Event counts & Per-structure energy & Misses I/O and sleep & Med & Low & Low \\
McPAT-class & Technology scaling & Config matches silicon & Block config & Technology knobs & Vendor gating drifts & Med & Low & Low \\
Hybrid calibration & Silicon alignment & Trace covers the phases & Sim + PMIC & Closes the gap & Per-SKU effort & Low & Low & Med \\
gem5 & Pre-silicon contention & Radio/storage models exist & Full-system events & OS contention & Slow; incomplete I/O & Med & Low & Med \\
Accel-Sim & Vision kernels & GPU-like spatial accel. & Kernel traces & Memory traffic & Setup cost & Low & Low & Low \\
PMC / RAPL & On-device profiling & Counter semantics stable & HW counters & Fast & Coarse; SKU drift & Low & Low & Low \\
ML surrogate & Many workloads / SKU & Train set covers regimes & Rich counters & Accuracy & Opaque failures & Low & Low & Low \\
SLM / LLM bench & On-device language & Runtime stays fixed & Tokens, DRAM, joules & End-to-end & Model churn & Low & Med & Med \\
RTL / gate & Sign-off blocks & Netlist available & Toggles & High fidelity & Vendor IP & Low & Med & Low \\
DRAM command trace & Bandwidth-bound & Commands visible & DRAM trace & Separates DRAM & Hard to compose & Med & Low & Med \\
Phase segmentation & Nonstationary days & Trace is representative & Long traces & Stops false averages & Needs coverage & High & Low & High \\
\bottomrule
\end{tabularx}
\end{table*}
}

Beyond Table~\ref{tab:lpcompute_models}, three \textbf{scenario patterns} dominate mobile IoT compute discussions~\cite{mittal2014embedded,shi2016edge}. \textbf{Always-on perception} (voice wake, step counting) stresses shallow sleep and small periodic bursts; models must capture always-on islands and interrupt storms~\cite{mittal2014embedded,sultan2014processor}. \textbf{Bursty local learning} stresses DRAM and tensor engines; analytical models need tiling metadata while trace-driven tools need representative inputs~\cite{chen2016eyeriss,parashar2019timeloop,mahajan2020accelsim,lu2024slm,husom2025sustainablellm}. \textbf{On-device language inference} (SLMs and quantized LLMs) adds decode-dominated phases where CPU cluster choice and LPDDR bandwidth often dominate MAC counts~\cite{xu2024devicellm,lu2024slm}. \textbf{Gateway aggregation} stresses sustained throughput and thermal coupling while forwarding traffic from many leaf nodes~\cite{shi2016edge,chang2018wsntopology,xie2021apollo}. Choosing a model family is therefore choosing which failures a team is willing to see early: interpretability-heavy stacks surface lever points but may lie about tails; fidelity-heavy stacks may hide where errors concentrate~\cite{patterson2024energy,reda2012survey,zhai2024mlsurvey}.

\par

\textbf{Compute-specific synthesis.} The recurring failure mode for low-power compute on mobile IoT is not arithmetic inefficiency but \emph{phase mis-attribution}: a kernel that is efficient under a steady benchmark migrates to a different core, a different sleep depth, and a different contention regime once a duty-cycled radio or background flush is added---and a fraction of the joules that the model attributes to ``compute'' actually belong to coordination~\cite{andreev2015connectivity,wunderlich2003simpoints,xie2021apollo}. The diagnostic discipline implied by Sections~\ref{sec:evaldim} and~\ref{sec:lpcompute} is therefore stricter for compute than the pillar's classical literature suggests: declare the \textbf{coordination branch} per phase, declare the \textbf{boundary} between core, accelerator, LLC, and DRAM so that double-counting is impossible, and declare the \textbf{model family} from Table~\ref{tab:lpcompute_models} whose assumptions match the trace's stationarity~\cite{patterson2024energy,reda2012survey,chandrasekar2012drampower}. In our experience, ``CPU-only'' and ``kernel-only'' wins that survive these three declarations also survive integration; those that do not are a distinct epistemic category---true on the bench, false in the product---and should be reported as such. Section~\ref{sec:lpcomm} has already established that lens for firmware opacity and air-interface tails; Section~\ref{sec:lpstore} continues to persistence, where suspend semantics and FTL phase calendars become the leading edge of the same composability problem~\cite{andreev2015connectivity,patterson2024energy,mittal2014embedded}.

\section{Review of Approaches: Energy-Aware Storage}

\label{sec:lpstore}

\begin{figure*}[!t]
\centering
\begin{tikzpicture}[
  scale=0.92, transform shape,
  font=\small,
  >={Stealth[length=2.2mm]},
  dim/.style={rectangle, draw=black!55, fill=blue!4, inner sep=2.2pt, minimum height=10mm, text width=24mm, align=center},
  main/.style={rectangle, draw=black!68, fill=gray!8, rounded corners=2.5pt, inner sep=4pt, text width=66mm, align=center},
  mod/.style={rectangle, draw=black!48, fill=yellow!9, inner sep=2pt, text width=22mm, align=center, minimum height=9mm},
  br/.style={rectangle, draw=black!48, fill=green!7, inner sep=2.5pt, text width=24mm, align=center, minimum height=10mm},
  arr/.style={draw=black!50, line width=0.55pt},
  dframe/.style={draw=black!55, line width=0.55pt, rounded corners=3pt, inner sep=2mm},
  fcap/.style={font=\small\itshape, text=black!58, align=center}
]
\node[dim, text width=36mm, minimum height=20mm, anchor=south west] (D3) at (0,0)
  {\textbf{Eight gates}\\[0.8mm]{\small defined once in Section~\ref{sec:evaldim}\\(Table~\ref{tab:eval_gates}, Fig.~\ref{fig:evaldim_logic})}\\[0.8mm]{\small read here as gates on \textbf{persistence} claims}};
\node[dframe, fit=(D3)] (D3box) {};

\node[main, anchor=west] (S) at ($(D3box.east)+(15mm,0mm)$) {\textbf{NAND / FTL / file-system plane (sealed mobile IoT)}\\[0.6mm]
{\small program/erase $\cdot$ wear leveling $\cdot$ journaling $\cdot$ secure erase / attestation paths}\\[0.4mm]
{\small OTA images, checkpoints, page cache writeback, bus contention with DRAM and accelerators}};
\path[arr] (D3box.east) -- node[above, font=\small, yshift=0.3mm] {constrain} (S.west);

\matrix (M3) [matrix of nodes, below=4mm of D3box.south west, anchor=north west, column sep=2.5pt, row sep=2.2pt, nodes={mod}]
{
  {FS byte/\\I/O proxy} & {P/E block\\energy} & {FTL / GC\\queue model} \\
  {OTA +\\radio burst} & {Journal/\\crypto I/O} & {Trace /\\SimPoint phase} \\
  {DRAM--\\NAND bus} & {Edge cache/\\sync acct.} & {PMIC\\I/O totals} \\
};
\node[below=1.2mm of M3.south, font=\small\itshape, text=black!58, align=center, text width=72mm] {Table~\ref{tab:lpstore_models}: storage-side model families};
\path[arr] (M3.north east) -- ($(S.south west)!0.35!(S.south east)$) node[pos=0.42, left, font=\small, xshift=-0.5mm] {estimate};

\matrix (B3) [matrix of nodes, anchor=west, column sep=0pt, row sep=2.8pt, nodes={br}] at ($(S.east)+(12mm,0mm)$)
{
  \textbf{Async.} coordination \\
  \textbf{Sync.} coordination \\
  \textbf{Frame-slotted} \\
  \textbf{Multi-ch.\ hetero.} \\
};
\node[fcap, above=1.5mm of B3.north, text width=28mm] {Four-branch temporal context (Secs.~\ref{sec:lpcomm}--\ref{sec:lpstore})};
\path[arr] (S.east) -- node[above, font=\small, yshift=0.3mm] {context} (B3.west);
\end{tikzpicture}
\caption{Energy-aware mobile IoT: low-power storage map (evaluation dimensions, model families, four-branch context).}
\label{fig:lpstore_logic}
\end{figure*}
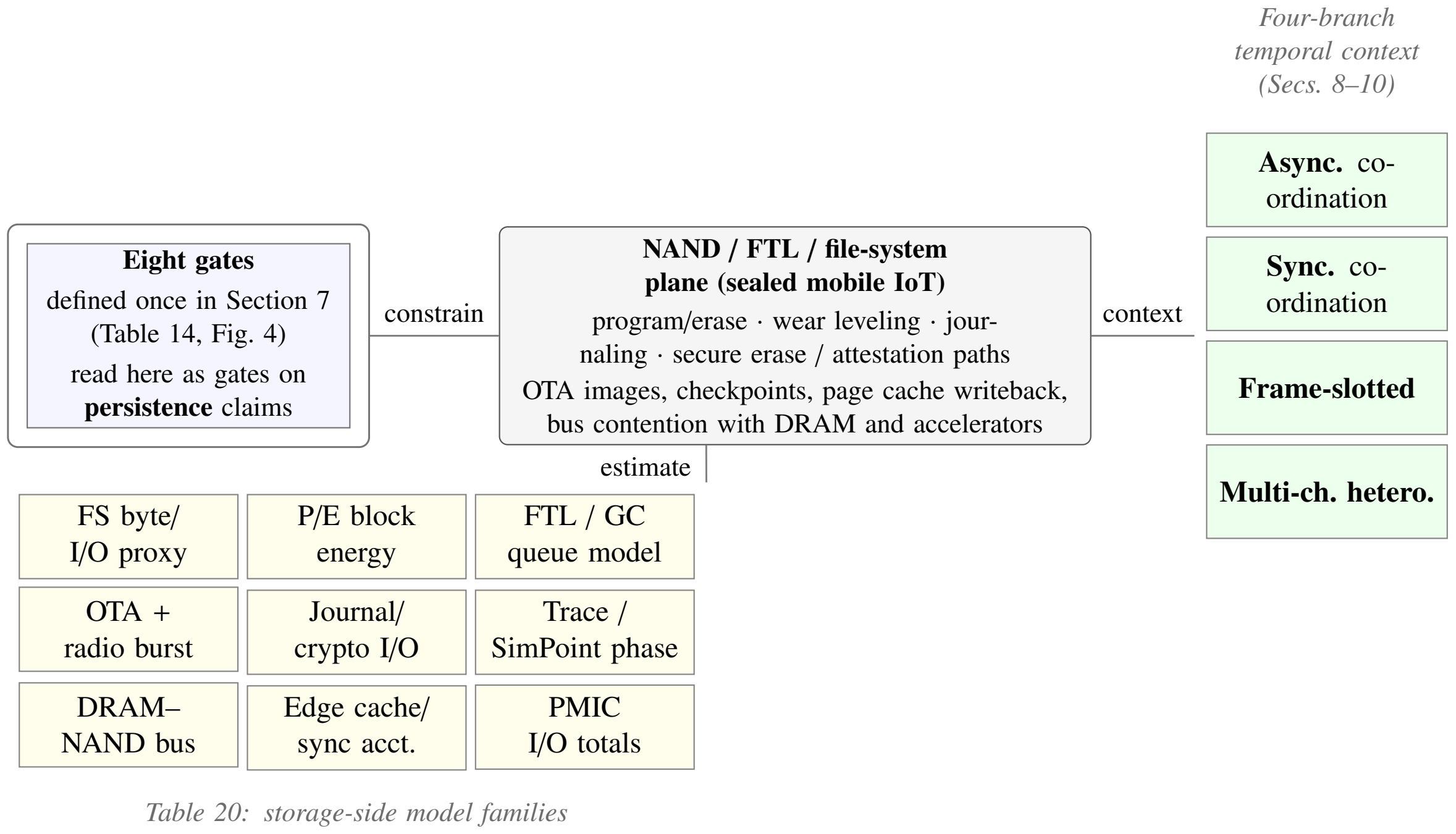

\textbf{Low-power storage} on mobile IoT endpoints is the problem of retaining firmware, models, logs, and media under tight energy envelopes when \textbf{non-volatile} activity is bursty, vendor-specific, and capable of blocking deep sleep long after the user interface appears idle~\cite{patterson2024energy,mittal2014embedded}. Those bursts contend for the same shallow retention and paging windows already shaped by Section~\ref{sec:comm_power}: maintenance I/O that overlaps a DRX-on duration or a scan campaign pays both NAND and radio-event energy, so persistence sizing cannot assume an idealized idle air interface~\cite{andreev2015connectivity,patterson2024energy,mittal2014embedded}. Flash translation layers (FTLs), wear leveling, garbage collection, and journaling create \emph{hidden} background traffic that competes with user-visible input/output (I/O) and with radio--compute pipelines: a ``small'' log append can trigger program/erase chains that pin buses and keep the system in shallow retention~\cite{patterson2024energy,mittal2014embedded}. Over-the-air updates, encrypted checkpoints for local learning, and tamper-evident records further couple storage to \textbf{security islands}, transport retries, and thermal headroom because charge pumps and NAND arrays are not thermally neutral~\cite{roy2003leakage,skadron2004hotspot,sultan2014processor}. Because edge caching and split inference change what must live locally, storage-centric claims must be checked against Section~\ref{sec:evaldim} together with Sections~\ref{sec:lpcomm} and~\ref{sec:lpcompute} rather than as isolated write-amplification factors~\cite{shi2016edge,jin2020review,barroso2023dasc}. This section treats storage as a \emph{phase-heavy} subsystem whose energy is parameterized by maintenance calendars, durability policies, and opaque FTL scheduling~\cite{patterson2024energy,reda2012survey}. Figure~\ref{fig:lpstore_logic} mirrors Sections~\ref{sec:lpcomm} and~\ref{sec:lpcompute}: the eight dimensions gate persistence claims, modeling families approximate the NAND/file-system plane, and the four coordination branches name the temporal regimes in which OTA, logging, and checkpoint bursts land. The exposition follows the same three blocks as Section~\ref{sec:lpcomm}: academia and industry (Subsection~\ref{sec:lpstore:status}); relations to evaluation dimensions and branches (Subsection~\ref{sec:lpstore:relations}); open problems, classic models, and a comparative table (Subsection~\ref{sec:lpstore:models}).

\subsection{State of the Art in Academia and Industry}
\label{sec:lpstore:status}

\textbf{Academia.} File-system bytes are unreliable flash-energy proxies under opaque FTL, write amplification, and GC~\cite{patterson2024energy,mittal2014embedded}. Host-side surveys connect OS/file-system placement to UFS/eMMC behavior and energy~\cite{chang2025flashhost}. \rev{The device-level reason this indirection exists is that NAND reliability degrades as cells shrink, so each generation spends more internal work---stronger error correction, more aggressive retry and read-refresh, more relocation---per host byte written~\cite{grupp2012bleaknand,cai2017ssderrorsurvey}. That internal work is invisible above the FTL, it is not proportional to host traffic, and it grows over a device's life. A host-side energy model calibrated on a fresh part therefore drifts on an aged one, which is a Gate~5 portability failure rather than a measurement error, and it is one reason two papers reporting ``joules per gigabyte written'' on the same nominal part can disagree.} OTA flows (download, verify, stage, swap) create maintenance calendars that can dominate quiescent days~\cite{patterson2024energy,andreev2015connectivity}. Checkpoints and model caches further couple storage to DRAM and radio~\cite{chandrasekar2012drampower,xie2021apollo,lu2024slm,shi2016edge}.

\textbf{Industry.} Sealed FTL, hardware-wrapped encryption, and mandated OTA cadences produce energy spikes uncorrelated with user traffic~\cite{mittal2014embedded,sultan2014processor}. OS battery tools rarely attribute NAND program energy per app~\cite{pathak2012whereenergy,patterson2024energy}. Reproducible claims must name filesystem, encryption, OTA channel, and whether PMIC totals include I/O conversion loss~\cite{patterson2024energy,reda2012survey}.

\subsection{Reading Storage Work Through the Gates and Branches}
\label{sec:lpstore:relations}

\rev{\textbf{Gate and branch crosswalk (not restated).} Definitions stay in Sections~\ref{sec:evaldim} and~\ref{sec:logic_framework}. Storage-specific stresses are Gate~1, when garbage collection and erase tails dominate average write throughput~\cite{patterson2024energy,mittal2014embedded}; Gate~4, when file-system bytes are treated as NAND energy; Gate~5, when an OTA download, decode, and program share one bus~\cite{xie2021apollo,chang2025flashhost}; and Gate~7, when a program burst heats a skin-contact package~\cite{skadron2004hotspot}. Do not rank a storage format without a maintenance calendar and a declared branch.}

\rev{\textbf{Why a byte ranking can reverse.} File-system bytes ignore the flash translation layer. Write amplification turns one host write into several NAND programs. Garbage collection then adds erase traffic that the application did not issue, often after the user-visible write has completed, and that erase can block deep sleep. A log format with fewer host bytes can therefore lose on NAND energy and on sleep eligibility if its amplification or collection is worse. The same format can win if collection is confined to a frame-slotted quiet window instead of an asynchronous idle interval. Maintenance scheduling---OTA verify, checkpoint rotation, secure erase---is part of the comparison, not a footnote: moving that work across a paging boundary changes both joules and the reachable sleep state~\cite{patterson2024energy,mittal2014embedded,chang2025flashhost}.}

\subsection{Scenarios, Open Issues, and Modeling Families}
\label{sec:lpstore:models}

\textbf{Open problems}~\cite{patterson2024energy,reda2012survey}: (1)~\textbf{Reproducible flash ground truth} when FTL policies and encryption modes differ by SKU~\cite{mittal2014embedded}. (2)~\textbf{Composable NAND + DRAM + I/O models} that avoid double counting when OTA, decode, and program overlap~\cite{chandrasekar2012drampower,xie2021apollo}. (3)~\textbf{Maintenance phase coverage} in published traces (OTA storms, log rotation) versus ``screen off'' averages~\cite{patterson2024energy,wunderlich2003simpoints}. Classic \textbf{families} include file-system byte proxies, analytical program/erase block models, FTL garbage-collection queue abstractions, coupled OTA--radio burst models, journal and crypto I/O overlays, trace segmentation for maintenance, DRAM--NAND bus interaction models, edge cache and synchronization accounting, and PMIC-level I/O totals~\cite{andreev2015connectivity,patterson2024energy,mittal2014embedded,shi2016edge,sultan2014processor,jin2020review,dayarathna2015datacenter}.

Table~\ref{tab:lpstore_models} summarizes these families analogously to Table~\ref{tab:lpcomm_models}~\cite{mittal2014embedded,reda2012survey}.

\begin{table*}[!t]
\renewcommand{\arraystretch}{1.15}
\caption{Classic storage energy and modeling families for mobile IoT: scenario fit, strengths, weaknesses, and energy-aware four-branch alignment.}
\label{tab:lpstore_models}
\centering
\tblsetup\begin{tabularx}{\textwidth}{@{}>{\raggedright\arraybackslash}p{0.10\textwidth}YYYY@{}}
\toprule
\textbf{Family} & \textbf{Scenario fit} & \textbf{Strengths} & \textbf{Weaknesses} & \textbf{Four-branch note} \\
\midrule
Host-side flash / cross-layer management~\cite{chang2025flashhost,patterson2024energy} & Phone/edge UFS/eMMC with FTL & Links OS knowledge to placement & Retail FTL still opaque & Critical when \textbf{multi}-bus hubs cache models \\
File-system / byte-count proxies~\cite{patterson2024energy,mittal2014embedded} & App-visible I/O budgeting & Easy to log; portable across OS APIs & Misreports flash energy under FTL & Misleading under \textbf{async} GC storms \\
Analytical P/E block models~\cite{patterson2024energy,mittal2014embedded} & Capacity planning, early NAND sizing & Interpretable erase/program tradeoffs & Vendor FTL hides timing & Useful when \textbf{frame} maintenance is periodic \\
FTL GC / queueing abstractions~\cite{patterson2024energy,mittal2014embedded} & SSD-class intuition on phones & Captures tail structure qualitatively & Parameters rarely published on retail & Stresses \textbf{sync} peaks if commits align \\
OTA + transport burst coupling~\cite{andreev2015connectivity,patterson2024energy} & Firmware update campaigns & Joins radio and flash in one story & Needs download + verify + program traces & Dominates \textbf{frame-slotted} quiet-window misuse \\
Journal / crypto I/O overlays~\cite{patterson2024energy,sultan2014processor} & Secure products, tamper-evident logs & Surfaces integrity costs & Hard to separate from user I/O & Cross-cuts all branches under policy load \\
Trace segmentation (SimPoint-style)~\cite{patterson2024energy,wunderlich2003simpoints} & Maintenance-heavy user days & Prevents single-number lies & Requires representative traces & Essential for \textbf{async} background days \\
DRAM--NAND bus interaction~\cite{mittal2014embedded,chandrasekar2012drampower,xie2021apollo} & Learning checkpoints, large models & Explains contention with accelerators & Needs bus-level traces & Critical in \textbf{multi}-subsystem hubs \\
Edge cache / split sync accounting~\cite{shi2016edge,jin2020review,barroso2023dasc} & Split inference, CDN-style edges & Surfaces boundary shifts & Infrastructure data sparse & Tied to \textbf{multi}-interface product stories \\
PMIC / I/O rail totals~\cite{patterson2024energy,schone2015energy} & Sealed-product budgeting & Honest when NAND is opaque & No structural lever for FTL teams & Baseline for every branch \\
\bottomrule
\end{tabularx}
\end{table*}

{
\begin{table*}[!t]

\renewcommand{\arraystretch}{1.08}
\caption{Analytical comparison of storage model families (illustrative judgments). Complements Table~\ref{tab:lpstore_models}.}
\label{tab:lpstore_anal}
\centering
\tblsetup\scriptsize
\begin{tabularx}{\textwidth}{@{}>{\raggedright\arraybackslash}p{0.14\textwidth}YYYYYccc@{}}
\toprule
\textbf{Family} & \textbf{Scenario} & \textbf{Assumptions} & \textbf{Observability} & \textbf{Strength} & \textbf{Weakness} & \textbf{Port.} & \textbf{Therm.} & \textbf{Comp.} \\
\midrule
Host-side flash & UFS/eMMC phones & Host sees placement & OS + some FTL hints & Links OS to NAND & Retail FTL opaque & Med & Low & Med \\
FS byte proxy & App I/O budgets & Bytes track energy & File-system counters & Easy to log & Misses FTL energy & High & Low & Low \\
Analytical P/E & Early NAND sizing & Vendor timing known & P/E counts & Interpretable & Timing hidden & Med & Low & Low \\
FTL GC queue & Tail structure & GC parameters known & Queue stats & Qualitative tails & Params unpublished & Low & Low & Med \\
OTA + radio & Firmware campaigns & Download, verify, program logged & Joint trace & Joins two pillars & Needs all three phases & Med & Med & High \\
Journal / crypto & Secure logs & Integrity path separable & I/O + crypto counters & Shows integrity cost & Mixed with user I/O & Low & Low & Med \\
Trace segmentation & Maintenance days & Day trace exists & Long PMIC trace & Stops screen-off means & Coverage & High & Low & High \\
DRAM--NAND bus & Checkpoints & Bus trace exists & Bus commands & Contention & Rare on retail & Low & Med & High \\
Edge cache accounting & Split inference & Cloud side reported & Device + edge & Boundary shift & Sparse infra data & Med & Low & High \\
PMIC I/O totals & Sealed budgeting & Rail includes conversion & I/O current & Honest & No FTL lever & High & Low & Low \\
\bottomrule
\end{tabularx}
\end{table*}
}

Beyond Table~\ref{tab:lpstore_models}, three \textbf{scenario patterns} dominate mobile IoT storage discussions~\cite{mittal2014embedded,alfuqaha2015iot}. \textbf{Always-on logging} (health, industrial) stresses small synchronous writes and secure rotation policies~\cite{patterson2024energy,sultan2014processor}. \textbf{Model lifecycle} (local learning, on-device SLM/LLM checkpoints) stresses burst programming, DRAM pressure, and thermal coupling~\cite{shi2016edge,chen2016eyeriss,xie2021apollo,lu2024slm,chang2025flashhost}. \textbf{OTA and fleet maintenance} concentrates downloads, signature verification, and atomic swap windows that can overlap peak radio use~\cite{andreev2015connectivity,patterson2024energy}. Choosing a modeling family is choosing whether teams see FTL tails, integrity overhead, or edge-boundary shifts early enough to change product policy~\cite{reda2012survey,jin2020review}.

\par

\textbf{Storage-specific synthesis.} Persistence is the pillar most likely to be \emph{omitted} from a published energy story, and most likely to invalidate it. Three structural reasons distinguish low-power storage from compute and communication. First, \emph{bytes are not energy}: file-system-visible writes pass through FTL policies whose timing and amplification factors are vendor-specific and rarely published, so even a perfectly logged application trace under-attributes flash energy unless DRAM--NAND bus traces or PMIC I/O rails close the gap~\cite{patterson2024energy,mittal2014embedded,chandrasekar2012drampower}. Second, \emph{maintenance is the dominant phase, not the exception}: OTA campaigns, encrypted journal rotations, and checkpoint storms can dominate daily joules on otherwise quiescent endpoints, which means a survey-grade comparison of storage mechanisms must report a \textbf{maintenance calendar}---not just a steady write throughput~\cite{patterson2024energy,wunderlich2003simpoints,sultan2014processor}. Third, \emph{storage closes the loop on the other two pillars}: suspending a write barrier blocks deep sleep that the modem would otherwise have entered; deferring an OTA verifies the wrong checkpoint after a thermal throttle; an edge-cached model shifts both flash wear and uplink budgets without changing local arithmetic counts~\cite{andreev2015connectivity,shi2016edge,jin2020review}. Reading Table~\ref{tab:lpstore_models} with these three peculiarities in mind turns it from a model inventory into a \emph{commitment list}: every persistence claim should declare the maintenance phase, the bus-level observability used, and the boundary at which retained bytes are counted---otherwise the claim is, at best, anecdotal under the eight-dimensional reading of Section~\ref{sec:evaldim}~\cite{mittal2014embedded,reda2012survey,rajandekar2015m2mmac}.

\section{Illustrative Quantitative Phase-Budget Example}
\label{sec:case_study}

\rev{This section is an example of how to \emph{use} the instruments. It is not an experiment and not a validation of the taxonomy. Shares in Table~\ref{tab:case_budget} are \textbf{illustrative} order-of-magnitude composites for the two archetypes in Section~\ref{sec:running_examples}. They are not measurements from a commercial device. \rev{No integer in Table~\ref{tab:case_budget} was measured on a device for this paper.} Each row is tied to the literature that motivated the magnitude; the integer itself is not taken from one trace.}

\rev{\textbf{Evidence labels used in this manuscript.} \emph{Measured:} a number reported by an instrumented study we cite. \emph{Modeled:} a number produced by a simulator or analytical model in a cited study. \emph{Estimated:} a derived figure (for example a share inferred from a cited range). \emph{Illustrative:} a composite we constructed so that a ranking can be checked. Table~\ref{tab:case_budget} is illustrative. The Pathak tail figures in Section~\ref{sec:logic_framework} are measured. Model-family judgments in Tables~\ref{tab:lpcomm_models}--\ref{tab:lpstore_models} are illustrative analytical comparisons, not new measurements.}

{
\begin{table}[!t]

\renewcommand{\arraystretch}{1.08}
\caption{Illustrative daily energy share (\%) by anatomy phase (device-local boundary). Evidence class for every share: illustrative. Not a specific SKU.}
\label{tab:case_budget}
\centering
\tblsetup\footnotesize
\begin{tabularx}{\columnwidth}{@{}>{\raggedright\arraybackslash}p{0.24\columnwidth}CC>{\raggedright\arraybackslash}X@{}}
\toprule
\textbf{Phase} & \textbf{Ex.~A} & \textbf{Ex.~B} & \textbf{Literature anchor (not the source of the integer)} \\
\midrule
Deep sleep / retention & 35 & 18 & PSM/DRX and screen-off idle~\cite{andreev2015connectivity,carroll2010smartphone} \\
Idle / listen / scan & 28 & 22 & Paging, advertising, discovery~\cite{andreev2015connectivity,bocus2025superiotriot} \\
RX / TX (payload) & 12 & 15 & Payload vs.\ procedure time~\cite{centenaro2016longrange,pathak2012whereenergy} \\
Protocol tails / modem--AP & 15 & 20 & Tail after transfer~\cite{pathak2012whereenergy,xie2021apollo} \\
Compute (inference/DSP) & 5 & 12 & On-device inference under a radio envelope~\cite{lu2024slm,chen2016eyeriss} \\
Storage (log/OTA/FTL) & 5 & 13 & OTA and FTL bursts~\cite{chang2025flashhost,patterson2024energy} \\
\midrule
Dominant branch & Frame + async & Multi-RAT + framed OTA & Table~\ref{tab:branch_decision} \\
Gates most stressed & 1, 4, 7 & 1, 5, 8 & Table~\ref{tab:eval_gates} \\
\bottomrule
\end{tabularx}
\end{table}
}

\rev{\textbf{Sensitivity (illustrative).} Two perturbations test whether the ranking is an artifact of the chosen integers. For Example~A, move 8 points between idle/listen and protocol tails (listen 20 or 36; tails 23 or 7). Listen plus tails stays above payload RX/TX (12) in both cases. For Example~B, move 8 points between storage and compute (storage 5 or 21; compute 20 or 4). Storage plus tails stays at least comparable to compute in both cases. The ranking claim is stable inside that band. It remains illustrative. A measured trace can replace every cell; the procedure below says how.}

\rev{\textbf{Why these day-level shares, and what they are not.} The integers are not a 24-hour trace. They are diluted relative to short measured transfers. Pathak et al.\ find a 3G tail of about 60\% of one microbenchmark, and most of their apps spend about 50--90\% of energy in I/O rather than in the CPU~\cite{pathak2012whereenergy}; a full day that also contains deep sleep therefore shows a smaller tail share (15\% and 20\% in the two columns), not that 60\% figure. Carroll and Heiser show cellular and display power dominating the application processor in several everyday scenarios~\cite{carroll2010smartphone}, which is why Example~A assigns only 5\% to compute while the radio phases carry the day. We did not find a public sealed-device trace that jointly logs modem state, CPU, and NAND over 24\,h, so Table~\ref{tab:case_budget} is not fitted to one. The comparison with those datasets is qualitative: tails and I/O can dominate payload and CPU on the timescales those papers measured.}

\rev{\textbf{How to replace Table~\ref{tab:case_budget} with a measurement.}} (i)~Collect PMIC (or equivalent) traces for $\geq$24\,h; (ii)~segment into the five anatomy phases plus compute/storage overlays; (iii)~label each segment with a dominant branch; (iv)~recompute shares and re-rank candidate mechanisms under Gates~1, 4, 5, and~8. If a mechanism improves payload RX/TX share but increases tails or storage, the Gate~1 ranking should reverse. Label the replacement table \emph{measured}.

\rev{\textbf{Takeaways (illustrative).}} (1)~Example~A: listen/scan plus tails exceed payload RX/TX; a higher BLE PHY rate does not by itself restore deep sleep if supervision keeps shallow retention~\cite{bocus2025superiotriot}. (2)~Example~B: storage and tails jointly rival compute; an OTA day can dominate battery energy while CPU counters look quiet~\cite{chang2025flashhost,patterson2024energy}. (3)~Additive CPU+radio+flash models fail Gate~5 when modem DMA and an NPU share DRAM~\cite{xie2021apollo}. These are checks on the example, not product guarantees.

\section{Model-Family Selection Workflow}
\label{sec:model_decision}

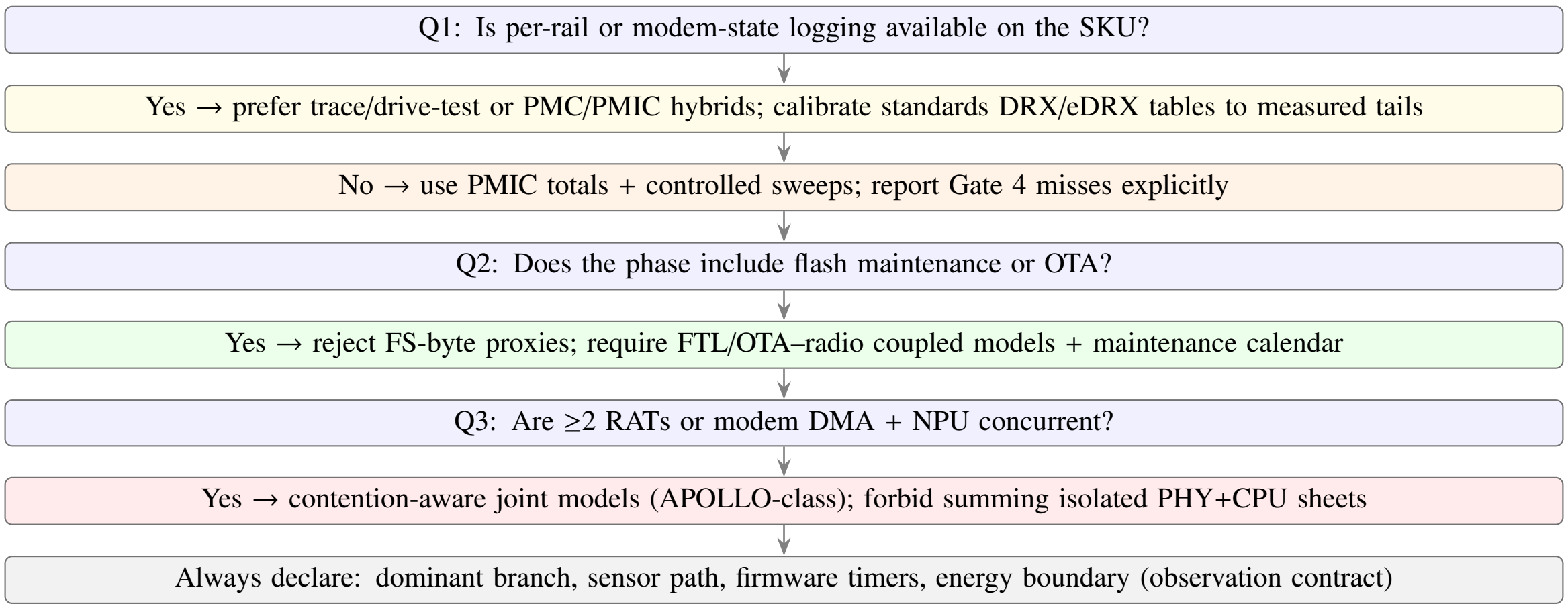
\begin{figure}[!t]
\centering
\begin{tikzpicture}[
  font=\footnotesize,
  >={Stealth[length=1.8mm]},
  box/.style={draw=black!55, rounded corners=2pt, align=center, inner sep=3pt, text width=0.92\columnwidth},
  arr/.style={->, draw=black!50, line width=0.5pt}
]
\node[box, fill=blue!6] (q1) {Q1: Is per-rail or modem-state logging available on the SKU?};
\node[box, below=3mm of q1, fill=yellow!10] (yes1) {Yes $\rightarrow$ prefer trace/drive-test or PMC/PMIC hybrids; calibrate standards DRX/eDRX tables to measured tails};
\node[box, below=3mm of yes1, fill=orange!10] (no1) {No $\rightarrow$ use PMIC totals + controlled sweeps; report Gate~4 misses explicitly};
\node[box, below=3mm of no1, fill=blue!6] (q2) {Q2: Does the phase include flash maintenance or OTA?};
\node[box, below=3mm of q2, fill=green!8] (yes2) {Yes $\rightarrow$ reject FS-byte proxies; require FTL/OTA--radio coupled models + maintenance calendar};
\node[box, below=3mm of yes2, fill=blue!6] (q3) {Q3: Are $\geq$2 RATs or modem DMA + NPU concurrent?};
\node[box, below=3mm of q3, fill=red!8] (yes3) {Yes $\rightarrow$ contention-aware joint models (APOLLO-class); forbid summing isolated PHY+CPU sheets};
\node[box, below=3mm of yes3, fill=gray!10] (out) {Always declare: dominant branch, sensor path, firmware timers, energy boundary (observation contract)};
\draw[arr] (q1) -- (yes1);
\draw[arr] (yes1) -- (no1);
\draw[arr] (no1) -- (q2);
\draw[arr] (q2) -- (yes2);
\draw[arr] (yes2) -- (q3);
\draw[arr] (q3) -- (yes3);
\draw[arr] (yes3) -- (out);
\end{tikzpicture}
\caption{Decision workflow for selecting communication/compute/storage model families under sealed-device observability constraints.}
\label{fig:model_decision}
\end{figure}

Tables~\ref{tab:lpcomm_models}--\ref{tab:lpstore_models} catalog families; Fig.~\ref{fig:model_decision} turns them into a decision workflow for practitioners under realistic observability.

\section{Critical Review: Comparison of Approaches and Joint-Optimization Lessons}
\label{sec:lessons}

A survey of energy-aware methods must do more than list families: it must say \emph{when} an approach is the wrong tool. Table~\ref{tab:approach_fail} condenses recurring failure modes observed across Sections~\ref{sec:lpcomm}--\ref{sec:lpstore}.

\begin{table*}[!t]
\renewcommand{\arraystretch}{1.1}
\caption{Critical comparison: common approach classes and when they mislead for sealed mobile IoT.}
\label{tab:approach_fail}
\centering
\tblsetup\footnotesize
\begin{tabularx}{\textwidth}{@{}>{\raggedright\arraybackslash}p{0.18\textwidth}>{\raggedright\arraybackslash}X>{\raggedright\arraybackslash}X@{}}
\toprule
\textbf{Approach class} & \textbf{Looks strong when} & \textbf{Misleads when} \\
\midrule
PHY / rate--power sheets & Coverage planning, static channels & Tails, multi-RAT, firmware sleep ladders dominate~\cite{andreev2015connectivity} \\
CPU/accelerator joules/op & Isolated kernel microbenchmarks & Modem DMA/LLC contention or thermal caps change the schedule~\cite{xie2021apollo,skadron2004hotspot} \\
File-system byte counts & App-visible I/O budgeting & FTL GC/OTA redefine energy and sleep eligibility~\cite{chang2025flashhost,patterson2024energy} \\
Edge offload ``saved UE energy'' & Device-only boundary, short traces & Uplink tails + cloud joules reverse the ranking~\cite{shi2016edge,jin2020review} \\
Learned schedulers / power models & Rich counters, stationary regimes & Firmware drift and branch shifts without retrain/certification~\cite{zhai2024mlsurvey,chang2025dqngran} \\
Standards DRX/eDRX tables alone & Naming reproducible knobs & Vendor sleep depth and FTL remain Gate~4/5 opaque~\cite{veedu2022redcap,joerke2025redcap} \\
\bottomrule
\end{tabularx}
\end{table*}

\rev{\textbf{Modeling-family weaknesses (critical reading of Tables~\ref{tab:lpcomm_models}--\ref{tab:lpstore_models}).} The family tables are not inventories of interchangeable tools. Communication: PHY link budgets miss MAC tails, mobility, and multi-RAT; WSN duty-cycle Markov models fit preamble-style listen poorly to cellular state machines; topology scaling is weakly calibrated to cellular PHY; DRX and PSM tables name standards knobs while vendor sleep ladders stay hidden; drive-test fits transfer poorly across SKUs; learned access is hard to certify; PMIC totals alone give no structural MAC lever. Compute: Wattch-style event models miss firmware sleep, accelerators, and I/O; McPAT-class configs may miss vendor gating; hybrid calibration is per-SKU and trace-sensitive; gem5 is slow and incomplete for radio and storage; Accel-Sim is costly and GPU-centric; PMC and RAPL domains are coarse and drift by SKU; ML surrogates are data-hungry and opaque under branch shifts; RTL tools rarely cover a full multi-radio day; DRAM command models need composability with core tools. Storage: host-side placement still faces an opaque retail FTL; file-system bytes misreport NAND energy; analytical P/E and GC queue models lack published retail timing; OTA--radio coupling needs download, verify, and program traces; journal and crypto overlays are hard to separate from user I/O; DRAM--NAND bus models need bus traces; edge-cache accounting often lacks infrastructure data; PMIC I/O totals give FTL teams no structural lever. Choosing a family is choosing which of these blind spots a study accepts.}

\rev{\textbf{Synthesis recommendations, not universal laws.} The six statements below are conclusions of this survey's synthesis. They are not established laws, and a measured trace can falsify any one of them for a given phase.}
\begin{enumerate}
\item \textbf{Optimize sleep eligibility, not the payload bit.} Airtime wins that lengthen modem--AP tails lose on Gate~1 after phase split~\cite{andreev2015connectivity,pathak2012whereenergy}.
\item \textbf{Joint calendars beat pillar-local calendars.} Inference, paging, and encrypted commits must co-schedule or storage/radio vetoes deep sleep~\cite{chang2025flashhost,shi2016edge}.
\item \textbf{Non-additivity is a first-class result.} Shared LLC/DRAM makes subsystem sums invalid; report couplings explicitly~\cite{xie2021apollo,chandrasekar2012drampower}.
\item \textbf{Boundary choice can invert ``green'' conclusions.} Device-only versus cloud-inclusive totals can reverse offload rankings~\cite{jin2020review,barroso2023dasc,xu2024devicellm}.
\item \textbf{Maintenance days need first-class metrics.} OTA/checkpoint storms deserve segmented traces, not screen-off means~\cite{wunderlich2003simpoints,patterson2024energy}.
\item \textbf{Standards timers are necessary but not sufficient.} Visible knobs enable replay; hidden sleep ladders and FTL still gate trust~\cite{veedu2022redcap,joerke2025redcap}.
\end{enumerate}

These principles motivate the observation contract and roadmap in Section~\ref{sec:open_problems}.

\section{Open Problems and Research Roadmap}
\label{sec:open_problems}

This section synthesizes the classification framework and cross-pillar comparison, then states open problems with a concrete research roadmap (Table~\ref{tab:roadmap}).

\rev{\textbf{Engineering problems versus fundamental questions.} Near-term items in Table~\ref{tab:roadmap} are engineering problems: they can be discharged with traces, calendars, and reporting templates (observation contract, maintenance-calendar reporting, multi-RAT sweeps, a public phase-labeled corpus, scan-budget logs, standards fields in radio papers). Fundamental questions cannot be closed by another measurement template. \emph{(P1) Composition,} linked to Gap~1 and Gate~5: when may subsystem models be summed, and what certificate shows non-additivity on a shared interconnect~\cite{xie2021apollo,chandrasekar2012drampower}? \emph{(P2) Measurement epistemology,} linked to Gap~2 and Gates~4 and~8: what minimal observation contract makes two papers' joules comparable without per-rail access on every SKU~\cite{patterson2024energy,reda2012survey}? \emph{(P3) Regime-aware evaluation,} linked to Gap~3 and Gates~1 and~2: how should benchmarks carry branch labels and maintenance calendars so that averages stop mis-ranking policies~\cite{wunderlich2003simpoints,rajandekar2015m2mmac}? Table~\ref{tab:problem_link} ties every roadmap row to a gap and to gates. Horizon (near, medium, long) is not the same distinction as engineering versus fundamental.}

\subsection{Summary of Classification Framework}
\label{sec:conclusion:taxonomy}

Three classification layers structure the survey. \textbf{Layer~A --- coordination branches} (Table~\ref{tab:fourway}): asynchronous, synchronous, frame-slotted, and multichannel heterogeneous regimes describe how time, spectrum, and on-chip activity align; they denote dominant \emph{phases}, not fixed labels. \textbf{Layer~B --- evaluation gates} (Table~\ref{tab:eval_gates}): eight dimensions (outcome, cost, evidence) test whether claims are comparable on sealed hardware. \textbf{Layer~C --- model families} (Tables~\ref{tab:taxonomy}, \ref{tab:lpcomm_models}--\ref{tab:lpstore_models}): analytical, calibrated, trace-driven, standards-parameterized, and learned families trade accuracy, interpretability, and observability differently. Table~\ref{tab:cross} explains why Layer~C models are not always composable. Studies should be tagged by (branch, gate, family) before numerical comparison.

\subsection{Comparative Synthesis Across Pillars}
\label{sec:conclusion:compare}

Sections~\ref{sec:lpcomm}--\ref{sec:lpstore} surveyed communication, compute, and storage with matched depth. Table~\ref{tab:pillar_synthesis} contrasts what each pillar optimizes, how it fails in integration, and which model families survive on phone-class products.

\begin{table*}[!t]
\renewcommand{\arraystretch}{1.12}
\caption{Cross-pillar comparative synthesis for energy-aware mobile IoT (family detail in Tables~\ref{tab:lpcomm_models}--\ref{tab:lpstore_models}). \textit{Cross-cutting lesson:} \rev{communication is a frequent default envelope, not the limiter of every phase (Table~\ref{tab:scope_limits})}; compute fills or violates that envelope when radio is the limiter; storage fixes persistence and maintenance that redefine ``idle''; no pillar table alone predicts battery life without branch context and Table~\ref{tab:cross} couplings.}
\label{tab:pillar_synthesis}
\centering
\tblsetup\begin{tabularx}{\textwidth}{@{}%
  >{\raggedright\arraybackslash}p{0.10\textwidth}%
  >{\raggedright\arraybackslash}X%
  >{\raggedright\arraybackslash}X%
  >{\raggedright\arraybackslash}X%
  >{\raggedright\arraybackslash}X@{}}
\toprule
\textbf{Pillar} & \textbf{Energy lever} & \textbf{Model families} & \textbf{Integration failure} & \textbf{Fit branch(es)} \\
\midrule
Comm. & Policy timers (DRX, scans, tails) & Standards DRX; drive traces; PMIC totals & Modem--AP shallow idle; multi-RAT on one rail & Framed; multi-RAT \\
Compute & DRAM/NPU traffic; DVFS depth & Hybrid calib.; gem5 contention; ML surrogate & Radio shallow sleep or throttle blocks gains & Sync.; multi-RAT \\
Storage & P/E bursts; FTL GC & OTA--radio coupled; bus/PMIC I/O traces & Bytes$\neq$energy; GC blocks deep sleep & Async.; framed OTA \\
\bottomrule
\end{tabularx}
\end{table*}

\textbf{Qualitative comparison.} Communication work is strongest when standards-visible timers and phase-split traces are reported; it is weakest when PMIC totals obscure multi-RAT share or PHY models omit tails~\cite{andreev2015connectivity,patterson2024energy}. Compute work is strongest when memory traffic and accelerator contention are modeled; it is weakest when CPU-only traces ignore modem DMA or thermal throttling~\cite{chen2016eyeriss,xie2021apollo,zhai2024mlsurvey}. Storage work is strongest when maintenance calendars and OTA bursts are explicit; it is weakest when file-system byte counts stand in for NAND program energy~\cite{patterson2024energy,chang2025flashhost}. \textbf{System composition}, not any single pillar, is the central object of study.

\subsection{Remaining Gaps}
\label{sec:conclusion:gaps}

Products still fail low-power goals when wireless tails dominate useful work, pillar-local optimizations break sleep alignment, sealed hardware hides attribution, workloads migrate across branches within a day, shared buses invalidate additive budgets, or security/carrier policy caps reachable batching~\cite{carroll2010smartphone,patterson2024energy,skadron2004hotspot}. Prior literature often silos pillars (Table~\ref{tab:survey_compare}), mixes sensors and boundaries, omits branch context (Table~\ref{tab:fourway}), under-models flash maintenance, and under-reports learning upload/checkpoint energy~\cite{alfuqaha2015iot,reda2012survey,chang2025flashhost,zhai2024mlsurvey}. Section~\ref{sec:lessons} turns these gaps into actionable joint-optimization rules.

\rev{\textbf{How the framework answers those gaps.} The four branches name the regime of a migrating day so averages are not compared across incompatible phases. The eight gates and the reporting template force a declared sensor path and energy boundary so two papers cannot claim the same percentage under different meters. The observation contract and Table~\ref{tab:approach_fail} make the modeling-family blind spots above first-class results rather than footnotes. Upload and checkpoint energy for on-device or federated learning belong in Gate~1 and Gate~8 with a maintenance calendar, not only in a CPU joules-per-token number.}

\subsection{Outlook and Research Roadmap}
\label{sec:conclusion:outlook}

\textbf{Near term (1--3 years):} adopt the six-item \textbf{observation contract} below; publish phase-split traces with declared branch labels; report maintenance calendars on storage-heavy days; begin multi-RAT sweeps where PMIC allows~\cite{bocus2025superiotriot,patterson2024energy}.

\textbf{Medium term (3--7 years):} deliver \textbf{composability certificates} and joint contention traces; FTL-aware retail-phone flash models; closed-loop schedulers across inference, paging, and encrypted logs; RedCap/eRedCap mobility studies under framed regimes~\cite{chang2025flashhost,raja2026redcapsurvey}.

\textbf{Long term (7+ years):} certifiable learning-based policies under firmware and workload drift; system-level on-device LLM energy that moves with partition; standardized maintenance and federated-round energy metrics~\cite{xu2024devicellm,mokadem2023fedenergy}.

\textbf{Synthesis.} Endpoint energy is often---not always---a communication-limited joint allocation (Table~\ref{tab:scope_limits}); fair comparison needs anatomy, gates, and branch labels (Tables~\ref{tab:branch_decision}, \ref{tab:eval_gates}).
Sections~\ref{sec:lpcomm}--\ref{sec:lpstore} were written in parallel so the pillars can be read side by side: radio tails and multi-RAT policy keep the AP shallow; shared accelerators, LLC, and DRAM with modem DMA break simple energy sums; FTL and OTA work redefine ``idle''~\cite{andreev2015connectivity,patterson2024energy,mittal2014embedded,carroll2010smartphone,pathak2012whereenergy,chandrasekar2012drampower,xie2021apollo,gu2020lpwatechsurvey}. Together this is one \textbf{composition} problem: small modeling slips (double-counting LLC, skipping DRX tails, equating file bytes with program energy) become large product errors when everything shares buses, heat limits, and PMIC rails~\cite{patterson2024energy,reda2012survey}.

\textbf{Three recurring tensions.} First, measurable quantities on sealed phones rarely coincide with quantities that remain portable across firmware releases~\cite{patterson2024energy,reda2012survey,kumar2023mlmicro}. Second, interpretable design knobs favored in reviews often differ from trace-calibrated models used in production, whose failures appear only under multi-axis stress tests~\cite{zhai2024mlsurvey,larsen2023greeneran}. Third, energy boundaries strongly affect offload conclusions unless uplink tails, cloud energy, and flash wear are reported jointly~\cite{shi2016edge,jin2020review,barroso2023dasc}. Standards expose some timers but not hidden modem or FTL behavior or RF heating that blocks deep sleep; the four-branch framework remains necessary~\cite{andreev2015connectivity,mittal2014embedded,skadron2004hotspot,rajandekar2015m2mmac}.

\par

\rev{\textbf{Observation contract.} The principal gap is procedural: a shared reporting checklist proposed to improve comparison without assuming per-rail observability on every SKU~\cite{patterson2024energy,reda2012survey}. It is not claimed to be sufficient by itself for measurement comparability across the entire literature. At minimum, each study should report:} (1)~the \textbf{dominant branch} per phase and how phases were split~\cite{wunderlich2003simpoints,rajandekar2015m2mmac}; (2)~the \textbf{sensor path} (PMC, PMIC, on-die, mix) and what it misses on that SKU~\cite{patterson2024energy,hackenberg2013rapl}; (3)~key \textbf{firmware and timer settings} (modem build, governor, DRX/eDRX/PSM, connection interval)~\cite{andreev2015connectivity,larsen2023greeneran}; (4)~the \textbf{maintenance plan} (OTA, logs, checkpoints) behind non-stationary days~\cite{patterson2024energy,mittal2014embedded}; (5)~the \textbf{thermal setup} (room, skin contact, airflow)~\cite{roy2003leakage,skadron2004hotspot}; (6)~the \textbf{energy boundary} (device only, gateway, or includes cloud)~\cite{shi2016edge,jin2020review,barroso2023dasc}. None of this is unusual; missing it is why cross-paper comparison is risky today.

\rev{\textbf{Reporting template.} The contract is a reproducibility checklist. One row per phase is enough:}

{
\begin{table}[!t]

\renewcommand{\arraystretch}{1.08}
\caption{Minimum reporting template for one phase (reproducibility checklist).}
\label{tab:report_template}
\centering
\tblsetup\begin{tabularx}{\columnwidth}{@{}>{\raggedright\arraybackslash}p{0.28\columnwidth}Y@{}}
\toprule
\textbf{Field} & \textbf{What to write} \\
\midrule
Dominant branch & One label and its energy share; secondary label if within 10 points (Section~\ref{sec:logic_framework}) \\
Sleep-envelope owner & Pillar that blocks the deepest sleep state \\
Limiter test & Communication-limited only if removing radio idle and tails changes energy by $\geq$20\% \\
Sensor path & PMIC, PMC, on-die, or mix, and what that path misses \\
Timers & Modem build, DRX/eDRX/PSM or connection interval \\
Maintenance & OTA, garbage collection, checkpoints inside the phase \\
Thermal regime & Short peak or sustained throttling; ambient or skin contact \\
Boundary & Device-only, gateway-inclusive, or cloud-inclusive \\
Evidence class & Measured, modeled, estimated, or illustrative \\
\bottomrule
\end{tabularx}
\end{table}
}

\textbf{Composability.} Studies should report when sub-models \emph{cannot} be summed---for example, under APOLLO-class contention, modem DMA, or shared DRAM commands~\cite{chandrasekar2012drampower,jin2020review,xie2021apollo}---and which couplings govern each claim. Non-additivity is often an informative result rather than a modeling failure.

\rev{\textbf{Joint control.} Closed-loop scheduling across inference, paging, and encrypted logs under thermal and security constraints remains open on sealed endpoints~\cite{andreev2015connectivity,skadron2004hotspot,sultan2014processor,shi2016edge}. The relevant general result is that coordinated allocation under changing conditions beats a static split: Can's wind--solar microgrid controller reallocates power between two sources from their instantaneous availability, with a PI loop on the shared bus~\cite{can2026windsolar}. That study is a renewable-microgrid simulation. It is not evidence about modem tails, NAND, DRAM, or phone SoCs. The only transfer we make is methodological: energy coordination should track the current regime rather than a fixed pillar budget. On an endpoint, the analogous state is the dominant branch, the sleep-envelope owner, and whether the thermal regime is a short peak or sustained throttling.}

\textbf{Learning and certification.} Learned schedulers and power models require \textbf{post-deployment explainability} for standards and safety review~\cite{zhai2024mlsurvey,chang2025dqngran,chang2023drlmultibs,kairouz2021federatedlearning,li2020federatedchallenges,mokadem2023fedenergy,kaur2023fliot,dang2024federatedenergy}. Federated and on-device training add radio uploads and flash checkpoint bursts that MAC-only or CPU-only studies under-report~\cite{mcmahan2017fedavg,bonawitz2019fedscalable,mokadem2023fedenergy}. Edge-LLM surveys similarly require energy boundaries that follow model partitioning, not MAC tuning alone~\cite{friha2024llmedge,xu2024devicellm}.

{
\begin{table*}[!t]

\renewcommand{\arraystretch}{1.08}
\caption{Each roadmap item classified as a near-term engineering problem (Eng.) or a fundamental question (Fund.), with the literature gap and evaluation gates it answers.}
\label{tab:problem_link}
\centering
\tblsetup\scriptsize
\begin{tabularx}{\textwidth}{@{}>{\raggedright\arraybackslash}p{0.28\textwidth}cc>{\raggedright\arraybackslash}Y>{\raggedright\arraybackslash}p{0.12\textwidth}@{}}
\toprule
\textbf{Open problem (Table~\ref{tab:roadmap})} & \textbf{Class} & \textbf{Gap} & \textbf{Why it is open} & \textbf{Gates} \\
\midrule
Shared observation contract & Eng. & 2 & Reporting, not a new mechanism & 4, 8 \\
Composability certificates & Fund. & 1 & P1; when sums are valid & 5 \\
Multi-RAT attribution on PMIC rails & Eng. & 2 & One rail hides the active interface & 1, 4 \\
RedCap/eRedCap under mobility & Eng. & 3 & Framed regime not yet measured in motion & 1, 2 \\
On-device LLM decode vs.\ prefill & Eng. & 1 & Thermal cap changes the schedule & 1, 7 \\
Interpretable surrogates for governors & Fund. & 2 & Gate~6; fidelity without a knob & 6 \\
Maintenance-calendar reporting & Eng. & 1 & OTA/logs omitted from ``idle'' & 1, 4 \\
FTL-aware flash energy on retail phones & Fund. & 1 & Opaque FTL; bytes are not joules & 4, 5 \\
Closed-loop inference / paging / logs & Fund. & 1, 3 & P3; joint schedule & 1, 3, 7 \\
Federated training energy per round & Eng. & 1 & Upload and checkpoint omitted & 1, 8 \\
Public phase-labeled trace corpus & Eng. & 2 & No shared artifact to compare & 4 \\
Coexistence scan budget on wearables & Eng. & 1 & Scan energy vs.\ battery and skin & 1, 3, 7 \\
Wear-aware placement for model caches & Eng. & 1 & Host placement vs.\ NAND energy & 1, 5 \\
Reporting templates aligned with 3GPP timers & Eng. & 2 & Visible knobs absent from papers & 3, 5 \\
\bottomrule
\end{tabularx}
\end{table*}
}

\textbf{Research roadmap.} Table~\ref{tab:roadmap} lists open problems by pillar and branch. Table~\ref{tab:problem_link} states which of those rows are engineering work and which are fundamental, and which gap and gates each row addresses.

\begin{table*}[!t]
\renewcommand{\arraystretch}{1.10}
\caption{Future research roadmap for energy-aware mobile IoT energy (priority: H = high, M = medium).}
\label{tab:roadmap}
\centering
\tblsetup\begin{tabularx}{\textwidth}{@{}>{\raggedright\arraybackslash}p{0.07\textwidth}X>{\raggedright\arraybackslash}p{0.10\textwidth}>{\centering\arraybackslash}p{0.035\textwidth}>{\centering\arraybackslash}p{0.055\textwidth}X@{}}
\toprule
\textbf{Pillar} & \textbf{Open problem} & \textbf{Dominant branch} & \textbf{Pri.} & \textbf{Hor.} & \textbf{Enabling artifact / metric} \\
\midrule
Cross-cutting & Shared observation contract (PMIC, modem, FTL opacity) & All & H & Near & Phase-split traces + declared sensor path \\
Cross-cutting & Composability certificates (when sub-models cannot sum) & Multi-RAT; framed & H & Med. & Contention-aware joint traces (APOLLO-class) \\
Communication & Multi-RAT attribution on sealed PMIC rails & Multichannel & H & Near--Med. & Per-interface sweeps + drive-test logs \\
Communication & RedCap/eRedCap energy under mobility and eDRX & Frame-slotted & M & Med. & Standards tables + silicon traces~\cite{raja2026redcapsurvey} \\
Compute & On-device LLM/SLM decode vs.\ prefill under thermal caps & Sync.; framed & H & Near--Med. & Joules/token + DRAM + radio boundary \\
Compute & Interpretable ML surrogates for product governors & Async.; multi-RAT & M & Med. & Error maps by sleep state / counter \\
Storage & Maintenance-calendar reporting (OTA, logs, checkpoints) & Async.; framed & H & Near & SimPoint-style day traces, not screen-off means \\
Storage & FTL-aware flash energy on retail phones & All & H & Med. & Bus-level or PMIC I/O + host models~\cite{chang2025flashhost} \\
Joint control & Closed-loop scheduling: inference, paging, encrypted logs & Framed; sync. & M & Med. & Policy simulators with security constraints \\
Learning & Federated / on-device training under energy budgets & Async.; multi-RAT & M & Long & Upload + checkpoint energy per round \\
Cross-cutting & Public mobile IoT trace corpus (radio+CPU+storage phases) & All & H & Near & Branch-labeled artifacts; metadata schema \\
Communication & Coexistence scan budget vs.\ battery on wearables & Multi-RAT & M & Near--Med. & Scan matrix + PMIC + thermal logs \\
Storage & Wear-aware flash placement for on-device model caches & Async.; framed & M & Med. & Host API + WA + recharge metric \\
Standards & Harmonize reporting templates with 3GPP timer tables & Framed & M & Med. & DRX/eDRX/PSM fields in every radio energy paper \\
\bottomrule
\end{tabularx}
\end{table*}

\section*{Limitations}
\label{sec:limitations}

\rev{This survey has methodological limits that bound the generality of the framework. \emph{Database selection:} primary searches used IEEE Xplore, ACM Digital Library, and Web of Science; other indexes and non-English venues are underrepresented. \emph{Language:} inclusion was restricted to English. \emph{Inclusion criteria:} energy, power, or battery life had to be an explicit outcome, so papers that treat power only as a secondary metric are missing. \emph{Screening counts:} vendor hit totals and title/abstract exclusion counts were not archived, so a complete PRISMA funnel cannot be reconstructed. \emph{Coding reliability:} only the 18-row representative matrix was double-coded; Cohen's kappa is not reported; the 186-work set has category labels but not a frozen branch/gate crosstab. \emph{Hardware variability:} sealed-SKU firmware and carrier policy change sleep ladders, so a model that fits one product may fail on the next. \emph{Publication bias:} failed product energy stories are under-published relative to successful laboratory claims. \emph{Platform churn:} RedCap, on-device SLM/LLM, and host-side flash stacks evolve faster than survey cycles; the instruments are meant to remain usable under that churn, not to freeze a 2026 device list. \emph{Thermal scope:} Section~\ref{sec:evaldim:thermal} and Table~\ref{tab:thermal_evidence} treat thermal behaviour as a constraint shared by all three pillars---sensing interface, compact model, throttling actuator, and sustained regime---rather than as a fourth hardware pillar with its own device stack and modeling-family survey; readers looking for die-level thermal architecture should consult the architectural thermal literature directly~\cite{skadron2004hotspot,huang2006hotspot,brooks2007power}.}

\section{Conclusion}
\label{sec:conclusion}

\rev{Four contributions should be read as the result of the survey, separate from the open-problem list. (1)~The coordination gap: sealed endpoints need a joint account of radio, compute, and storage because pillar-local joules often fail after integration (Section~\ref{sec:intro}, Table~\ref{tab:survey_compare}). (2)~An operational taxonomy: four branches with a dominant-share rule for mixed regimes (Table~\ref{tab:branch_decision}). (3)~Eight evaluation gates plus the three boundary names, so outcome, cost, and evidence are not collapsed into one percentage (Table~\ref{tab:eval_gates}). (4)~A reproducibility checklist and one-row reporting template for studies that cannot see every rail (Table~\ref{tab:report_template}). The phase-budget numbers are an illustration of that template, not a fifth contribution and not a measurement. We propose (branch, gate, family, coupling) as fields in that checklist, not as a standard.}

\section*{CRediT authorship contribution statement}
\textbf{Yuchao Chang:} Conceptualization, Methodology, Investigation, Writing -- original draft, Writing -- review \& editing, Project administration.
\textbf{Haoran Wei:} Investigation, Validation, Writing -- review \& editing.

\section*{Declaration of competing interest}
The authors declare that they have no known competing financial interests or personal relationships that could have appeared to influence the work reported in this paper.

\section*{Data availability}
No new datasets were generated or analyzed for this survey. All cited sources are publicly available through the publishers or repositories referenced in the bibliography. \rev{The two artifacts that support the classification---the category assignment of the 186-work core synthesis set (Table~\ref{tab:lit_corpus}) and the double-coded representative matrix (Table~\ref{tab:lit_matrix})---are printed in full in the manuscript and are also supplied as a machine-readable table with the submission. Raw database hit logs and item-level coding-disagreement logs were not archived during the review and therefore cannot be released; this is recorded in the Limitations section rather than reconstructed after the fact.}

\section*{Funding}
This research did not receive any specific grant from funding agencies in the public, commercial, or not-for-profit sectors.

\section*{Declaration of generative AI and AI-assisted technologies in the manuscript preparation process}
During the preparation of this work the author(s) used AI-assisted writing tools in order to improve organization and language. After using this tool/service, the author(s) reviewed and edited the content as needed and take(s) full responsibility for the content of the published article.

\section*{Acknowledgments}
The authors thank colleagues who provided informal feedback on early drafts. Any remaining errors are our own.

\bibliographystyle{elsarticle-num}
\bibliography{manuscript}

\section*{Vitae}
% Author biographies are optional for SUSCOM; retained for Editorial Manager if requested.

\noindent
\begin{minipage}[t]{0.22\textwidth}
\centering
\includegraphics[width=\linewidth,height=1.15in,keepaspectratio]{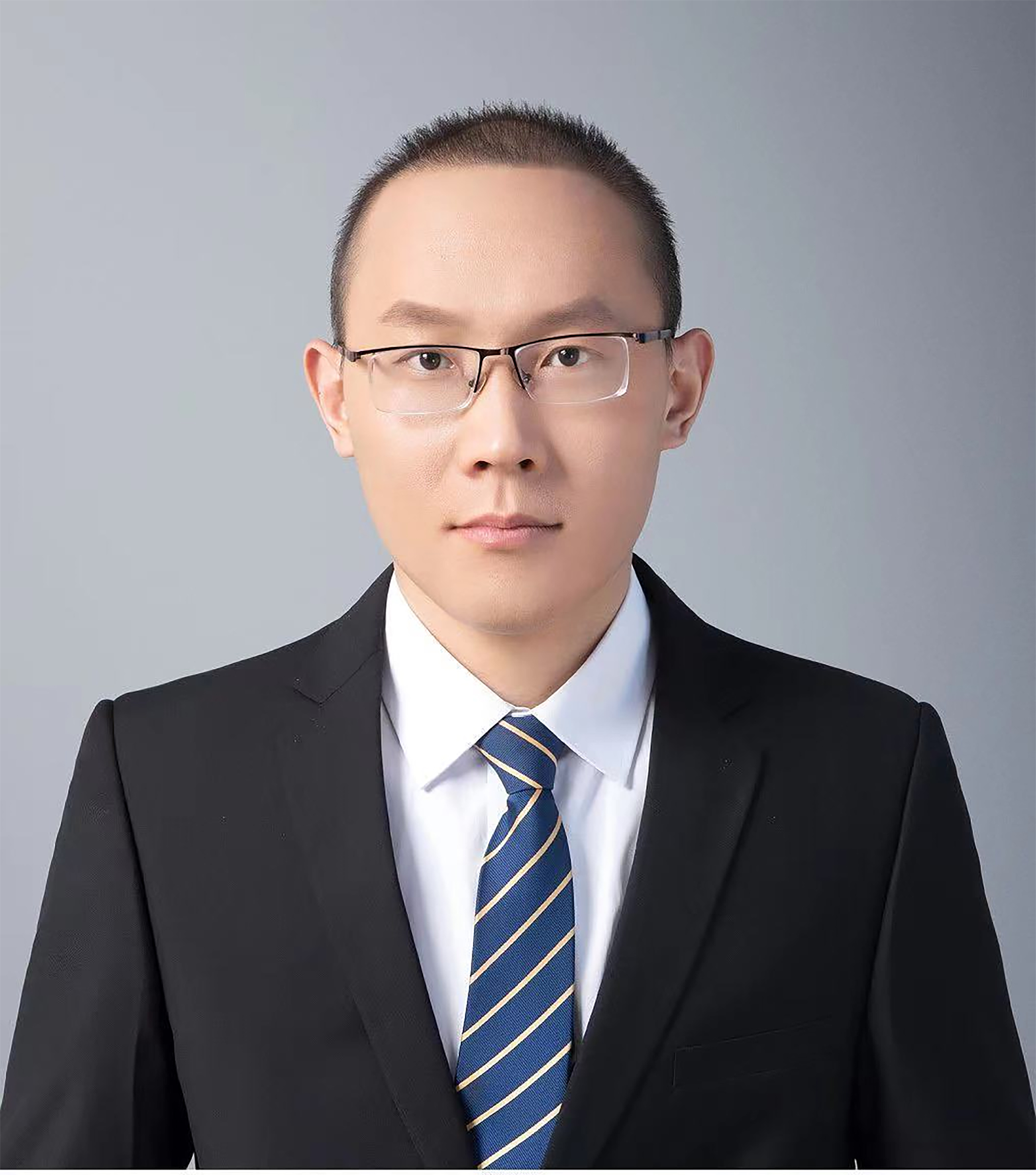}\\[-0.2em]
{\footnotesize Fig.~V1}
\end{minipage}\hfill
\begin{minipage}[t]{0.76\textwidth}
\textbf{Yuchao Chang} received the Ph.D.\ degree from the University of Chinese Academy of Sciences in 2019. He is with the Performance and Power Group, OPPO, Shanghai, China. He was a visiting Ph.D.\ student at the University of Texas at Dallas (2018--2019). His research interests include green communications and computing and AI-assisted optimization for electronic devices. He serves as an editor and guest editor for several journals.
\end{minipage}

\vspace{0.9em}
\noindent
\begin{minipage}[t]{0.22\textwidth}
\centering
\includegraphics[width=\linewidth,height=1.15in,keepaspectratio]{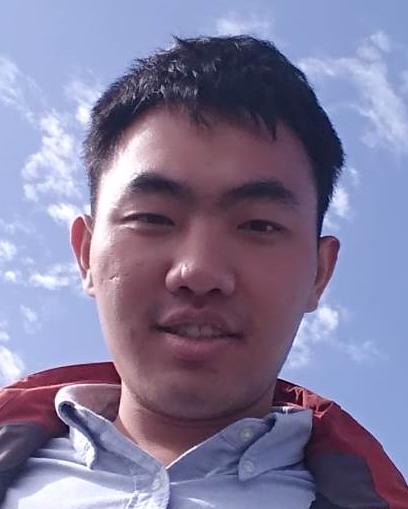}\\[-0.2em]
{\footnotesize Fig.~V2}
\end{minipage}\hfill
\begin{minipage}[t]{0.76\textwidth}
\textbf{Haoran Wei} received the Ph.D.\ degree in electrical engineering from the University of Texas at Dallas in 2020. He is a Staff Engineer at Rivian and Volkswagen Group Technologies, Palo Alto, USA. His research interests include multimodal signal processing and deep learning on mobile devices. He serves as an Associate Editor for the Journal of Real-Time Image Processing.
\end{minipage}

\end{document}